\documentclass{aa}  
\usepackage[usenames, dvipsnames]{color}
\usepackage{graphicx}
\usepackage{float}
\usepackage{placeins}
\usepackage{multicol}
\usepackage{subfigure} 
\usepackage{multirow} 
\usepackage{booktabs} 
\usepackage{longtable} 
\usepackage{txfonts}
\usepackage{amsmath}
\usepackage{hyperref}
\begin{document}

   \title{The X-shooter Spectral Library (XSL): Data Release 3\thanks{The spectra are available at the CDS via anonymous ftp to \url{cdsarc.u-strasbg.fr} (130.79.128.5) or via \url{http://cdsweb.u-strasbg.fr/cgi-bin/qcat?J/A+A/}.}}

   \subtitle{}
     
      \author{Kristiina Verro\inst{\ref{inst1}}
             \and S.~C. Trager\inst{\ref{inst1}}
            \and R.~F. Peletier\inst{\ref{inst1}}
            \and  A. Lan\c{c}on\inst{\ref{inst4}}
            \and  A. Gonneau\inst{\ref{inst2},\ref{inst4}}
            \and A. Vazdekis\inst{\ref{inst8},\ref{inst9}}
            \and \\P. Prugniel\inst{\ref{inst5}}
            \and Y.-P. Chen\inst{\ref{inst6}}
            \and P.~R.~T. Coelho\inst{\ref{inst11}}
            \and P. S{\'a}nchez-Bl{\'a}zquez\inst{\ref{inst10}}
            \and L. Martins\inst{\ref{inst12}}
            \and \\ A. Arentsen\inst{\ref{inst4}}
            \and M. Lyubenova\inst{\ref{inst3},\ref{inst1}}
            \and J. Falc{\'o}n-Barroso\inst{\ref{inst8},\ref{inst9}}
            \and M. Dries\inst{\ref{inst1}}
                }

\institute{Kapteyn Astronomical Institute, University of Groningen, Landleven 12, 9747 AD Groningen, the Netherlands \\
\email{verro@astro.rug.nl} \label{inst1}
    \and
    Observatoire Astronomique de Strasbourg, Universit\'e de Strasbourg, CNRS, UMR 7550, 11 rue de l'Universit\'e, \\F-67000 Strasbourg, France\label{inst4}
    \and
        Institute of Astronomy, University of Cambridge, Madingley Road, Cambridge CB3 0HA, United Kingdom \label{inst2}
        \and
    ESO, Karl-Schwarzschild-Str. 2, D-85748 Garching bei  M\"unchen, Germany \label{inst3}
    \and
    CRAL-Observatoire de Lyon, Universit\'e de Lyon, Lyon I,  CNRS, UMR5574, France \label{inst5}
    \and
    New York University Abu Dhabi, Abu Dhabi, P.O. Box 129188, Abu Dhabi, United Arab Emirates\label{inst6} 
    \and 
    Instituto de Astrof\'isica de Canarias, V\'ia L\'actea s/n, La Laguna, Tenerife, Spain\label{inst8}
    \and
    Departamento de Astrof\'isica, Universidad de La Laguna, E-38205 La Laguna, Tenerife, Spain\label{inst9}
    \and
    Departamento de F\'isica de la Tierra y Astrof\'isica, UCM, 28040 Madrid, Spain\label{inst10}
    \and
    Universidade de S\~{a}o Paulo, Instituto de Astronomia, Geof\'isica e Ci\^{e}ncias Atmosf\'ericas, Rua do Mat\~{a}o 1226, 05508-090, S\~{a}o Paulo, Brazil\label{inst11}
    \and
    NAT - Universidade Cidade de São Paulo, Rua Galvão Bueno, 868, São Paulo, Brazil\label{inst12}
    }

  \date{Received 7 October 2021; Accepted ?? ? ????}

 
\abstract{We present the third data release (DR3) of the X-shooter Spectral Library (XSL). This moderate-to-high resolution, near-ultraviolet-to-near-infrared (350--2480\,nm, $R\sim10\,000$) spectral library is composed of 830 stellar spectra of 683 stars. DR3 improves upon the previous data release by providing the combined de-reddened spectra of the three X-shooter segments over the full 350--2480\,nm wavelength range. It also includes additional 20 M-dwarf spectra from the ESO archive. We provide detailed comparisons between this library and Gaia EDR3, MILES, NGSL, CaT library, and (E-)IRTF. The normalised rms deviation is better than $D=0.05$ or 5\% for the majority of spectra in common between MILES (144 spectra of 180), NGSL (112/116), and (E-)IRTF (55/77) libraries. Comparing synthetic colours of those spectra reveals only negligible offsets and small rms scatter, such as the median offset(rms) $0.001\pm0.040$\,mag in the $(\mathrm{box1-box2})$ colour of the UVB arm, $-0.004\pm0.028$\,mag in $(\mathrm{box3-box4})$ of the VIS arm, and $-0.001\pm0.045$\,mag in $(\mathrm{box2-box3})$ colour between the UVB and VIS arms, when comparing stars in common with MILES. We also find an excellent agreement between the Gaia published $(\mathrm{BP-RP})$ colours and those measured from the XSL DR3 spectra, with a zero median offset and an rms scatter of 0.037\,mag for 449 non-variable stars. The unmatched characteristics of this library, which combine a relatively high resolution, a large number of stars, and an extended wavelength coverage, will help us to bridge the gap between the optical and the near-IR studies of intermediate and old stellar populations, and to probe low-mass stellar systems.}


   \keywords{catalogs --
                Hertzsprung-Russell and C-M diagrams --
                stars: general 
                }

   \maketitle
%

\section{Introduction}

Stellar spectral libraries are the cornerstones of stellar population synthesis models, which in turn are instrumental tools in studying the fundamental properties of unresolved stellar systems. The determination of properties such as the initial mass functions, star formation rates, star formation histories, total stellar masses, and stellar metallicities and abundance patterns of galaxies from their spectral energy distributions (SEDs) has advanced our understanding of galaxy formation and evolution tremendously \citep[see the reviews by][]{Tinsley1980, Renzini2006, Conroy2013}. 

\begin{table*}
    \caption{Main characteristics of some recent empirical libraries.}
    \centering
    \begin{tabular}{lrrrr}
    \hline\hline
         Empirical & \multicolumn{1}{c}{\# of spectra}  & \multicolumn{1}{c}{$\lambda$ start} & \multicolumn{1}{c}{$\lambda$ end} & \multicolumn{1}{c}{$\sim R$} \\
         libraries &  & \multicolumn{1}{c}{(nm)} & \multicolumn{1}{c}{(nm)} & \multicolumn{1}{c}{($\lambda / \Delta \lambda$)}\\
         \hline
         \textbf{XSL$^{1,2,3}$} & \textbf{830} & $\mathbf{350\phantom{.0}}$&$\mathbf{2\,480\phantom{.0}}$ & \textbf{10\,000}  \\
         MaStar$^{4}$ & 8\,646 & $362.2$&$1\,035.4$ & 1\,800\\
         LEMONY$^{5}$ & 1\,273 & $380\phantom{.0}$&$900\phantom{.0}$ & 1\,500 \\
         E-IRTF$^{6}$ & 284 &$700\phantom{.0}$&$2\,500\phantom{.0}$ & 2\,000 \\
         MILES$^{7,8}$ & 985 & $352.5$&$750\phantom{.0}$ & 2\,100 \\
         IRTF$^{9}$ & 210 & $800\phantom{.0}$&$2\,500\phantom{.0}$ & 2\,000 \\
         CO-library$^{10}$& 220 & $2\,110\phantom{.0}$&$2\,370\phantom{.0}$ & 2\,500 \\
         ELODIE$^{11,12,13}$ & 1\,962 & $389.2$&$680\phantom{.0}$& 10\,000 \\
         HST-NGSL$^{14}$ & 374 & $167.5$&$1\,025\phantom{.0}$ & 1\,000\\
         INDO-US$^{15}$ & 1\,273 & $346\phantom{.0}$&$946.4$ & 5\,000 \\
         STELIB$^{16}$ & 249 & $320\phantom{.0}$&$950\phantom{.0}$ & 1\,600 \\
         CaT$^{17}$ & 706 & $834.8$&$902\phantom{.0}$ & 6\,000 \\
         L\&W$^{18}$ & 182/142/108 & $500\phantom{.0}$&$2\,500\phantom{.0}$ & 150/1\,100 \\
         \hline
         
    \end{tabular}
    \label{tab:overview_libraries}
    \tablebib{ (1) \citet{Chen2014}; (2) \citet{DR2}; (3) this paper; (4)\citet{MaStar2019}; (5) \citet{LEMONY}; (6) \citet{EIRTF}; (7) \citet{MILES1}; (8) \citet{MILES}; (9) \citet{IRTF}; (10) \citet{COlib}; (11) \citet{ELODIE2001}; (12) \citet{ELODIE2004}; (13) \citet{ELODIE2007}, ELODIE is also available at $R=42\,000$, with the flux normalised to the pseudo-continuum; (14) \citet{NGSL2006}; (15) \citet{INDOUS2004}; (16) \citet{STELIB2003}; (17) \citet{CaT}; (18) \citet{LW2000}, $\lambda$ coverage for optical, NIR and combined spectra respectively, with lower optical resolution.}
\end{table*}

A suitable empirical stellar library needs multiple goals to be met \citep{Trager2014}. The library should cover all phases of stellar evolution at all masses at all metallicities (and ideally all abundance ratios); the spectral library should cover a broad wavelength range at moderate-to-high spectral resolutions, as different evolutionary phases contribute to different wavelengths of the population model and high resolution is needed to study the smallest galaxies; observations across the wavelength range need to be simultaneous to account for stellar variability; the spectra need careful flux and wavelength calibration and accurate stellar atmospheric parameters. A variety of empirical stellar spectral libraries are publicly available, some of which are listed in Table \ref{tab:overview_libraries}.  

These goals are not straightforward to reach, as they combine state-of-the-art observational techniques, as well as a comprehensive theoretical understanding of stellar SEDs and evolutionary phases. Empirical libraries cannot cover the Hertzsprung-Russell (HR) diagram as extensively as theoretical libraries do, and they are limited by resolution and wavelength range. Observational problems, such as flux calibration issues or telluric line contamination, can be minimised, although they are always present. A possibility is to use theoretical stellar spectra, which can be calculated for an arbitrary set of stellar parameters with high spectral resolution, for a wide wavelength range, and which are without observational problems. However, theoretical models often have many approximations and simplifications, such as assuming local thermodynamic equilibrium and plane-parallel geometry. Theoretical libraries can perform comparably well to the empirical libraries when modelling integrated spectra of stellar populations in the ultraviolet and optical, especially when longer wavelength ranges and the continuum shape are important \citep{Martins2019, Coelho2020}. Theoretical models have difficulties reproducing the observed features in stellar spectra due to incomplete atomic line lists. This is worse for cooler stars due to incomplete molecular line lists \citep{MartinsCoelho2007, Kurucz2011, Coelho2014, Knowles2019, Coelho2020,Lancon2021}, and thus theoretical cool star spectra cannot be used at present to make accurate predictions for absorption line indices in the near-infrared (NIR). NIR empirical libraries are therefore still superior to theoretical libraries at the present time. An empirical stellar spectrum shows the star as it exists in the real Universe. Issues arising from incomplete line lists or theoretical approximations are absent, although stellar parameter estimation is always model dependent. 

The next step in fulfilling the goals discussed above is the X-shooter Spectral Library \citep[XSL:][]{Chen2014,DR2}. XSL is a near-ultraviolet (NUV) to NIR moderate-resolution ($R\sim10\,000$) spectral library, which aims to cover the entire HR diagram as uniformly as possible, with an emphasis on cool stars. The relatively high resolution makes it possible to study the kinematics of low-mass systems and discs of galaxies with the latest generation of high spectral resolution integral-field spectrographs, such as VIRUS-W \citep{VIRUSW1, VIRUSW2}, MEGARA \citep{MEGARA,MEGARA2}, and WEAVE \citep[Jin et al., in prep.]{WEAVE1}.
The extended wavelength to the NIR range is particularly valuable. The need for good empirical stellar libraries in this wavelength region is imperative, due to the recent advances in the infrared instruments on large telescopes such as X-shooter \citep{Vernet2011} and KMOS \citep{KMOS1, KMOS2}, which have made it possible to extend the stellar population studies of unresolved galaxies into the NIR. Furthermore, the ELT and JWST will focus on the NIR. The combination of a relatively high resolution, the large number of stars, and the extended wavelength coverage of the XSL will help us to join  optical and  NIR studies of intermediate and old stellar populations. XSL is aimed at becoming a benchmark stellar library in the optical and NIR.
All XSL data releases are available also on our website\footnote{\url{http://xsl.astro.unistra.fr}}. 

This paper is structured as follows. Section \ref{Sec:content} reviews the previous data releases and describes the content of this new data release. The process of adding 20 additional M-dwarf spectra to XSL and the determination of their stellar parameters are described in Sect.~\ref{Sec:Mdwarfs}. Section \ref{Sec:merging} gives details about how we combine the three XSL spectral segments to produce spectra with the full X-shooter wavelength range. Section \ref{Sec:flux_loss} gives details on the re-calibration of the spectral shape of some of the XSL stars, which were not corrected for flux loss in XSL Data Release 2 \citep[hereafter DR2]{DR2}. We make relevant quality assurance comparisons in Sects. \ref{Sec:interpolated_lib} and \ref{Sec:literature}.
\section{DR3 content}
\label{Sec:content}

\begin{figure}
    \centering
    \includegraphics[width = 0.5\textwidth]{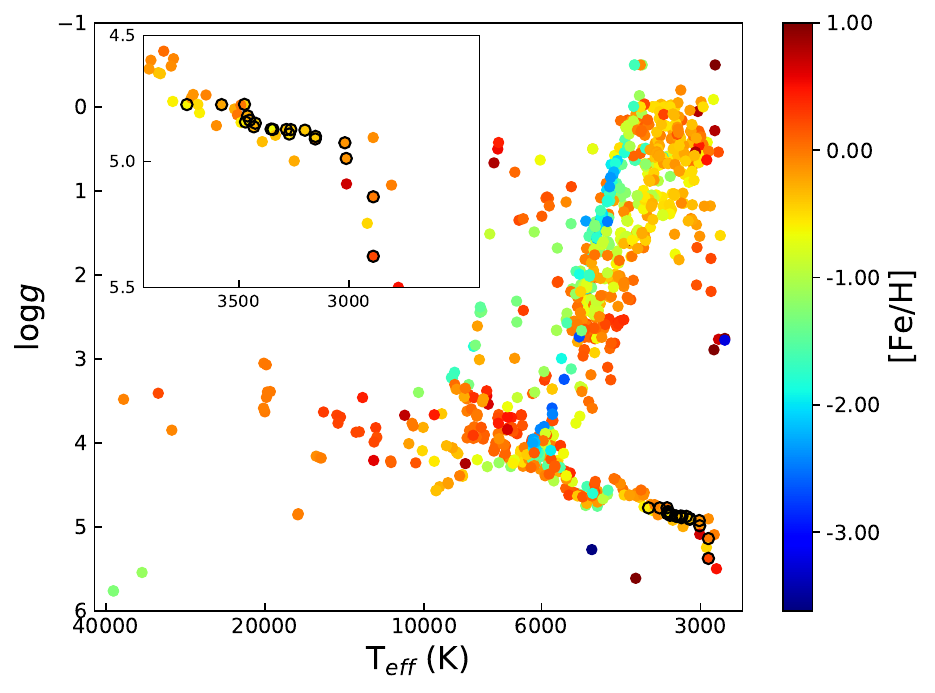}
    \caption{Stellar atmospheric parameters, compiled from \citet[][]{Arentsen2019, Gonneau2017}; this study (the new M dwarfs) and in a few cases, other literature sources. The new DR3 M-dwarf stars are marked with open symbols.}\label{fig:MainHRD}
\end{figure}
\begin{figure}
    \centering
    \includegraphics[width = 0.45\textwidth]{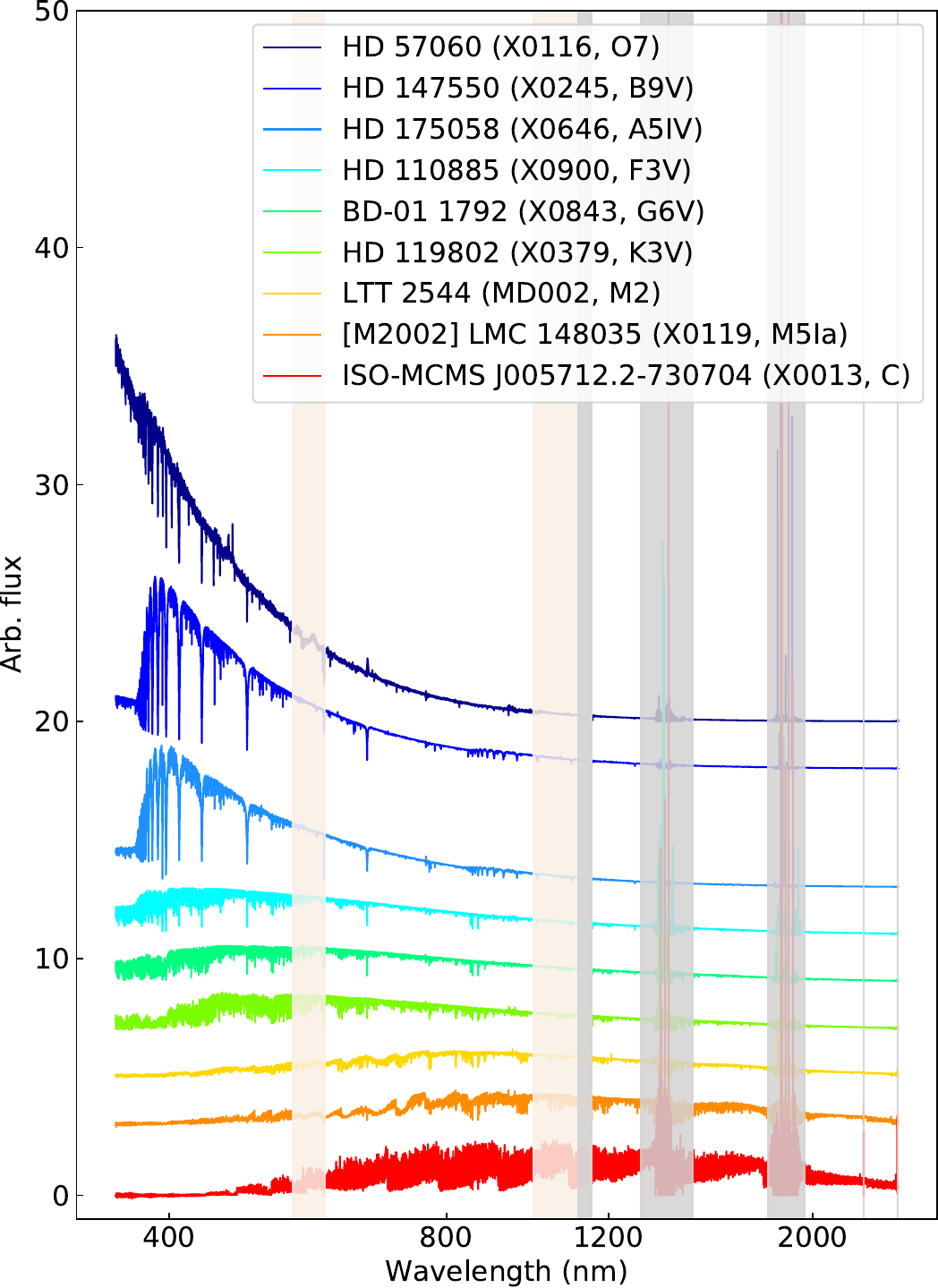}
    \caption{Typical XSL spectra in the OBAFGKMC sequence. Shaded bands mask the regions of dichroic contamination and deepest telluric regions in the NIR arms.}
    \label{fig:OBAFGKM}
\end{figure}

In this third XSL Data Release (DR3), we provide 830 spectra of 683 stars, corrected for Galactic extinction and merged to a full wavelength range of 350--2480\,nm. The DR3 dataset consists of DR2 spectra and the spectra of 20 archival M-dwarf stars. It should be noted that 82 spectra (9\% of DR3) are not corrected for Galactic dust extinction, mostly because 138 spectra (17\% of the XSL DR3) are not corrected for slit flux losses. A summary of the content and associated data of DR3 is shown in Table \ref{tab:overview_dr3}, the HR diagram is shown in Fig.~\ref{fig:MainHRD}, comments on some individual spectra are given Appendix \ref{app:comments_stars}, and the DR3 header keyword dictionary is given in Table \ref{table:key_dico}. A few examples of DR3 spectra can be found in Fig.~\ref{fig:OBAFGKM}.

We would like to give a word of caution: between 545 and 590\,nm and to a lesser extent 994 and 1150\,nm, the spectra suffer from artefacts from known instabilities of the transmission of the dichroic plates that separate the arms of the X-shooter spectrograph, and therefore they have poor flux calibration. DR3 spectra are given in the rest frame, so the exact location of the dichroic contamination varies with radial velocity. We give a description in Sect.~\ref{Sec:merging}.

\begin{table}
    \centering
    \caption{Summary of DR3 content and associated data.}
    \label{tab:overview_dr3}
    \begin{tabular*}{0.45\textwidth}{lr}
    \hline\hline
          \multicolumn{1}{c}{Source or issue}& \multicolumn{1}{c}{\# of spectra} \\
         \hline
        DR2 & 813 \\
        ESO archive & 20 \\
        Parameters from \citet{Arentsen2019} & 754 \\
        C-star param. from \citet{Gonneau2017} & 35 \\
        \hline
        Missing NIR & 27 \\
        Missing UVB & 20 \\
        Not corrected for flux loss & 138 \\
        Not de-reddened & 85\\
         \hline
    \end{tabular*}

\end{table}

\subsection{Previous XSL data products}

The observations were carried out with the X-shooter three-arm spectrograph on ESO’s VLT \citep{Vernet2011} in two phases, a pilot study and a Large Programme, spanning six semesters in total. The  XSL  target  stars  were  selected  to  cover as much of the HR diagram as possible, with a wide range of chemical compositions. The sources for the XSL target list were   existing spectral libraries and the first-generation PASTEL catalogue \citep{PASTEL2010}. The XSL sample has a strong overlap with the MILES spectral library \citep[][]{MILES1, MILES}, NGSL \citep[][]{NGSL2006}, and the ELODIE library \citep[][]{ELODIE2001,ELODIE2004, ELODIE2007}.  
We observed nearly\footnote{The very brightest stars observed in the beginning of the XSL pilot programme did not have wide-slit observations taken due to exposure-length constraints.} every target with a narrow slit and wide slit. We first took the narrow-slit spectrum to achieve the required spectral resolution. The narrow-slit widths for the UVB, VIS, and NIR arms of X-shooter were $0\farcs5$, $0\farcs7$, and $0\farcs6$, resulting in resolving powers of $R=9\,793$, 11\,573 and 7\,956, respectively. We took wide-slit spectra of the targets immediately after the narrow-slit exposure, using a $5\arcsec$ wide slit, to flux calibrate the spectra.

In Data Release 1, \citet[hereafter DR1]{Chen2014} presented 246 spectra of 237 unique stars, observed during the pilot programme, over the wavelength range of the two optical arms of X-shooter (300--1024\,nm). 

Data Release 2 provided 813 spectra of 666 unique stars. The DR2 spectra are available in three spectral ranges, corresponding to the three arms of the spectrograph:
UVB, 300--556\,nm; VIS, 533--1020\,nm; and NIR, 994--2480\,nm. The DR2 data are homogeneously reduced and calibrated. Of the DR2 spectra, 85\% are flux calibrated; that is, they are corrected for all stable transmission factors of the acquisition chain and for wavelength-dependent losses due to the narrow slits used in the observations. The remaining 15\% of the DR2 spectra are not corrected for slit losses, because the matching wide-slit observations were faulty. DR2 spectra are corrected for atmospheric extinction and telluric absorption, as well as for radial velocity, and are provided in the rest frame (with wavelengths in air). For warmer stars ($T_\mathrm{eff}> 5000\,\mathrm{K}$), typical S/N values are about 70, 90 and 96 for the UVB, VIS, and NIR arms, respectively. The S/N for the cool stars ($T_\mathrm{eff}< 5000\,\mathrm{K}$) varies between 0 and 100 with wavelength across the UVB arm. The resolving power of each XSL spectra is constant with wavelength within one spectral segment. Systematic differences in the SEDs of \citet{DR2} and previous stellar spectral libraries are small. For example, the average differences between the synthetic colours measured on XSL spectra and on spectra from MILES or (E-)IRTF are below 1\%.

Furthermore, \cite{Arentsen2019} provided a uniform set of stellar atmospheric parameters, effective temperatures, surface gravities, and iron abundances for 754 spectra of 616 XSL stars. \citet[hereafter L20,]{Lancon2021} quantifies the match between the XSL spectra and the PHOENIX theoretical spectra of \citet{Husser2013}, to study the systematic differences between empirical and theoretical spectra.

\subsection{Additional M-dwarf stars}

The stars present in XSL DR2 do not provide sufficient coverage of main-sequence M dwarfs for stellar population synthesis purposes. On one hand, the use of polynomial interpolators in stellar population modelling \citep[e.g.][]{ELODIE2001, Koleva2009,Prugniel2011,Wu2011, Verro2021SPOP} relies on good coverage of the stellar library; on the other hand, \citet{Coelho2020} have shown that poor parameter coverage affects the predicted colours and determined galaxy ages. In DR3, XSL has been extended with 20 M dwarfs from the ESO X-shooter archive. These stars were used in creating the MIX stellar library (a mixture of MILES and XSL) of \citet{Dries2018} and were selected from the ESO programs 385.D-0200 (Chromospheric structure of low-mass stars, PI A. Reiners), 088.D-0556, and 092.D-0300 (Observations of M dwarf secondaries, PI V. Neves). \citet{Dries2018} only selected spectra for which literature values for their effective temperature and their surface gravity were available at the time (mid 2017). At the time of the publication of the current paper, nearly all of these stars have already been characterised by \citet{Kuznetsov2019}. Their X-shooter M-dwarf dataset has 153 stars, significantly more than the selected 20. The rest of the spectra have potential to become part of XSL, but this is beyond the scope of the current data release. We reduced and calibrated the 20 M-dwarf spectra selected by \citet{Dries2018} with the same procedures as the DR2 stars \citep[see the description in Sect. 3 of][]{DR2}. Due to having only narrow-slit observations, the flux-loss correction is done with a cubic spline function, as described in Sect.~\ref{Sec:flux_loss}.
We determined the stellar parameters for these stars in the same way as \citet{Arentsen2019} for the majority of XSL stars. These new stars are listed in Table \ref{tab:MD_parameters}, shown separately in Fig.~\ref{fig:MainHRD}, and described further in Sect.~\ref{Sec:Mdwarfs}.

\subsection{XSL over the full X-shooter wavelength range}
The DR2 spectra consist of three segments observed simultaneously. Combining these three segments is not a trivial task. Spectra between 545--556\,nm in the UVB, 556--590\,nm in the VIS, and 994--1150\,nm in the NIR suffer from artefacts from known instabilities of the transmission of the dichroic plates that separate the UVB and VIS arms, and the VIS and NIR arms of the  X-shooter spectrograph. In addition, the DR2 relative scaling of arm spectra is sometimes inadequate: the UVB and NIR spectra need scaling factors relative to the VIS spectrum. This is due to arm-dependent flux losses and arm-dependent sky-subtraction issues. We determined scaling factors for the UVB and NIR spectra relative to the VIS spectrum ($S_\mathrm{UVB}$ and $S_\mathrm{NIR}$) to enforce a smooth transition from the UVB to the VIS spectrum and the VIS to the NIR spectrum. This process, during which the amount of dust extinction is also determined, is described in Sect.~\ref{Sec:merging}.

\section{Stellar parameters and radial velocities of M dwarfs}
\label{Sec:Mdwarfs}
\subsection{Stellar parameters}

 We used the full spectrum fitting package ULySS \citep{Koleva2009,Prugniel2011, Sharma2016} equipped with the MILES stellar library to fit the XSL, UVB, and VIS spectra for stellar parameters, effective temperature ($T_\mathrm{eff}$), surface gravity ($\log g$, where $g$ is in $\mathrm{cm\,s^{-2}}$), and metallicities ($[\mathrm{Fe} / \mathrm{H}]_\mathrm{e}$), similarly to what was done in \cite{Arentsen2019} for other XSL stars.
 
 To be exact, we measure an equivalent metallicity, $[\mathrm{Fe} / \mathrm{H}]_\mathrm{e}$, which is the $[\mathrm{Fe} / \mathrm{H}]$ of the reference spectrum that best fits our target\footnote{We followed the equivalent metallicity term and the $[\mathrm{Fe} / \mathrm{H}]_\mathrm{e}$ labelling suggested by \citet{Baratella:submitted}.}. If the reference spectrum has a different abundance pattern than the target, $[\mathrm{Fe} / \mathrm{H}]_\mathrm{e}$ will be biased with respect to $[\mathrm{Fe} / \mathrm{H}]$. However, XSL, as well as the reference library MILES, are mainly composed of stars from the local neighbourhood and therefore follow the same abundance trend (alpha-enhanced at low metallicity). Therefore, when we use the MILES interpolator to fit XSL spectra, we do not have an average $[\mathrm{Fe} / \mathrm{H}]_\mathrm{e}$ bias, and $[\mathrm{Fe} / \mathrm{H}]$ is close to $[\mathrm{Fe} / \mathrm{H}]_\mathrm{e}$. The $[\mathrm{Fe} / \mathrm{H}]$ of \cite{Arentsen2019} used throughout this paper are also equivalent metallicities, $[\mathrm{Fe} / \mathrm{H}]_\mathrm{e}$.
 
 As in \cite{Arentsen2019}, we used the VIS solution alone if $T_\mathrm{eff}\mathrm{(UVB)} < 3800\,\mathrm{K}$, otherwise the weighted average of the UVB and VIS estimates we adopt as stellar parameters, weighted by the inverse of the square of the internal errors. Table \ref{tab:MD_parameters} lists the inferred stellar parameters. The errors provided by ULySS, with values of a few to a few tens of kelvins, describe internal errors only, and do not encompass any larger systematic errors.

\begin{figure}
    \centering
    \includegraphics[width = 0.5\textwidth]{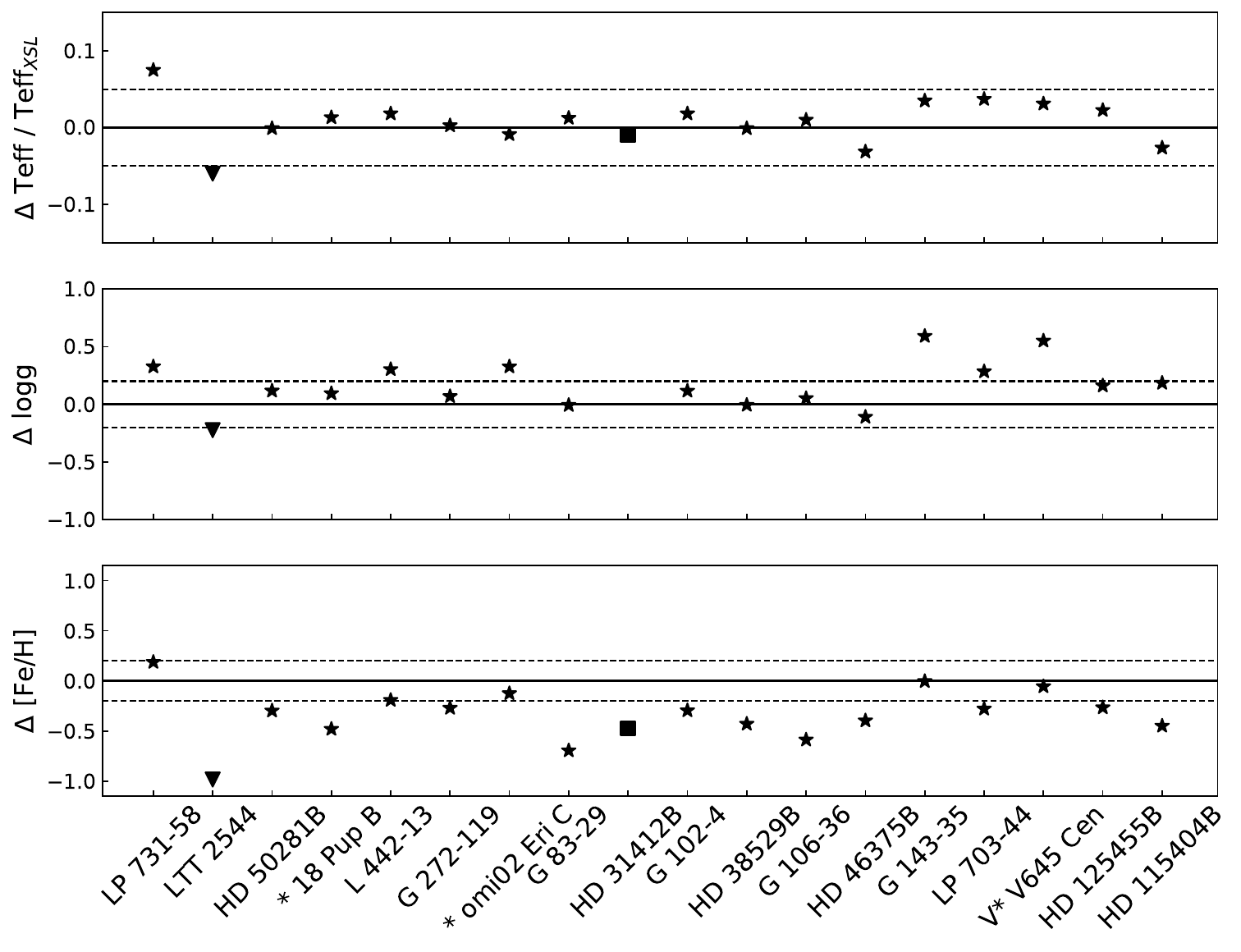}
    \caption{Estimated stellar parameter comparison with literature values ($ \Delta X = X_\mathrm{XSL} - X_\mathrm{lit}$) for the archival M dwarfs. The star symbols show data from \cite{Kuznetsov2019}, filled squares from \cite{Terrien2015}, and triangles from \cite{RAVE2017}. GJ 768.1 B and LP 659$-$4 do not have literature parameters. The dashed lines represent 5\%, 0.2\,dex, and 0.2\,dex differences in $T_\mathrm{eff}$, $\log g$, and $[\mathrm{Fe} / \mathrm{H}],$ respectively.}
    \label{fig:MDLiterature}
\end{figure}

We suggest considering the differences between the literature stellar parameters and those determined here as a rough but realistic error estimate, even though there could be systematic differences between them. Stellar parameters are available for the majority of these M dwarfs in \citet{Kuznetsov2019}. We used the values from the global 586--844\,nm fit of their study. We used the \citet{Terrien2015} stellar parameters for HD 31412B and the \citet{RAVE2017} parameters for LTT 2544. As seen from Fig.~\ref{fig:MDLiterature}, our temperature estimates are in excellent agreement with the literature values, with differences below 5\% for the majority. The surface gravities are also in good agreement. \citet{Kuznetsov2019} estimated their uncertainty to be about 0.2\,dex, which is shown with the dashed line in Fig.~\ref{fig:MDLiterature}. Most of our estimates are within this error.

Metallicity estimation is more problematic, with systematic differences between our estimates and the literature values. This bias may arise from our measurements using the MILES interpolator, although \citet{Sharma2016} did improve the MILES interpolator for the cool dwarf stars with additional FEROS spectra.

As seen in Fig.~\ref{fig:MainHRD}, our parameter estimations place these 20 M-dwarf stars on the same narrow cool dwarf sequence as traced by the \citet{Arentsen2019} parameters. Our methods are nearly identical but both are potentially biased towards the \citet{Sharma2016} stellar parameters. L20 shows that when using the PHOENIX theoretical spectra to determine stellar parameters, the cooler ($<6000\,\mathrm{K}$) dwarf sequence of the HR diagram becomes significantly broader in $\log g$. However, they believe that these systematic trends reflect the differences between the observed stars and the PHOENIX models and do not interpret the width of the sequence as being physical.

The determination of the physical properties of M dwarfs is challenging and depends heavily on the methods used. For the sake of consistency with the rest of XSL, we prefer our estimates over the literature values.

\subsection{Correction to rest-frame velocity}

The barycentric radial velocities, $cz$, are estimated separately in each of the three arms. The separate arm estimation is necessary because excess differences of $10\,\mathrm{km\,s^{-1}}$ between arms exist. While ULySS fitting provided stellar parameters for these stars, the underlying MILES library has a wavelength range limited to the XSL UVB and part of VIS. Hence, it cannot be used to determine the radial velocity of the NIR arm. We used ULySS solutions for UVB and VIS arms and the Penalized PiXelFitting (pPXF) method \citep{Cappellari2004,Cappellari2017}, with the PHOENIX library as templates, for the NIR spectra. We used the following grid for the PHOENIX templates:\\ \mbox{$T_\mathrm{eff} \in [2900, 3000, 3100, 3200, 3300, 3400]\,\mathrm{K}$}, \mbox{$\log g \in [4.0, 4.5, 5.0] \log(\mathrm{cm\,s^{-2}})$} and  \mbox{$[\mathrm{Fe} / \mathrm{H}]\in[0.0, 0.5]\,\mathrm{dex}$}.

Using two methods to determine $cz$ allows us to estimate the uncertainty in the measurements by comparing pPXF values of VIS and UVB with those determined from ULySS fitting used in parameter estimation: the average difference is $0.5\,\mathrm{km\,s^{-1}}$ in UVB and $3\,\mathrm{km\,s^{-1}}$ in VIS. These values can be taken as representative of the uncertainties in the rest-frame wavelengths of the M dwarfs.

\section{Merging the XSL spectra}
\label{Sec:merging}

\begin{figure}
\centering     
\subfigure[]{\includegraphics[width=0.5\textwidth]{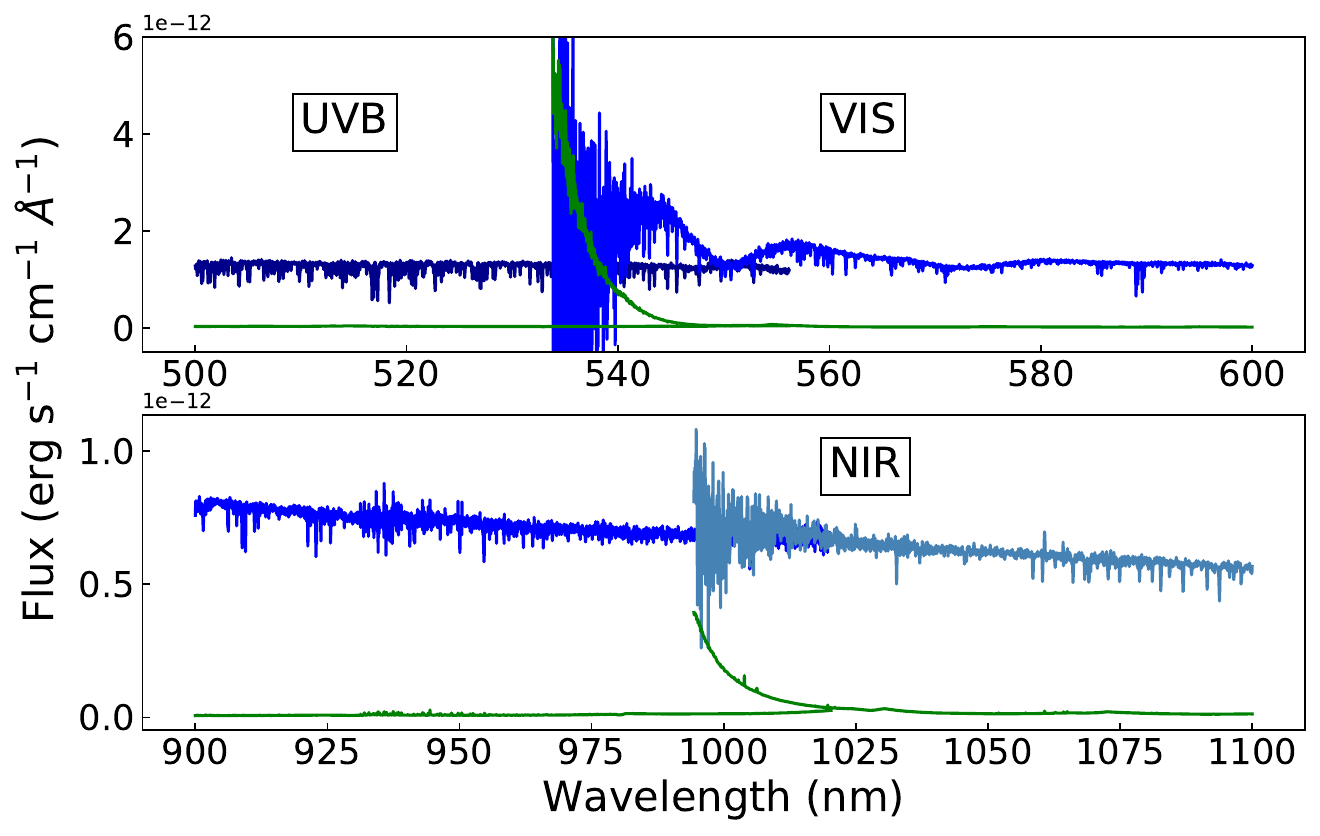}}
\subfigure[]{\includegraphics[width=0.5\textwidth]{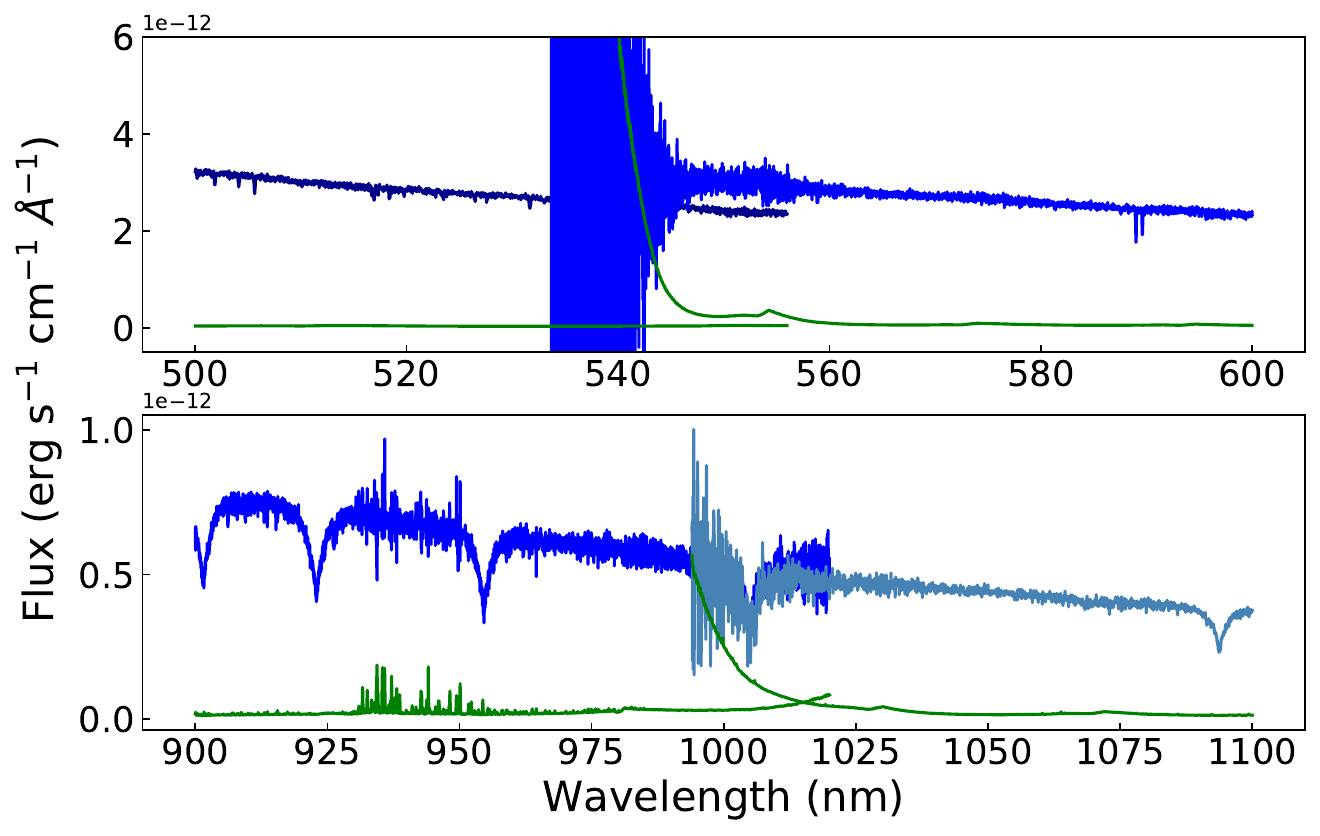}}
\caption{UVB--VIS and VIS--NIR overlap regions of XSL stars (a) HD 6229 (X0007, $T_\mathrm{eff} = 5055\,\mathrm{K}$, $\log g = 2.1$, $[\mathrm{Fe} / \mathrm{H}]=-1.2$) and (b) HD 167946 (X0460, $T_\mathrm{eff} = 10080\,\mathrm{K}$, $\log g = 4.1$, $[\mathrm{Fe} / \mathrm{H}]=-0.5$), together with the error spectra (in green).}
    \label{fig:merging}
\end{figure}

\begin{figure}
\centering     
\includegraphics[width=0.5\textwidth]{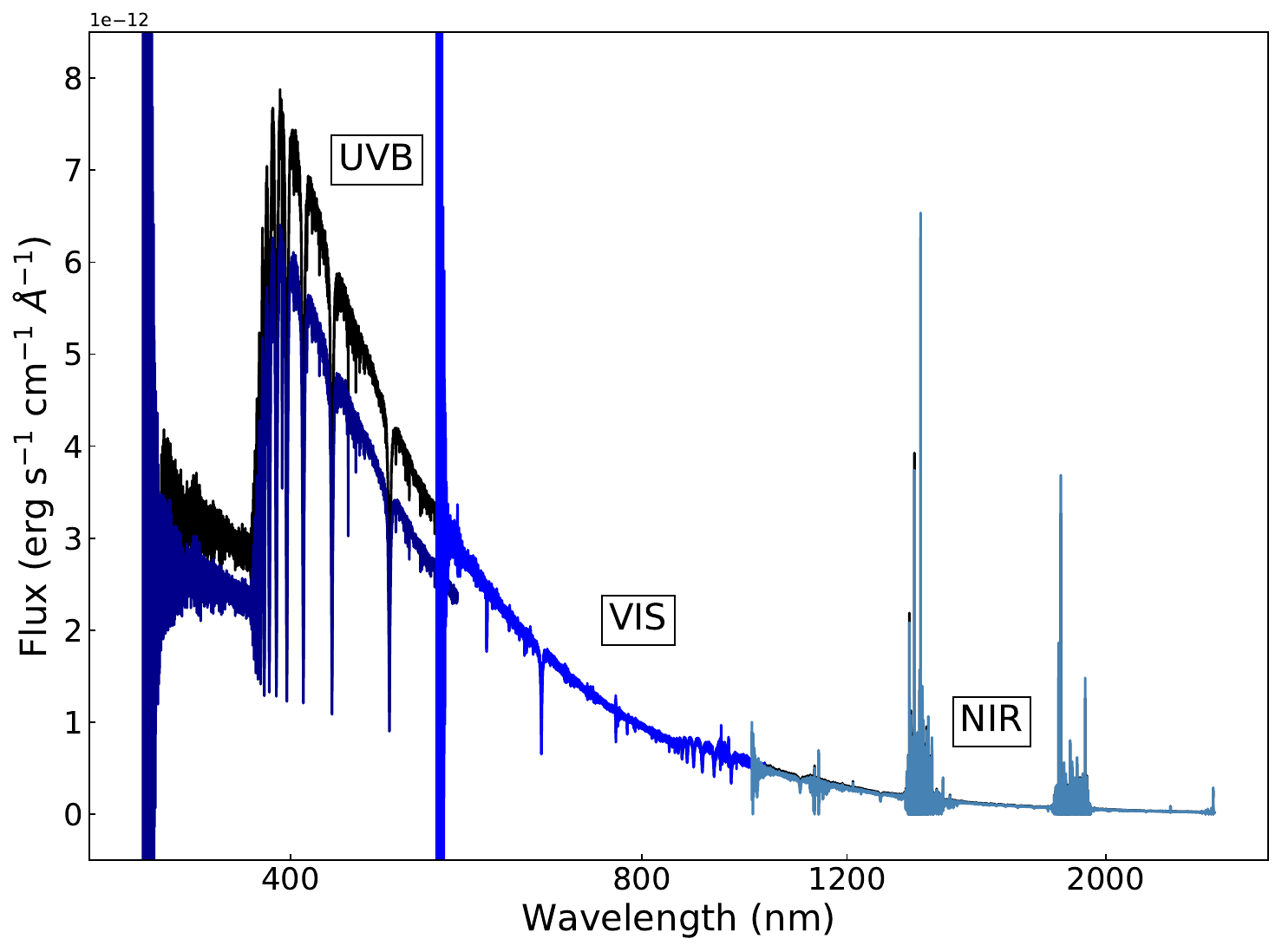}
\caption{DR2 UVB, VIS, and NIR spectra (in shades of blue) of XSL star HD 167946 (X0460) compared to the DR3 spectrum (black: before de-reddening, presented in log scale in wavelength).}
    \label{fig:full_merging}
\end{figure}

The relative scaling of arm spectra determined in DR2 is at times inadequate, due to arm-dependent flux losses and arm-dependent sky-subtraction issues. Merging them without any calibration would result in a unnatural SED, as illustrated in Figs.~\ref{fig:merging} and \ref{fig:full_merging}. Our solution is to re-calibrate the individual DR2 spectra with the use of scaling factors to scale the UVB and NIR arm spectra relative to the VIS arm. Here, we describe how the scaling factors are found, and how in doing so, we also find the amount of dust extinction. The stars in the XSL are located in a variety of environments: in the solar neighbourhood, in star clusters, in the Galactic bulge, and in the Magellanic Clouds. Hence, they can have a high degree of reddening in their spectrum due to extinction by interstellar matter. Interstellar extinction results in a change in the shape of an observed spectrum and should be corrected for.

\subsection{Scaling factors for UVB and NIR spectra and dust extinction}
Determining the scaling factors for UVB and NIR spectra relative to the VIS spectrum ($S_\mathrm{UVB}$ and $S_\mathrm{NIR}$), together with Galactic dust extinction (described by $A_V$), is a three-step process.
\label{Sec:fitting}
\subsubsection{Initial determination of the scaling factors}
Firstly, we find the initial $S_\mathrm{UVB}$ and $S_\mathrm{NIR}$ values by visual inspection. We scale the UVB and NIR spectra relative to the VIS until the continuum shapes match. Due to the dichroic contamination and flux losses at the edges of the spectral arms, visual matching is difficult. We inspect the whole spectrum for this purpose, zooming in and out as necessary. We mark the boundaries of the scaling factor inside of which we cannot visually better determine the match. The actual visual scaling factor is the average value inside the boundaries, and the boundaries themselves serve as a measure of trust in the visual scaling factor.
\subsubsection{Initial determination of $A_V$}
Secondly, we measure the initial values for the Galactic dust extinction for as many XSL spectra as possible. We do this by comparing the XSL spectrum to a theoretical stellar spectrum with similar parameters and by finding the best fit to the \citet{Cardelli1989} extinction law. We use the PHOENIX theoretical spectra as templates for $T_\mathrm{eff} < 12~000\,\mathrm{K}$ stars and the spectra generated using Tlusty BSTAR2006 atmosphere models \citep{Tlusty2007} for hotter stars. The template grid varies with temperature regime: the PHOENIX models have steps of $100\,\mathrm{K}$ and steps of $200\,\mathrm{K}$ for models $>7000\,\mathrm{K}$, while the Tlusty models switch from a step of $500\,\mathrm{K}$ to a step of $1000\,\mathrm{K}$ at $15\,000\,\mathrm{K}$. We do not perform any interpolation and we mask out the telluric and the dichroic contamination areas. L20 showed that $A_V\mathrm{(VIS)}$ is a better estimate of extinction than $A_V\mathrm{(UVB)}$, both of which are better than $A_V\mathrm{(NIR)}$. Taking into account these considerations, we use a 400--1340\,nm section of the spectrum to estimate the Galactic dust extinction.

\subsubsection{The spectral interpolator}
\label{Sec:interpolator}
Thirdly, we create an XSL spectral interpolator using the de-reddened merged spectra and the stellar parameters from \cite{Arentsen2019}. The interpolator allows us to compare an observed spectrum with an interpolated spectrum of the same stellar parameters.

An interpolator creates a synthetic spectrum at a given set of parameters (e.g.\ $T_\mathrm{eff}$, $\log g,$ and $[\mathrm{Fe} / \mathrm{H}]$) from a library of empirical or theoretical spectra using either neighbouring spectra (in which case the interpolator is called a `local' interpolator) or some some substantial fractions, or perhaps all of the entire library (in which case the interpolator is called a `global' interpolator). A local interpolator \citep[e.g.][]{Vazdekis2003, Sharma2016,Dries2018} interpolates spectra using its local neighbourhood: library stars in the vicinity of the point for which we want to create a spectrum are weighted and combined to create a representative spectrum for that point. A global interpolator \citep[e.g.][]{Prugniel2001, Koleva2009,Prugniel2011, Wu2011} fits polynomials of $T_\mathrm{eff}$, $\log g,$ and $[\mathrm{Fe} / \mathrm{H}]$ at each wavelength point to the whole or a large subset of the spectra in the library. Here, we use a combination of the two.

The combination of the global and the local interpolator uses the best characteristics of each in areas of the HR diagram where most desirable. For example, XSL has 40 M-dwarf stars and 40 hot stars with $T_\mathrm{eff}>10\,000\,\mathrm{K}$; due to the use of polynomials, a global interpolator would produce unreliable spectra in these regions without extrapolation support (with added theoretical or semi-empirical spectra). As the local interpolator averages stellar spectra in a box of parameters around the desired point, it works better in lower density regions of the HR diagram. However, a local interpolator is dependent on a smaller number of stars than a global interpolator, and hence issues in the spectra or in the parameter estimation of single stars are more likely to have an impact on the output spectrum, although the weight of these can be manually lowered. We ignore the metallicity parameter for stars $T_\mathrm{eff}$ < 4000~K and $T_\mathrm{eff}$ > 9000~K. In this range, the uncertainty in the metallicity is relatively large  \citep{Arentsen2019} and the coverage of the XSL is low.

A global interpolator is used with warm stars (4200--7000\,K), of which we have many, so individual stars have less weight and the output spectrum is less affected by problems in individual library spectra. A global interpolator assumes a smooth evolution of stellar spectra with stellar parameters, which is true on the main sequence and on the red giant branch, and thus it produces smooth sequences of stellar spectra along an isochrone. An overlap region between the different interpolation schemes, where spectra are generated by both methods and linearly combined, ensures a smooth transition from local to global interpolation scheme and vice versa. The exact parameter regions are given in Table \ref{tab:interp_parmaeter_ranges}. 

The description of XSL based spectral interpolator can be found in \citep{Verro2021SPOP}. The interpolator used here differs from the one described in \citep{Verro2021SPOP} by how we handle the very cool giant stars. In \citep{Verro2021SPOP}, the cool giants, ($T_\mathrm{eff} < 4000~K$) are not included in the global/local interpolation scheme, but they are incorporated in the stellar population models by using average spectra of `static' giant/O-rich TP-AGB/C-rich TP-AGB stars, binned by broad-band colour, and relying on empirical relations to dictate where an average spectrum of a star of a certain colour should occur. Here, we rely on the stellar parameter estimation of \citet{Arentsen2019} to include as many cool giants into the local interpolation scheme as possible.

\begin{table}
    \centering
    \caption[]{Parameter ranges for the type of interpolator used. }

    \begin{tabular*}{0.45\textwidth}{lc}
    \hline\hline
         Parameter region & Type of interpolator \\
         \hline
        $T_\mathrm{eff} < 4000\,\mathrm{K}$ & local \\
        $4000 < T_\mathrm{eff} < 4200/4500\,\mathrm{K}$ & both \\
        $4200\,\mathrm{K} < T_\mathrm{eff} < 7000\,\mathrm{K}$ (giant) & global \\
        $4500\,\mathrm{K} < T_\mathrm{eff} < 7000\,\mathrm{K}$ (dwarf) & global \\
        $7000\,\mathrm{K} < T_\mathrm{eff} < 9000\,\mathrm{K}$ &  both \\
        $T_\mathrm{eff} > 10\,000\,\mathrm{K}$ & local \\
         \hline
    \end{tabular*}
    
    \label{tab:interp_parmaeter_ranges}
    \tablefoot{ The dwarf/giant separation is at $\log g = 4.0$. The spectra are generated by both methods and linearly combined in the overlap regions.}
\end{table}

\begin{figure}
    \centering
    \includegraphics[width = 0.45\textwidth]{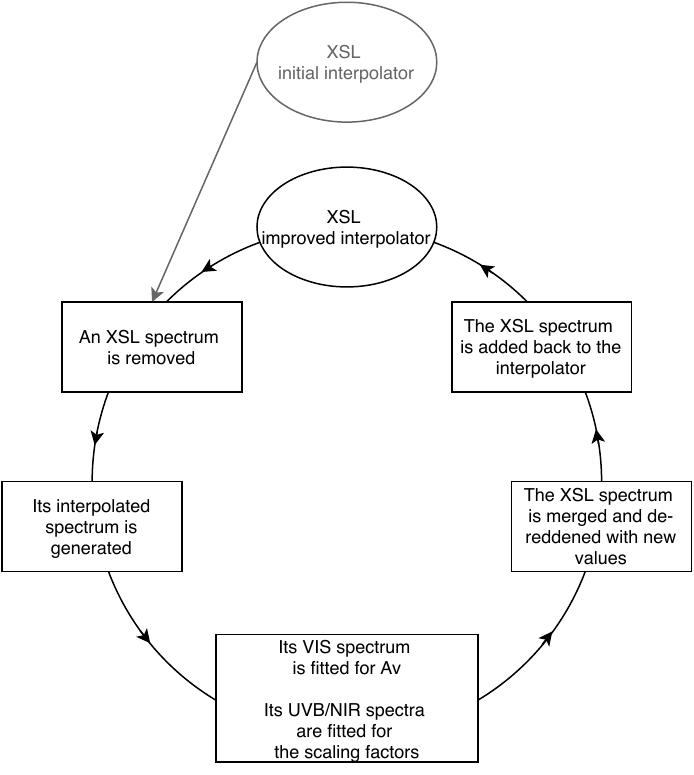}
    \caption{Flow chart of how the multiplicative scaling factors and extinctions were determined for XSL.}
    \label{fig:flow}
\end{figure}

We use the interpolator to re-determine the multiplicative scaling factors and extinction in the following way. We create the initial interpolator using spectra that we have de-reddened and merged using the scaling factors and extinction values in the previous two steps. We use this interpolator to create a spectrum of a star with the same parameters as an XSL star, but without using this star itself in calculating the output spectrum of the interpolation. Then, we find the multiplicative scaling factors and the $A_V$ by minimising the $\chi^2$ of the residuals between the observed spectrum and the interpolated spectrum. We do this in a specific order: first, we use the the VIS spectrum (600--900\,nm) to determine Galactic extinction only, and then we use the extinction corrected UVB and NIR spectra to fit for the scaling factors relatively to the VIS spectrum. We do this at low resolution, $R=2000$, to concentrate on the continuum shape. After determining these new values for an XSL spectrum, we add it back to the interpolator and the process continues for the next XSL spectrum. As one spectrum in the library affects the others through the interpolator, both the library and its interpolators improve through this process. This process is illustrated in Fig.~\ref{fig:flow}. XSL spectra that differ from its interpolated counterpart the most are fitted first. Some spectra are fitted multiple times in this process. Variable and peculiar stars are not used to create the interpolator, but they are fitted with an interpolated spectrum. XSL stars for which the initial $A_V$ could not be determined were not included in the initial interpolator but fitted for the $S_\mathrm{UVB}$, $S_\mathrm{NIR,}$ and $A_V$ values, and added to the refined interpolator last. Due to the difficulties arising from the large variety of types of stars in the dataset, we perform visual checks of all fits, and in some cases, we disregard the fitted values. Comparisons between the `interpolated library' and the XSL DR3 are shown in Sect.~\ref{Sec:interpolated_lib}. Some examples of fits of interpolated spectra to the final XSL DR3 spectra are shown in Fig.~\ref{fig:interp_xsl_spectra}. 

There is no one perfect method for determining parameters of stellar spectra spanning across the HR diagram. There are also some caveats to using the interpolator to determine $S_\mathrm{UVB}$, $S_\mathrm{NIR,}$ and $A_V$ values. The interpolator is less stable at the boundaries of the parameter space, where the stellar parameters are less well established, and the density of library stars is low. That can mean the interpolated spectrum is not so good, and the determined $S_\mathrm{UVB}$, $S_\mathrm{NIR,}$ and $A_V$ values can be biased. As we determine $A_V$ from the VIS spectrum first, and then the $S_\mathrm{UVB}$ and $S_\mathrm{NIR}$ values from the extinction corrected UVB and NIR spectra, the latter are influenced by the accuracy of determining the first. XSL stars do not have $[\alpha/\mathrm{Fe}]$ estimates, and inferring these is beyond the scope of this paper. We used BSTAR (Tlusty) and PHOENIX solar-scaled theoretical stellar spectra for the initial determination of the Galactic dust extinction. The majority of the XSL stars should follow the typical distribution of $[\alpha/\mathrm{Fe}]$ versus $[\mathrm{Fe}/\mathrm{H}]$ in the Milky Way. This would mean that the low-metallicity XSL stars would favour $\alpha$-enhanced PHOENIX models, instead of solar-scaled. This was also shown by L20 -- less than 15\% of the metal-rich XSL stars ($[\mathrm{Fe}/\mathrm{H}]> -0.5$), and about 75\% of the metal-poor stars ($[\mathrm{Fe}/\mathrm{H}]< -1.5$) prefer $\alpha$-enhanced models. For a given XSL spectrum, the change from solar to super-solar $[\alpha/\mathrm{Fe}]$ PHOENIX models leads mostly to a decrease in $T\mathrm{eff}$, a decrease in $\log g,$ and a decrease in metallicity. Therefore, using PHOENIX scaled solar models to determine the initial $A_V$ values might lead to a systematic bias of the interpolator towards lower $A_V$ values in the low metallicity regime.
Furthermore, the iterative method could introduce a drift in the $S_\mathrm{UVB}$, $S_\mathrm{NIR}$ and $A_V$ values at each iteration, where a spectrum with poorly determined values is introduced back to the interpolator, where it influences the determination of $S_\mathrm{UVB}$, $S_\mathrm{NIR,}$ and $A_V$ values of another XSL spectrum to be less accurate, and so on. However, the immediate visual inspection of the individual results, sometimes preferring visual inspection $S_\mathrm{UVB}$, $S_\mathrm{NIR}$ values, or literature $A_V$ values, as well as comparisons with various literature data in Sect. \ref{Sec:literature}, has ensured that such systematic trends are small, if present. 

\subsubsection{Adopted values}
\label{subsec:av}

Dust extinction, as parametrised by $A_V$ with a \citet{Cardelli1989} reddening law, is determined by using the interpolator for 581 spectra. The conventional interpolation methods fail for the very cool giants (especially in the NIR region), C-stars, and for different types of stars due to inaccuracies in the parameter estimation \citep{Arentsen2019}, as well as due to the time-variable SEDs of variable stars. We use L20 values for 41 spectra (A. Lan\c{c}on, priv. comm). For consistency, we use the L20 $A_V \mathrm{(VIS)}$ fit where the stellar parameters from \citet{Arentsen2019} are fixed. We use \citet{SMCLMC2019} values for 50 LMC and SMC stars, and correct them using respective extinction curves from \citet{Gordon2003}. In a few cases, we use other literature values, but there are nonetheless 85 spectra that we have not de-reddened. Among them are spectra that are uncorrected for narrow-slit flux loss, C-stars, and some peculiar stars. The UVB scaling factor is determined for 540 spectra through this routine and 270 by visual inspection (the majority of which are cool stars with noisy UVB spectra). The NIR scaling factor is determined for 554 spectra (and 249 by visual inspection). The relevant header keyword describing the origin of these parameters are AV\_ORI, S\_U\_ORI, and S\_N\_ORI for $A_V$, $S_\mathrm{UVB,}$ and $S_\mathrm{NIR,}$ respectively.

\subsection{Combining the three spectral segments}
The UVB and VIS spectra overlap, as do the VIS and NIR spectra, allowing us to merge the spectra without a gap between the segments. As mentioned above, the spectra between 545 and 556\,nm in the UVB, 556 and 590\,nm in the VIS, and 994 and 1150\,nm in the NIR suffer from dichroic artefacts, and the DR2 relative scaling of arm spectra is sometimes inadequate. We show overlap regions of XSL star HD 6229 (X0007) and HD 167946 (X0460) in Fig.~\ref{fig:merging} to illustrate this effect. We use the multiplicative UVB and NIR scaling factors to `lift' UVB and NIR spectra to the same flux level as the VIS spectrum. We then combine the UVB and VIS or the VIS and NIR spectra to a weighed mean spectrum in the overlap area, with the inverse of the variance spectrum compared to the weight. We smooth the spectrum in the overlap region to the lower resolution of the spectral arms: $\sigma_\mathrm{UVB} = 13\,\mathrm{km\,s^{-1}}$ and $\sigma_\mathrm{NIR} = 16\,\mathrm{km\,s^{-1}}$ in the UVB--VIS and the VIS--NIR overlap region, respectively. Then, we de-redden the spectra assuming a \citet{Cardelli1989} (or \citet{Gordon2003}) extinction law. Fig.~\ref{fig:full_merging} illustrates the outcome of the merging process of arm separated DR2 spectra of the star HD 167946 (X0460) to the arm combined DR3 (reddened) spectrum. 

The dichroic artefact is not described by the error spectrum, and so it cannot be minimised by the weighed mean combination. The severity varies from spectrum to spectrum, and it is worse between UVB and VIS than between VIS and NIR. It also reaches further into the VIS spectrum than the UVB overlap region. This area in the spectrum should be used with extreme caution. Table \ref{tab:dichroic_bands} shows commonly used photometric bands and absorption line indices that might be affected by the dichroic contamination. However, we provide the spectra in their rest frames, so the exact dichroic contamination region in the spectrum depends on the radial velocity of the star.

\begin{table}
    \centering
    \caption{Common photometric bands and Lick indices that might be affected by dichroic contamination.}
    \begin{tabular*}{0.45\textwidth}{ll}
    \hline\hline

           \multicolumn{2}{c}{UVB--VIS dichroic: 545--590\,nm }\\
           \hline
           Bandpasses & Cousins $V$; SDSS $r$; HST $F606W$, $F550M$; \\
           & Gaia BP, $G$ \\
            &\\
           Lick indices & aTiO$^{1}$, Fe5709$^{2}$, Fe5782$^{2}$, 
            NaD$^{2}$, TiO1$^{2}$\\
           \hline
           \multicolumn{2}{c}{VIS--NIR dichroic: 994--1150\,nm } \\
           \hline
           Bandpasses & SDSS $z$; HST $F110W$, $F850LP$, $FR914M$; \\
           &Gaia $RP$, $G$\\ 
           &\\
           Lick indices & CN/ZrO$^{3}$, FeH$^{3}$, FeH0.99$^{4}$, CI1.07$^{5}$, \\
           & CN1.10$^{5}$, NaI1.14$^{5}$, Na1.14$^{6}$, FeI1.16$^{5}$\\
           
         \hline
         
    \end{tabular*}
    \tablebib{ (1) \citet{Spiniello2014}; (2) \citet{Trager1998}; (3) \citet{CARTER86}; (4) \citet{Conroy2012}; (5) \citet{Rock2015}; (6) \citet{LaBarbera2016}. The exact region of dichroic contamination depends on the radial velocity of the star.}
    \label{tab:dichroic_bands}
\end{table}

\section{Re-calibration of the spectral shape}
\label{Sec:flux_loss}

The majority of spectra provided in DR2 are corrected for slit flux loss. However, 15\%  of  the  DR2 spectra are not, because of a lack of good-quality wide-slit observations. The new M-dwarf stars were also not observed through a wide slit, so these spectra are also not corrected for flux losses. We re-calibrate the spectra shape of an additional 64 spectra for flux loss, using a multiplicative cubic spline function, determined from the ratio of the XSL spectrum to the corresponding interpolated spectrum (as described in Section \ref{Sec:fitting}), assuming stellar parameters from \citet{Arentsen2019} or from this study (M-dwarf stars). This cubic spline simultaneously absorbs any flux calibration errors, flux loss, Galactic extinction, as well as applying a smoother transition between the spectral arms (but it does not correct for the dichroic contamination pattern if it is severe). Examples of XSL spectrum, the corresponding interpolated spectrum and the corrective spline function are shown in Fig.~\ref{fig:interp_xsl_spectra_slitcorr}. 
The inverse of the function is a multiplicative correction for the spectrum. As seen from the example spectra, the function itself usually has a similar shape to reddening, and therefore it mostly has the effect of de-reddening a spectrum. The smaller scale behaviour of the function corrects for sky-subtraction issues and flux-calibration errors. For example, the correction function of MD008 is relatively flat: this star does not exhibit much reddening from Galactic dust, which is to be expected as it is in the solar neighbourhood at a distance of 30 pc. Furthermore, we have estimates of extinction for the two giants, X0805 (CL* NGC 1978 LE 09) and X0706 (NGC 6838 1039), from other XSL DR3 spectra. For these, we would expect the polynomial to correct for relatively high dust extinction, $A_V =  0.35$~mag, and 0.62 mag, respectively.
We use a slightly different approach for the new M dwarfs. We first assume that there is no flux loss in the VIS spectrum and include them in the interpolator scheme. We find the extinction value in the same way as described in Sect.~\ref{Sec:fitting}, and then we fit a cubic polynomial to absorb any flux losses. These new M-dwarf stars are needed to fill the gaps in this parameter region in the HR diagram shown on Fig \ref{fig:MainHRD} to make the creation of an interpolated spectrum possible. Without an interpolated spectrum, we cannot correct for flux losses in this manner.

\begin{figure}
    \centering
    \includegraphics[width = 0.5\textwidth]{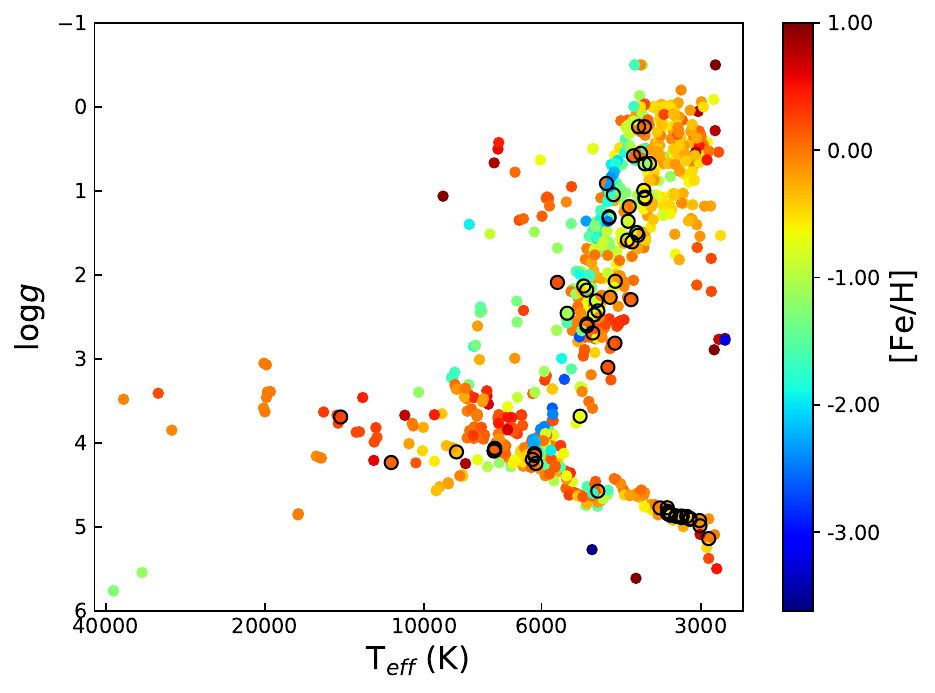}
    \caption{Same HR diagram as in Fig.~\ref{fig:MainHRD}, but now the stars with re-calibrated spectral shape are marked with open symbols.}
    \label{fig:loscorr}
\end{figure}

The re-calibrated spectra we provide are marked in Fig.~\ref{fig:loscorr} with empty circles. We provide both re-calibrated and original spectrum in the DR3 fits files. The re-calibration depends on the interpolator and the estimated stellar parameters, and these spectra should be used with caution. The relevant header keyword is SPL\_COR. 

\begin{figure*}
        {\includegraphics[width=\hsize]{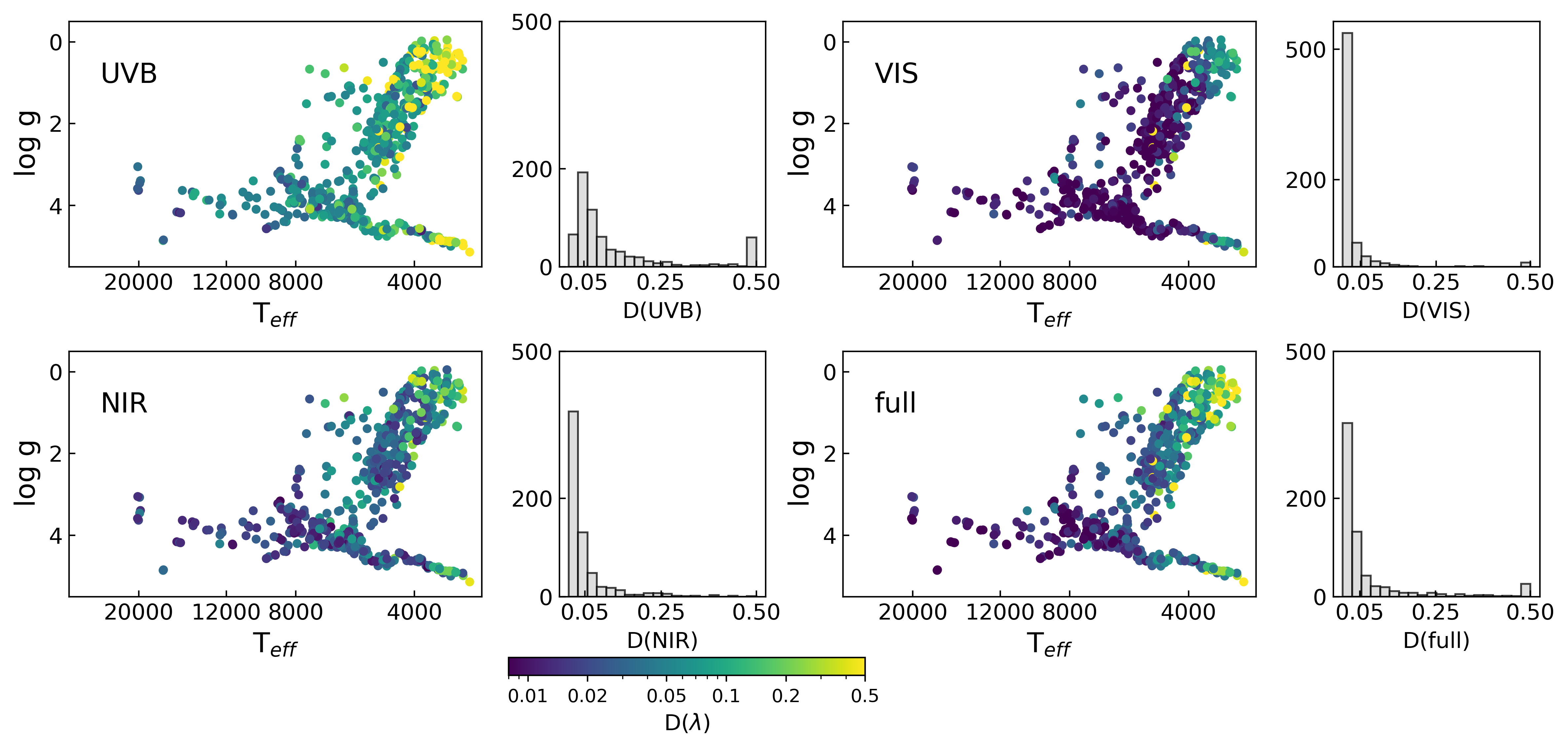}
        }
    \caption{ $D$-statistic comparing the interpolated and the XSL spectrum for the full wavelength range, and UVB, VIS, and NIR parts separately. The colour bar is logarithmic. Histograms show the distributions of $D$-statistic calculated within these spectral ranges at a resolving power of $R=500$. For better visualisation, the $D>0.5$ points are placed into the $D=0.5$ bin on the histograms.} 
    \label{fig:D_interpol} 
\end{figure*}

\section{Interpolator goodness of fit}
\label{Sec:interpolated_lib}

The XSL `interpolated library' is a set of 656 spectra generated with the spectral interpolator to match the XSL spectra according to their stellar parameters. This is the same dataset used in Sect.~\ref{Sec:fitting}. As these spectra are used to de-redden the XSL spectra, as well as to determine scaling factors of the UVB and NIR spectra relatively to VIS spectra, we describe here the goodness of fit between the XSL DR3 spectra and the interpolated spectra. The differences between the XSL library star ($F_\lambda^\mathrm{XSL}$) and its counterpart ($F_\lambda^\mathrm{interp}$) are described by a normalised rms deviation:
\begin{equation}
\label{Eq:D}
D = \sqrt{ \frac{1}{N}_\lambda \sum_\lambda  \left( \frac{F_\lambda^{interp} - F_\lambda^{xsl}}{F_\lambda^{xsl}}\right)^2 },
\end{equation}
where $N_\lambda$ is the number of wavelength pixels of the sum. For useful comparisons with the PHOENIX theoretical spectra, we calculate the $D$-statistic at a resolution of $R=500$, as in L20. We are only interested here in the quality of the shapes of the SEDs of the combined and extinction-corrected XSL spectra, and therefore low-resolution comparisons suffice. To give information about the fit in the different parts of the spectrum, we also calculate the $D$-statistics for the UVB, VIS, and NIR arm spectra separately. Certain problematic wavelength regions (such as areas with residuals from telluric lines, the ends of the X-shooter arms that have higher noise and that suffer from larger flux-calibration errors, emission lines) are masked out using the same masks as in L20. This ensures that $D$-statistic is not affected by pixels where we expect poor matches because of artefacts. Due to its denominator, $D$ is sensitive to regions of very low flux, such as in the UVB arm spectra of cool stars. To avoid this, denominator is replaced by a constant equal to 15\% of the maximal value of the UVB, estimated from 50 pixels bluer from 550\,nm. We note, however, that this does not work in cases where the whole UVB is noisy; S/N in the UVB arm varies strongly for cool stars and is in most cases less than $\mathrm{S/N}=20$, as shown in \citet[][Fig. 11]{DR2}.

The $D$-statistic across the full spectra and the UVB, VIS, and NIR arms separately are shown in Fig.~\ref{fig:D_interpol}. The HR diagrams illustrate the dependence of the goodness of fit on the spectral type, and the histograms show the distribution of the $D$-statistic in each spectral range. The $1\sigma$ flux-calibration errors correspond to $D\simeq 0.03$. Considering additional uncertainties from the extinction correction and arm merging, we characterise good matches as those with $D\leq0.05$ ($<5\%$). For better visualisation, very large $D$ values ($D>0.5$) are set to $D = 0.5$ in these figures. These high values can indicate very low S/N ratios, especially for the cool stars in the UVB arm. High $D$ values can also indicate poorer reproduction of the star by the interpolation due to uncertain stellar parameters, peculiarity of the spectrum, residual telluric lines, or due to the interpolation scheme itself. 

The average residuals between the XSL library star and its interpolated counterpart in the UVB arm are roughly 5\% or $D = 0.05$. This spectral region is the most difficult to interpolate for most spectral types, due to the multitude of spectral features compared to the VIS and NIR arms. In addition, the blackbody continuum shape changes rapidly with stellar parameters and cool stars have near-zero flux values in the UVB arm. For the VIS and NIR spectra, the matches are of the order of a few percent. In general, the SEDs of the interpolated spectra agree with the SEDs of XSL within the $1\sigma$ flux-calibration errors of XSL.

\section{Comparison with literature spectra and colours}
\label{Sec:literature}
Multiple comparisons between XSL and the literature have been made in the past. \citet{Chen2014} compared the DR1 spectra with literature spectra taken from NGSL \citep{NGSL2006}, MILES \citep{MILES1,MILES}, the calcium-triplet library CaT \citep{CaT, CaT2, CaT3}, UVES POP \citep{UVESPOP}, and ELODIE \citep{ELODIE2001,ELODIE2004,ELODIE2007}. There is good agreement in the line shapes and depths between XSL and these libraries. 

\citet{DR2} compared synthetic broad-band colours within the DR2 UVB and VIS arms to those of the MILES library \citep{MILES1,MILES}. In the NIR, they compared the XSL colours with those of the 2MASS survey \citep{Cohen2003, Skrutskie2006} and with synthetic colours of the IRTF/E-IRTF spectra \citep{IRTF,EIRTF}. An excellent match was found, with synthesised broad-band colours agreeing with those of the MILES library and of the combined IRTF and E-IRTF libraries to within $\sim1\%$. 

XSL DR3 has a combined wavelength range of 350--2480\,nm\ and is corrected for Galactic dust extinction. We use the extinction values from the \citet{Bayestar2019} 3D extinction maps and L20 to validate the extinction correction. To test the quality of merging the separate arm spectra, we compare XSL spectra with Gaia colours. To have more in-depth quality assurance, we compare the synthetic broad-band colours, absorption-line indices, and goodness of fit described by the $D$-statistic with those of MILES de-reddened spectra. The MILES library is widely used for stellar population purposes, and the spectra cover the UVB and half of the VIS spectral range. This comparison helps us to understand the quality of merging the UVB and the VIS arms, as well as the extinction correction. The NGSL spectra reach further into XSL VIS wavelength range. We provide broad-band colours and goodness of fit comparisons with this library, although using the reddened spectra. We forgo the absorption-line index comparisons with the NGSL, because the line-spread function (the residual velocity and the instrumental velocity dispersion) of the NGSL spectra varies irregularly from star to star and with wavelength \citep{NGSL_reso2012}. Instead, we compare the Ca\,\textsc{ii} triplet absorption-line indices, important features for stellar population applications, with those measured from the CaT library. In the NIR, we compare the synthetic broad-band colours, absorption-line indices and goodness of fit of XSL DR3 spectra with IRTF/E-IRTF reddened spectra.
\subsection{Comparison of XSL $A_V$ values with those from L20 and the \citet{Bayestar2019} 3D maps}
\begin{figure}
    \centering
    \includegraphics[width = 0.4\textwidth]{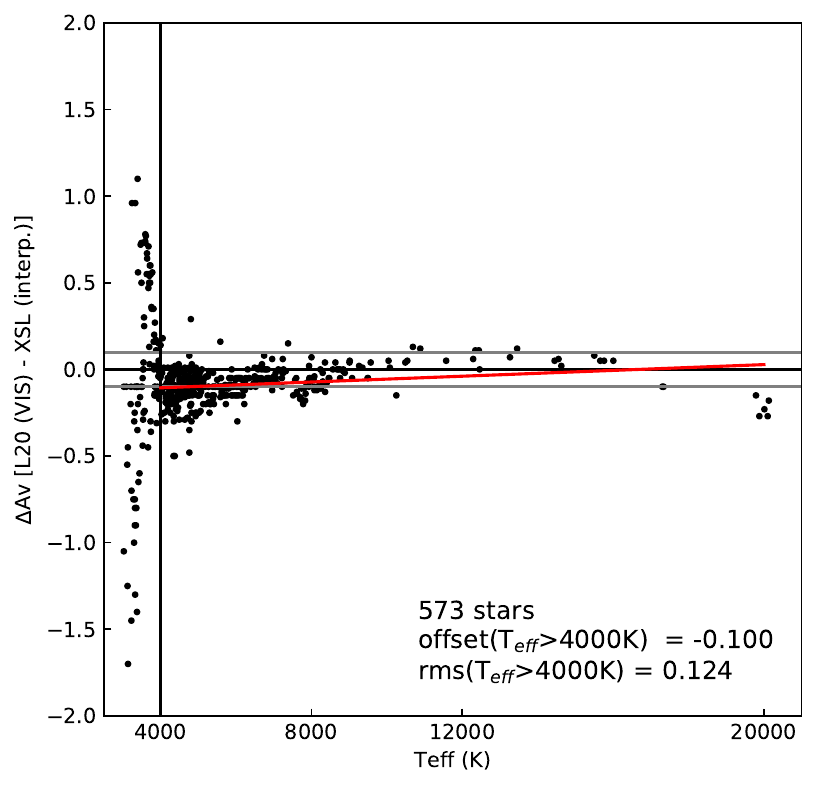}
    \caption{Differences between our $A_V$ estimates and those from the L20 VIS fit. The offset and the rms are given for stars warmer than $4000\,\mathrm{K}$. The red line denotes a linear fit to stars warmer than $4000\,\mathrm{K}$.}
    \label{fig:Av_L20}
\end{figure}

L20 mapped the differences between synthetic and the empirical stellar and line spectra over the HR diagram. We compare the $A_V$ determined in the process with our values. We use the L20 $A_V \mathrm{(VIS)}$ fits where the stellar parameters from \citet{Arentsen2019} are fixed. The comparison is shown in Fig.~\ref{fig:Av_L20}. The match is good for warm stars, with the rms being $0.12\,\mathrm{mag}$ for stars warmer than $4000\,\mathrm{K}$. Below this temperature, it is notoriously difficult to model stellar spectra, due to variability, large surface convection, and formation of dust. We note that there is a trend along the temperature axis. The initial $A_V$ values used for the first version of the interpolator were also determined by fitting with a set of PHOENIX templates. There can be a bias towards preferring the PHOENIX SED shapes. For this reason, we also provide a comparison between XSL DR3 $A_V$ values and those inferred from a Galactic 3D dust map. 

\begin{figure}
    \centering
    \includegraphics[width = 0.4\textwidth]{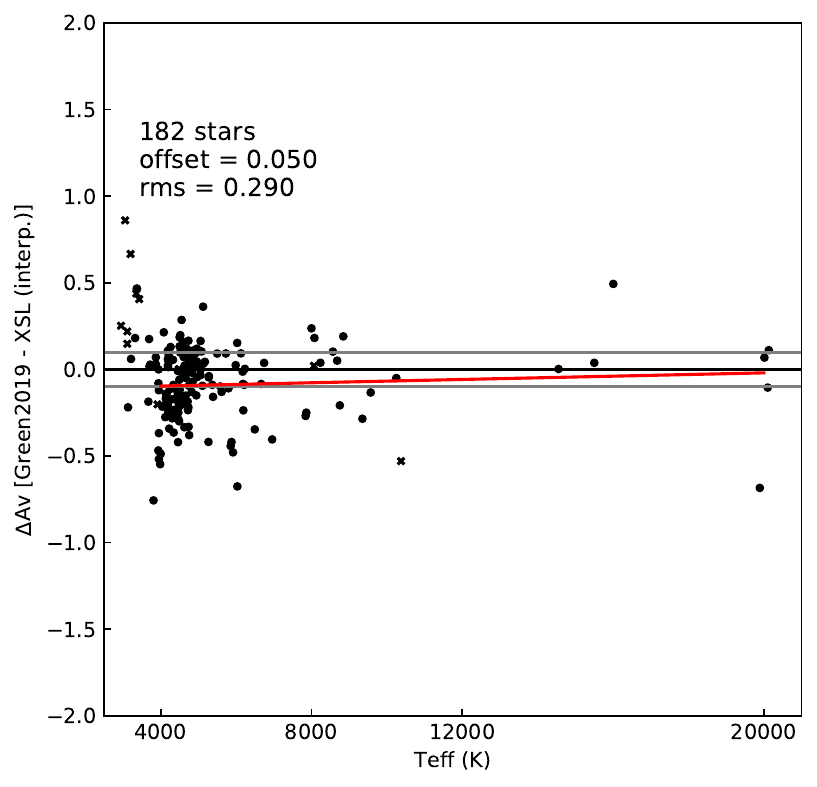}
    \caption{Differences between our $A_V$ estimates and those from the \citet{Bayestar2019} 3D dust maps. Stars with extinction values determined with our interpolator are marked with filled black circles, and those with L20 values are marked with black crosses. The red line denotes a linear fit to stars warmer than $4000\,\mathrm{K}$.}
    \label{fig:Av_dustmap}
\end{figure}

\citet{Bayestar2019} provides a 3D map of Milky Way dust reddening, called Bayestar2019. The map is based on Gaia parallaxes and stellar broadband optical and NIR  photometry from Pan-STARRS1 and 2MASS. It covers the sky north of a declination of $-30^{\circ}$, and reaches out to a distance of several kiloparsecs. We use the \citet{GaiaEDR3} coordinates for our stars and the conversion $E(B-V) = 0.884 \times \mathrm{(Bayestar19)}$, assuming $R_V = 3.1$. We have 182 stars in the mapped regions. The comparison is shown in Fig.~\ref{fig:Av_dustmap}. The reddening uncertainties of Bayestar2019 are roughly $0.1\,\mathrm{mag}$, but the rms of the difference is $0.29\,\mathrm{mag}$. There are many reasons for this discrepancy. Our methods rely on comparing a large section of a spectrum to a synthetic spectrum, on stellar parameters from \citet{Arentsen2019}, on the initial values determined using theoretical stellar spectra, and the assumption of $R_V = 3.1$ and the Cardelli extinction law. Bayestar2019 relies on photometry of 799 million stars. It is challenging to simultaneously determine stellar type (the intrinsic stellar colours and luminosity), distance, and reddening on the basis of photometry alone, although the Gaia parallax measurements have significantly advanced the 3D mapping of Galactic dust. Due to the multitude of different assumptions of the \citet{Bayestar2019} method, as well as L20 and our study, it is hard to know the most accurate values. However, a trend is seen in both Fig. \ref{fig:Av_L20} and Fig. \ref{fig:Av_dustmap}, with our estimates being $0.1\,\mathrm{mag}$ mag higher than both L20 and \citet{Bayestar2019} estimates around 4000--5000 K; the difference decreases with temperatures above that. This is within the suggested uncertainties but points to a bias in the XSL interpolator. The comparison of the three methods points toward a conservative uncertainty of $0.2\,\mathrm{mag}$ for stars warmer than $4000\,\mathrm{K}$.

\subsection{Comparison with Gaia colours}
\begin{figure}
    \centering
    \includegraphics[width = 0.5\textwidth]{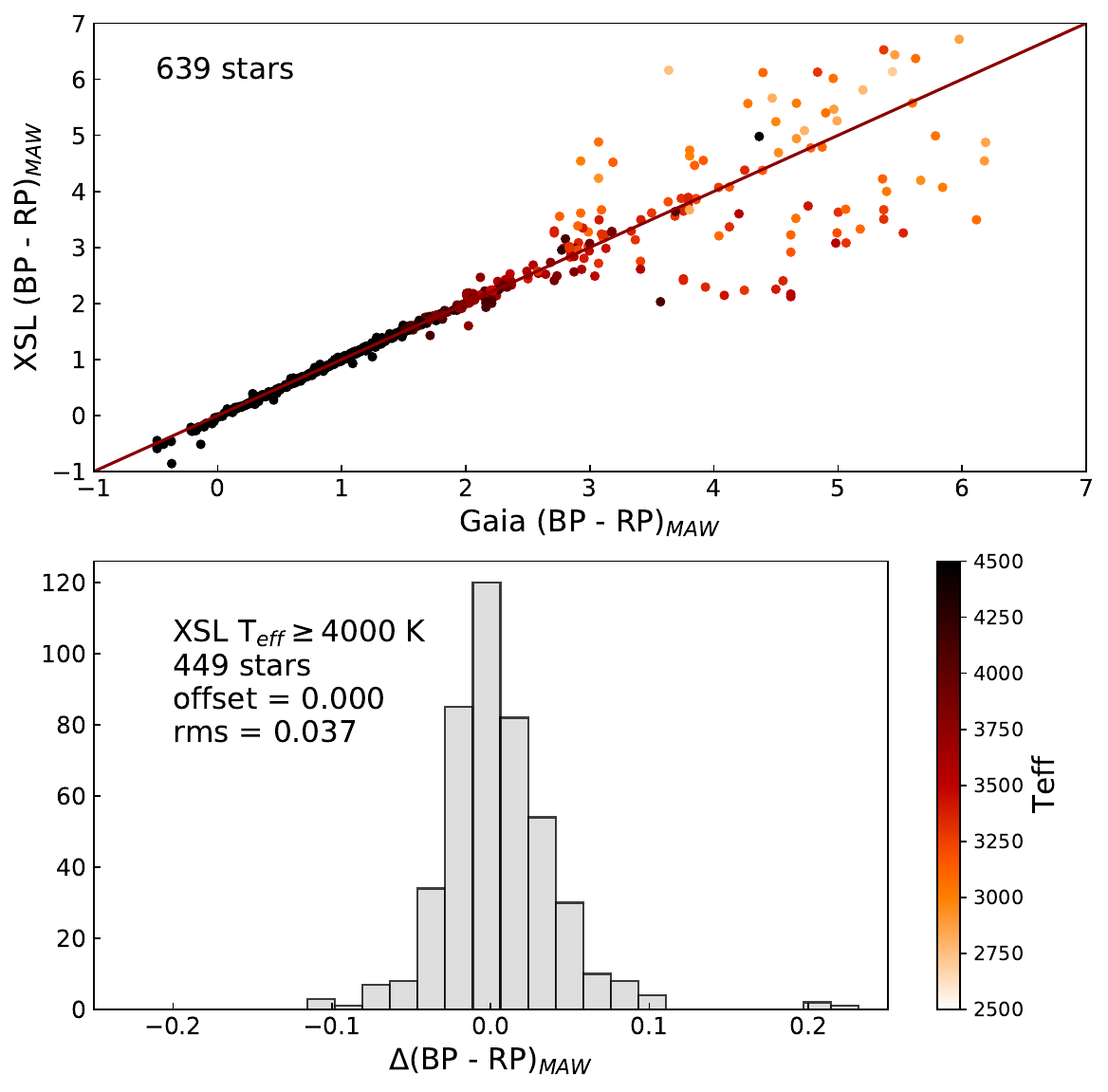}
    \caption{Differences between the published Gaia colours and XSL colours measured from the spectra. We use response curves for the Gaia photometric system presented by \citet[`MAW' in the figure]{GaiaColors2018}. We exclude stars with temperatures $T_\mathrm{eff} < 4000\,\mathrm{K}$ and variable stars from the histogram and rms calculations, but they are still included in the colour-colour comparison for illustrative purposes.}
    \label{fig:Gaiacompar}
\end{figure}

\noindent The Gaia Early Data Release 3 \citep[Gaia EDR3, ][]{GaiaEDR3} is the most homogeneous collection of photometry at the time of the publication of this paper. We use Gaia EDR3 BP and RP photometry of the XSL stars to compare with colours measured from the reddened XSL spectra. The majority of XSL spectra (807) have Gaia measurements, but we use 639 here as comparisons (flux corrected, non-peculiar spectra). 
For spectrophotometric colours, we use response curves for the Gaia photometric system presented by \citet[hereafter `MAW']{GaiaColors2018}, together with the corresponding zero points. The differences between the Gaia published and XSL spectrophotometric colours are shown in Fig.~\ref{fig:Gaiacompar}. M-giant stars with temperatures lower than about $4000\,\mathrm{K}$ experience photometric variability with typical periods of 100--1000 days, so the difference between our and Gaia colours in the red end [$(\mathrm{BP-RP})_\mathrm{MAW} > 2$] is expected. We therefore exclude stars with temperatures $T_\mathrm{eff} < 4000\,\mathrm{K}$, as well as other known variable stars, from the histogram and rms calculations to show the narrow relation of the warmer stars; however, the cooler stars are left in the colour-colour relation in Fig.~\ref{fig:Gaiacompar} for illustrative purposes. We compare 449 spectra, which are corrected for flux loss, belonging to non-variable warm stars. The colours measured from the XSL DR3 spectra and from Gaia mission match well, with a zero median offset and an rms scatter of $0.037\,\mathrm{mag}$. 
\subsection{Comparison with MILES}
The MILES empirical spectra library \citep{MILES1,MILES} is currently the benchmark library for studies of intermediate and old stellar populations. It consists of 985 stars and covers the 3525--7500\,\AA \ spectral range\ with a full width at half maximum (FWHM) resolution of 2.5\,\AA. The XSL sample is designed to have a strong overlap with the MILES library. In XSL DR3, 205 spectra of 173 stars have a MILES counterpart, from which 180 spectra belong to non-variable stars and are corrected for slit losses and Galactic dust extinction. \citet{DR2} defined artificial broad-band filters to compare the synthetic photometry of the two datasets. Here, we use three quality measures: the $D$-statistic defined by Eq. \ref{Eq:D}, synthetic colour comparisons, and various absorption-line indices. 

\begin{figure}
    \centering
    \includegraphics[width = 0.5\textwidth]{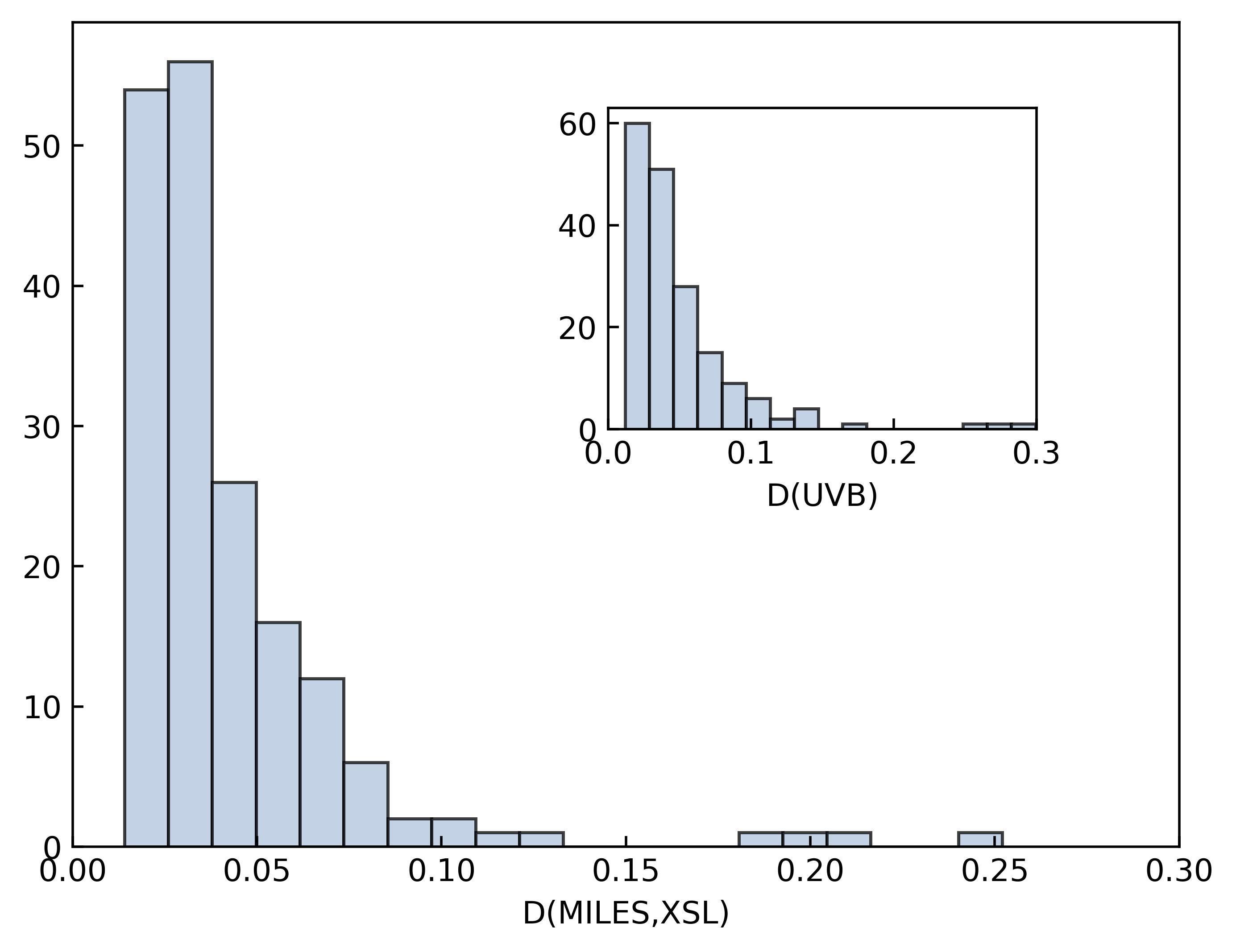}
    \caption{Histogram of the $D$-statistic comparison between MILES spectra and their XSL counterparts, for the full wavelength range (with problematic areas masked out) and for the UVB region separately (inset).}
    \label{fig:D_MILES_XSL}
\end{figure}{}

\begin{figure*}
\centering     
\subfigure[]{\includegraphics[width=0.95\textwidth]{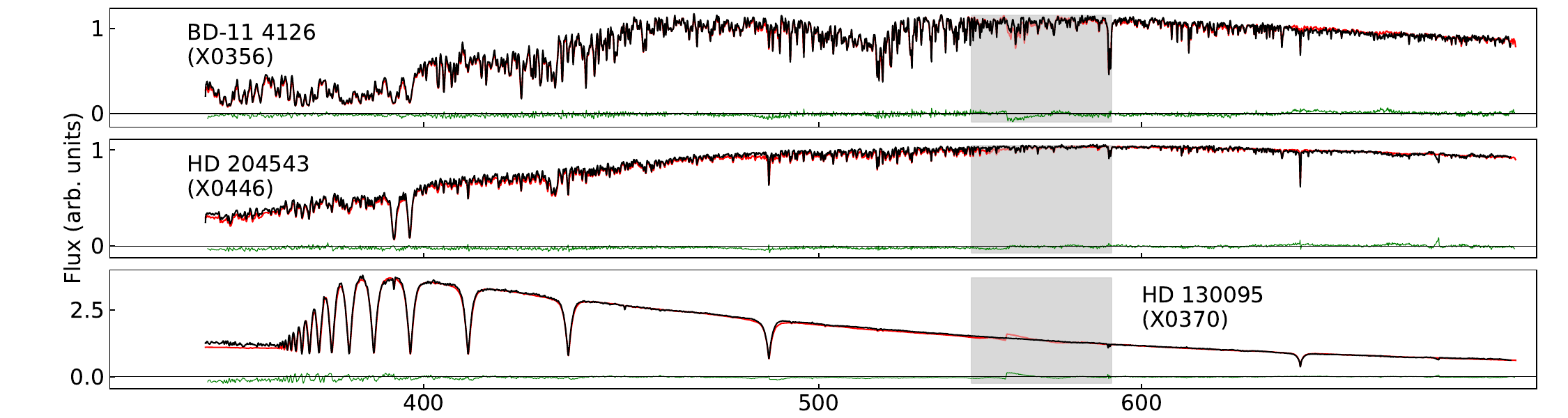}}
\subfigure[]{\includegraphics[width=0.95\textwidth]{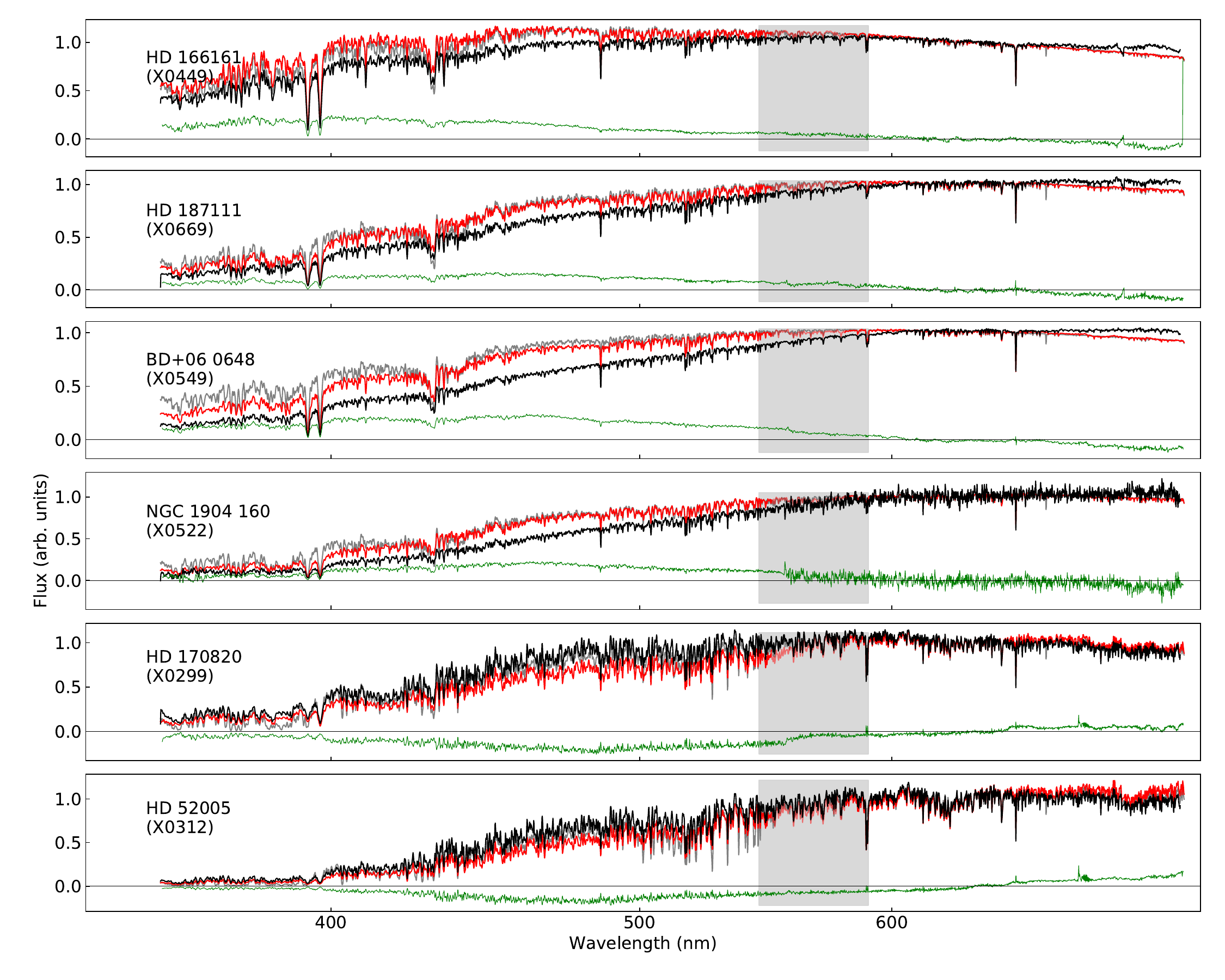}}
\caption{(a) Three stars in common between XSL (red) and MILES (black), smoothed to the MILES resolution. Residual spectra are in green. \textit{Top}: BD-114126. \textit{Middle:} HD 204543. \textit{Bottom:} HD 130095. Wavelength scale is logarithmic.
(b) Six stars that have the largest mismatch between MILES (black) and XSL (red), compared with PHOENIX theoretical star of similar stellar parameters (grey). The grey shaded areas on both sub-figures mark the UVB and VIS merging regions with possible dichroic contamination.}
    \label{fig:MILESstars}
\end{figure*}

In the MILES case, the $D$-statistic is calculated without the 15\% noise floor and is measured at MILES resolution. The comparison is made with the X-shooter dichroic region (550--580\,nm) masked out, as well as the edges of MILES spectrum: redward of 710\,nm and blueward of 390\,nm. Figure~\ref{fig:D_MILES_XSL} shows the distribution of the $D$-statistic for the full wavelength range and for the UVB region separately. There are no remarkable differences between the full wavelength $D$-statistic distribution and the one from the UVB-only comparison, which suggests that the small differences in the overall spectrum, instead of the UVB scaling factor alone, contribute to the $D$-statistic. If we look at the full MILES wavelength region, the comparison is equal to or better than $D=0.05$ or 5\% for the majority (144 spectra of 180). An additional 30 spectra are equal to or better than $D=0.1$, and six spectra have larger mismatch between the two libraries. Figure~\ref{fig:MILESstars} shows three stars in common between the two spectral libraries that match well. We investigate the source of the mismatch between the five spectra with the highest $D$-statistics on the bottom panel of Fig.~\ref{fig:MILESstars} by overplotting the XSL and MILES stellar spectra, as well as the PHOENIX theoretical spectra closest to XSL stars in the stellar parameter space.
Figure~\ref{fig:MILESstars} reveals that the differences are likely mainly due to the extinction correction: the XSL spectra match better with the theoretical spectra in all cases. This might mean a poor MILES extinction correction, or it might (also) mean that XSL is biased towards the shape of the PHOENIX spectra, which are used in the process of determining the extent of Galactic dust extinction in individual stars. 

\begin{table}
    \centering
    \caption{Artificial filters used for comparison with MILES spectral library and the NGSL.}
    \begin{tabular}{l c c}
    \hline\hline 
         Filter & $\lambda_{min}$ [nm]& $\lambda_{max}$ [nm]\\
         \hline
         box1 & 390  & 450 \\
         box2 & 460  & 520 \\
         box3 & 540  & 620 \\
         box4 & 630  & 710 \\
         \hline
    \end{tabular}
    \label{tab:MILES_boxes}
\end{table}

\begin{figure}
\centering     
\subfigure{\includegraphics[width=0.5\textwidth]{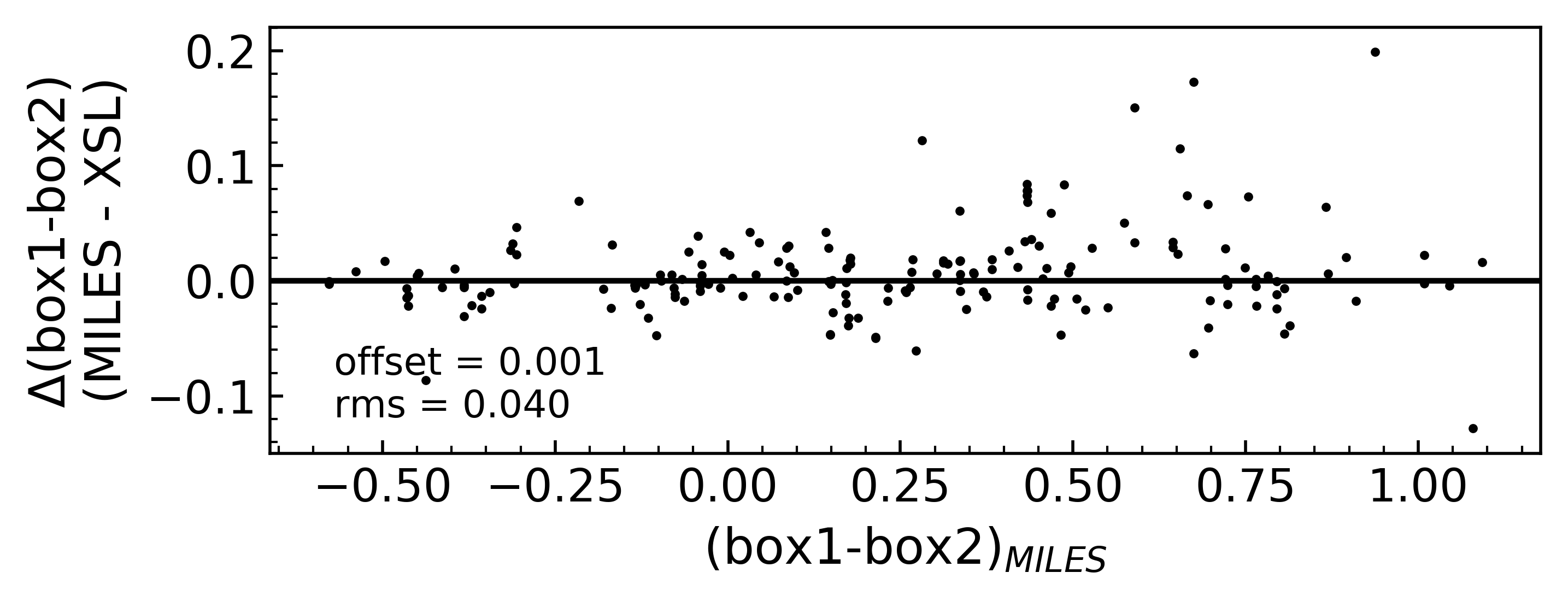}}
\subfigure{\includegraphics[width=0.5\textwidth]{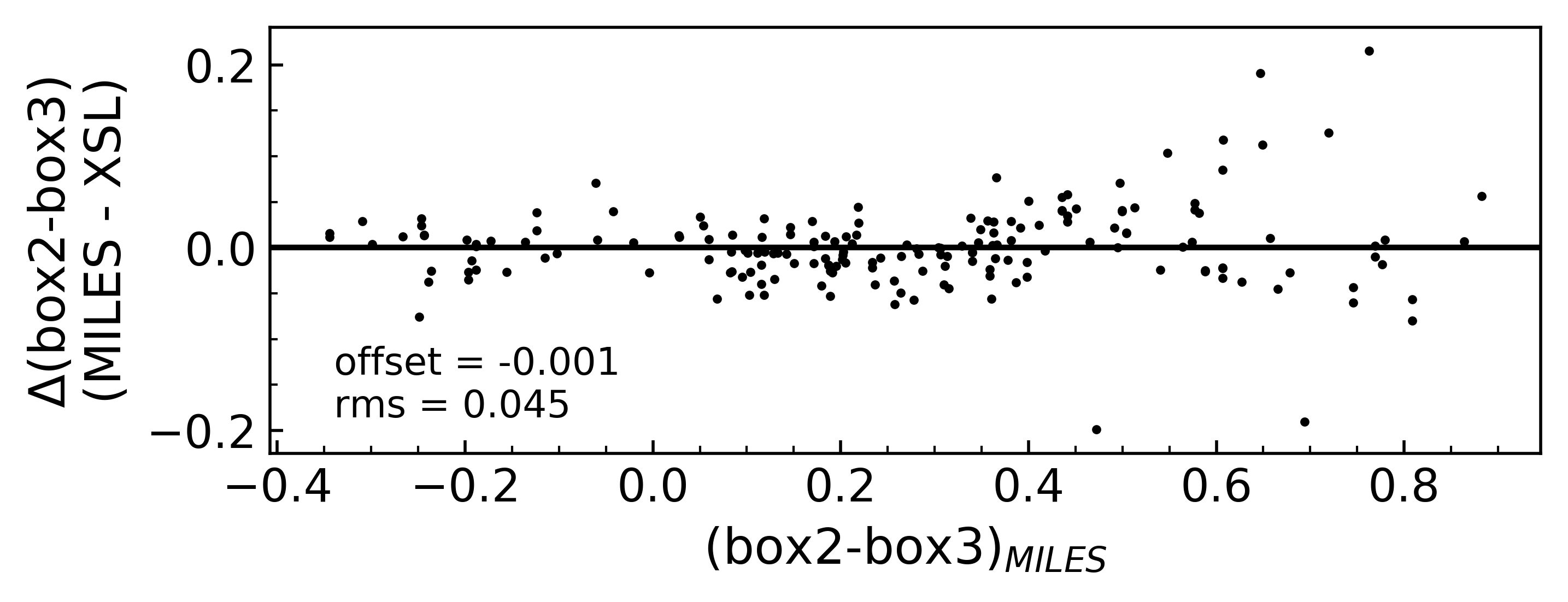}}
\subfigure{\includegraphics[width=0.5\textwidth]{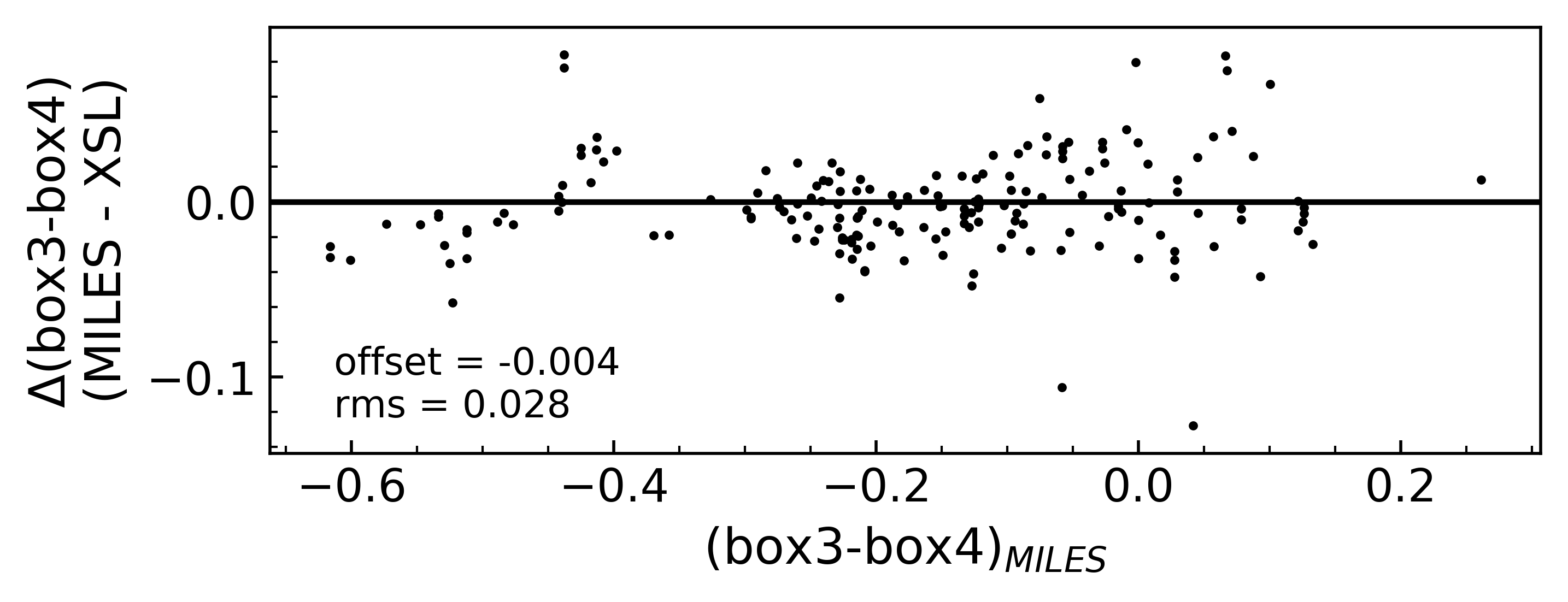}}
\caption{Comparison  of  synthetic  colours  between  XSL  and  MILES. }
    \label{fig:MILES_XLS_boxes}
\end{figure}

To complement the DR2-MILES comparison, we also measure the synthetic photometry defined in Table \ref{tab:MILES_boxes} with a box-function response, as in \citet[Table 6]{DR2}.
In DR2, only colours contained within a single X-shooter arm were calculated and compared with values of reddened MILES spectra. In Fig.~\ref{fig:MILES_XLS_boxes}, we compare the $(\mathrm{box1}-\mathrm{box2})$ colours in X-shooter's UVB arm, the $(\mathrm{box3}-\mathrm{box4})$ colours in X-shooter's VIS arm, and the $(\mathrm{box2}-\mathrm{box3})$ colours that span X-shooter's UVB and VIS arms of the de-reddened DR3 and de-reddened MILES spectra. The former two colours aim to measure the goodness of fit of the extinction corrected spectra, and the latter aims to describe the smoothness of the XSL UVB and VIS arm merging. Overall, the colours of the de-reddened merged XSL and MILES spectra are in excellent agreement and of the same order as the comparison done in DR2. The $(\mathrm{box1}-\mathrm{box2})$ colours have a median offset(rms) $0.001\pm0.040\,\mathrm{mag}$, similar to $-0.005\pm0.056$ mag measured in DR2. In the VIS, the $(\mathrm{box3}-\mathrm{box4})$ colour offset is $-0.004\pm0.028\,\mathrm{mag}$, compared with  $-0.01\pm0.03\,\mathrm{mag}$ from DR2. The average offset in $(\mathrm{box2}-\mathrm{box3})$ colours is $-0.001\pm0.045\,\mathrm{mag}$, demonstrating the quality of the UVB and VIS arm merging.

\begin{figure}
    \centering
    \includegraphics[width = 0.5\textwidth]{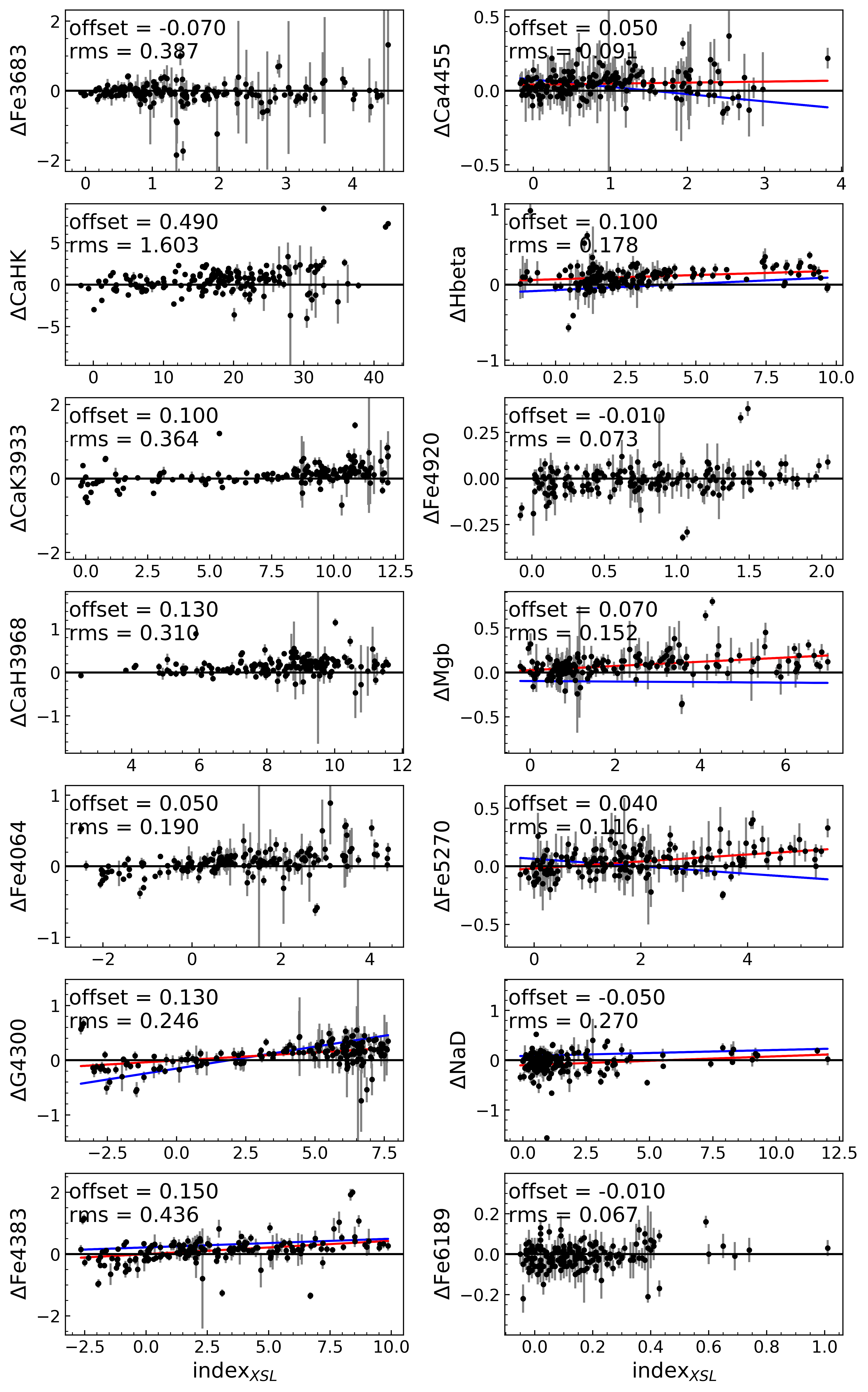}
    \caption{Selection of 5\,\AA\ LIS indices measured from MILES and XSL spectra. Here, $\Delta\mathrm{index} = \mathrm{index}_\mathrm{XSL} - \mathrm{index}_\mathrm{MILES}$. The grey bars are uncertainties drawn from a Monte Carlo sampling of the XSL spectra. The blue lines in some panels are from \citet{LEMONY} and the red lines denote a linear fit to our data.}
    \label{fig:MILES_XLS_Lick}
\end{figure}

When comparing spectra, the $D$-statistic describes the average deviation over the full wavelength range, colour differences provide a rough description of the shape differences of the SEDs, and differences in absorption-line indices describe local, spectral-line-level differences. We calculate a number of absorption-line indices across the MILES wavelength range in the 5\,\AA-Lick index system (LIS 5\,\AA), shown in Fig.~\ref{fig:MILES_XLS_Lick}. We perform a Monte Carlo error analysis for the absorption-line indices measured from XSL spectra, using the XSL error spectrum and calculating 100 realisations. There seem to be small systematic differences between certain absorption-line indices measured from the XSL and MILES spectra. These differences arise from absorption-line index width (10\,nm) flux calibration differences, which can be seen in the residual spectrum in Fig.~\ref{fig:MILES_IRTF_XSL}. Where possible, we overplot the linear relations between MILES and LEMONY Lick indices from \citet{LEMONY} on Fig.~\ref{fig:MILES_XLS_Lick}. Their Lick and our absorption-line indices are not strictly comparable, as we measure indices on the LIS 5\,\AA\ system, while \citet{LEMONY} use the original Lick/IDS system. These relations nevertheless show that similar systematic differences with MILES library exist. 
\subsection{Comparison with NGSL}

\begin{figure}
    \centering
    \includegraphics[width = 0.5\textwidth]{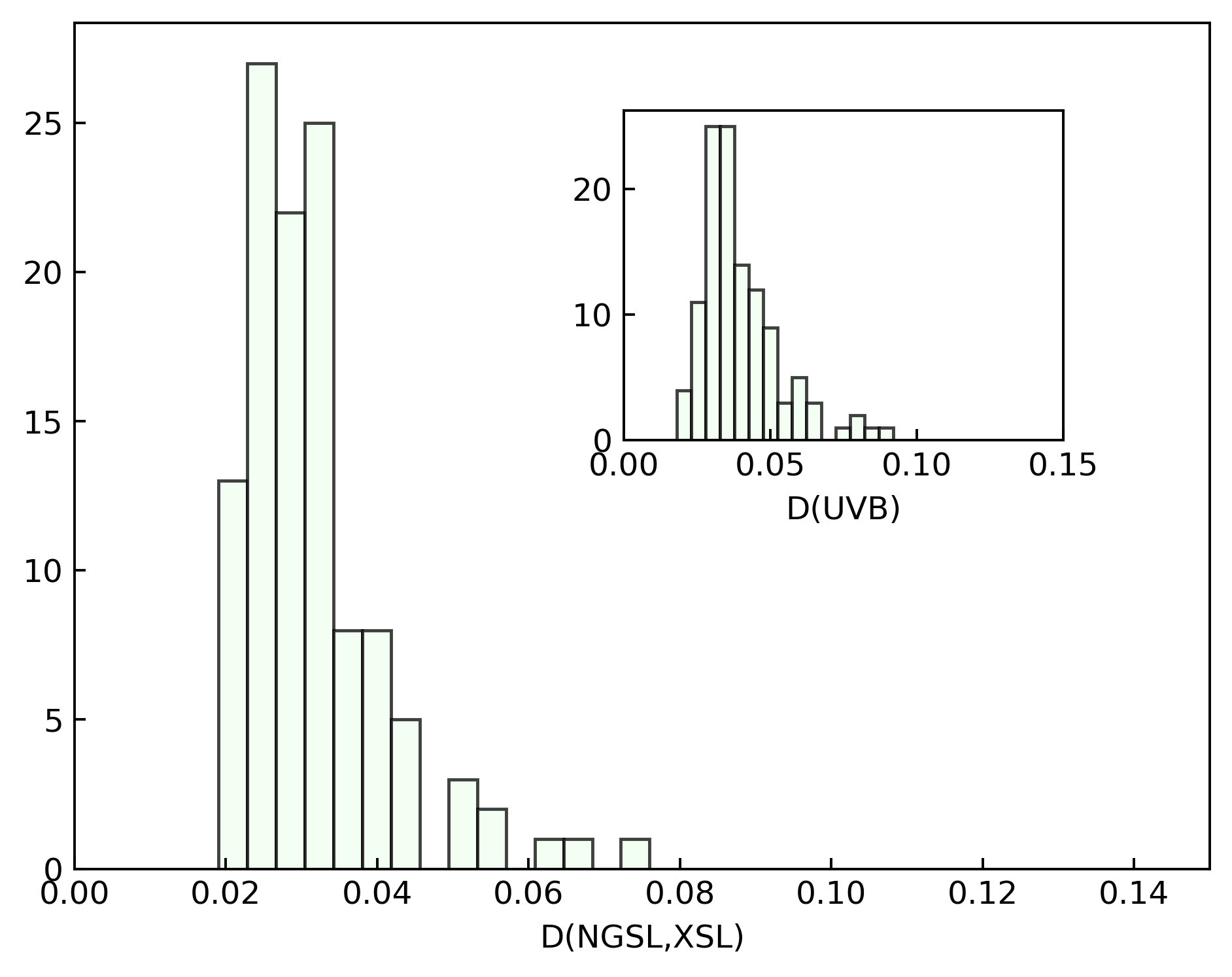}
    \caption{Histogram of the $D$-statistic comparison between NGSL spectra and their XSL counterparts for the full wavelength range (with problematic areas masked out) and for the UVB region separately (inset).}
    \label{fig:D_NGSL_XSL}
\end{figure}{}
\begin{figure}
\centering     
\subfigure{\includegraphics[width=0.5\textwidth]{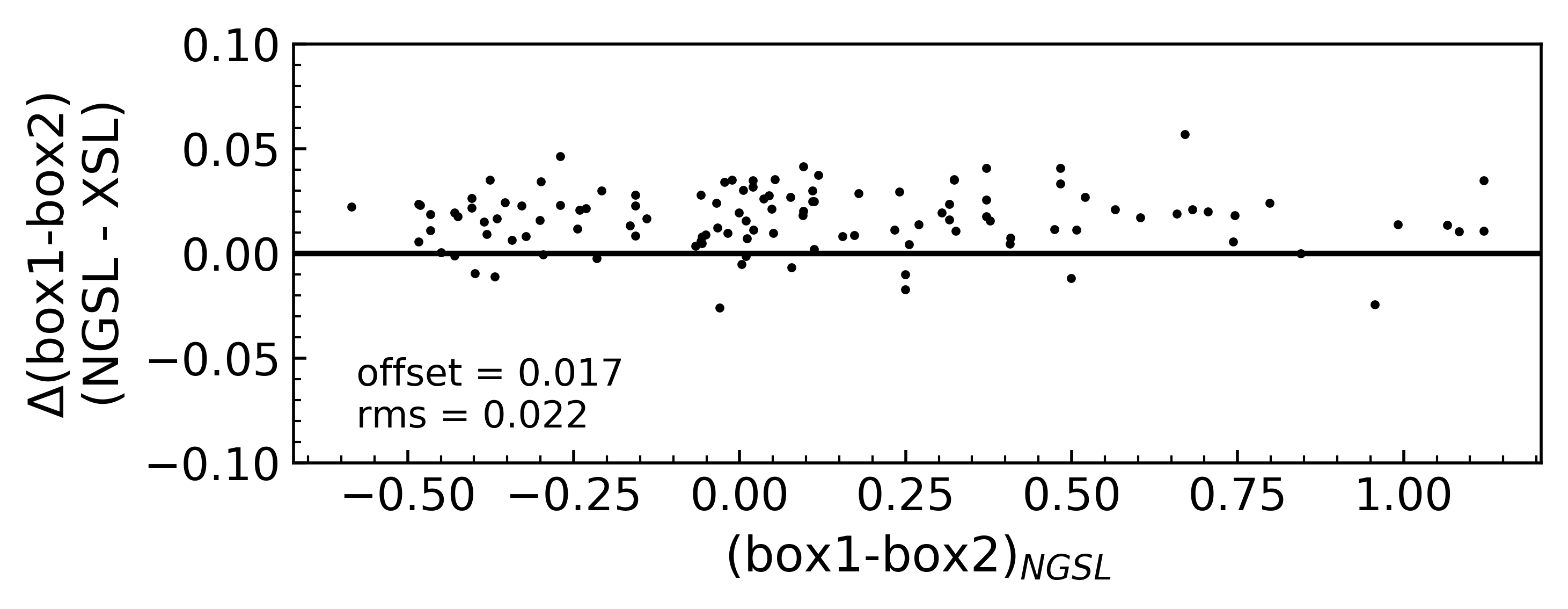}}
\subfigure{\includegraphics[width=0.5\textwidth]{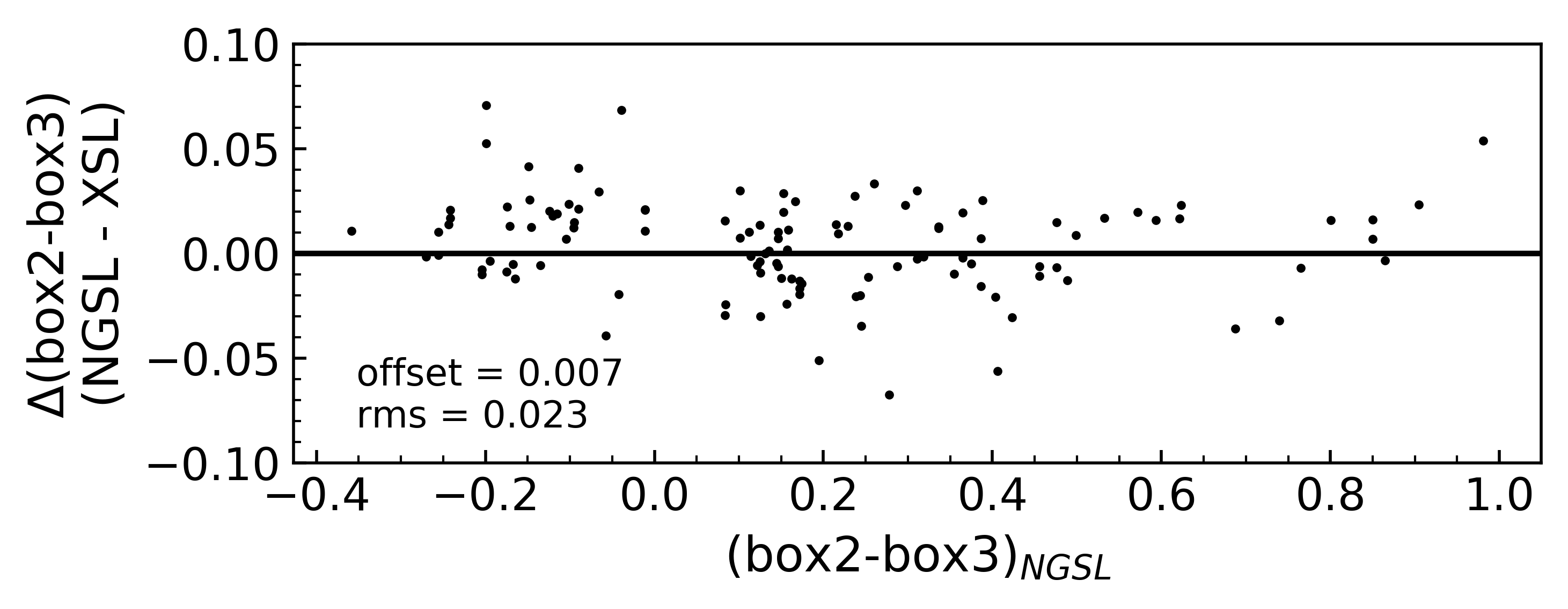}}
\subfigure{\includegraphics[width=0.5\textwidth]{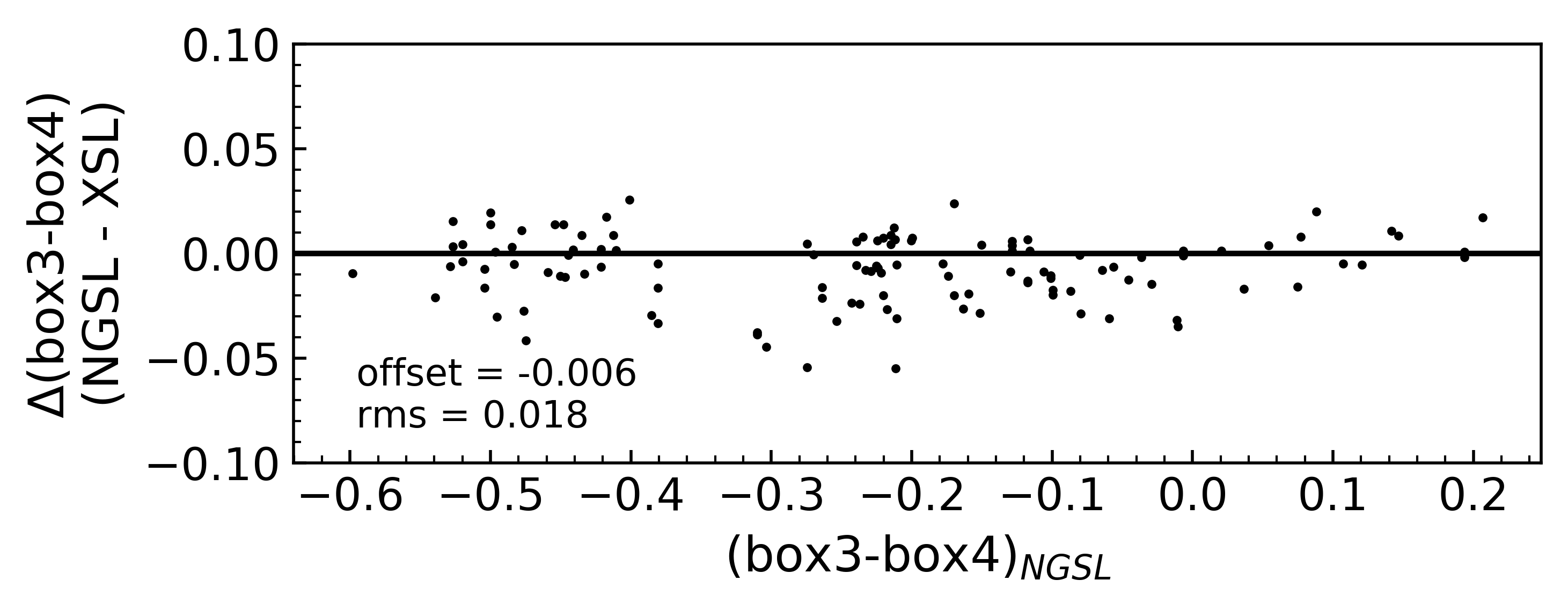}}
\caption{Comparison of synthetic colours between XSL and NGSL.}
    \label{fig:NGSL_XLS_boxes}
\end{figure}

The STIS Next Generation Spectral Library \citep[NGSL;][]{NGSL2006} consists of spectrophotometry of 374 stars of different spectral types, covering the spectral range 200--1000\,nm, at a resolving power R$\sim$1000 \citep{NGSL_reso2012}, taken with the Space Telescope Imaging Spectrograph on the \textsl{Hubble Space Telescope}. We use the NGSL version 2 spectra\footnote{\url{https://archive.stsci.edu/prepds/stisngsl/}}, and we compare these spectra to the reddened version of XSL spectra (i.e. the XSL spectra before the de-reddening described in Sect.~\ref{Sec:interpolator}). 

Here, we investigate the differences between spectra via the $D$-statistic and artificial synthetic photometry, but we forgo the absorption-line comparisons. The line-spread function of the NGSL spectra varies irregularly from star to star and with wavelength \citep[Fig.~1--3]{NGSL_reso2012}. There are 167 spectra of 135 stars in common between the two spectral libraries. The slit-throughput correction is thought to be unreliable ($\mathrm{offset} > 0.9\,\mathrm{pixel}$) for some  NGSL spectra \citep{NGSL2006}. After cleaning the dataset of variable stars, XSL spectra with slit flux losses, and unreliable NGSL spectra, 116 spectra can be compared.

Figure \ref{fig:D_NGSL_XSL} shows the $D$-statistic calculated between 380 and 980\,nm (and 380 and 540\,nm for the UVB) at the NGSL resolution (R $=1000$). For 112 spectra of the 116 spectra in common, the comparison is equal to or better than $D = 0.05$. For four spectra, the match is a slightly worse than that. Figure \ref{fig:NGSLstars} shows three spectra that match well between the two libraries. Figure \ref{fig:NGSLstars} also shows the four spectra with the highest $D$-statistic values, all of which show flux calibration and/or XSL UVB scaling factor issues. 

Comparison of the artificial synthetic photometry defined in Table \ref{tab:MILES_boxes}, measured on NGSL and XSL DR3 spectra is shown in Figure \ref{fig:NGSL_XLS_boxes}. The colours of the 116 stars are in excellent agreement, with the offset(rms) of $0.007\pm0.026\,\mathrm{mag}$ and $-0.006\pm0.018\,\mathrm{mag}$ for $(\mathrm{box2}-\mathrm{box3)}$ and $(\mathrm{box3}-\mathrm{box4),}$ respectively. However, there is a systematic offset $0.017\pm0.023\,\mathrm{mag}$ in the $(\mathrm{box1}-\mathrm{box2)}$ colour difference, which we do not see in the MILES comparison. The scatter in all colours is smaller than in the comparison with MILES, as comparing reddened spectra eliminates uncertainties arising from the extinction correction.

\subsection{Comparison with CaT library}

\begin{figure}
    \centering
    \includegraphics[width = 0.5\textwidth]{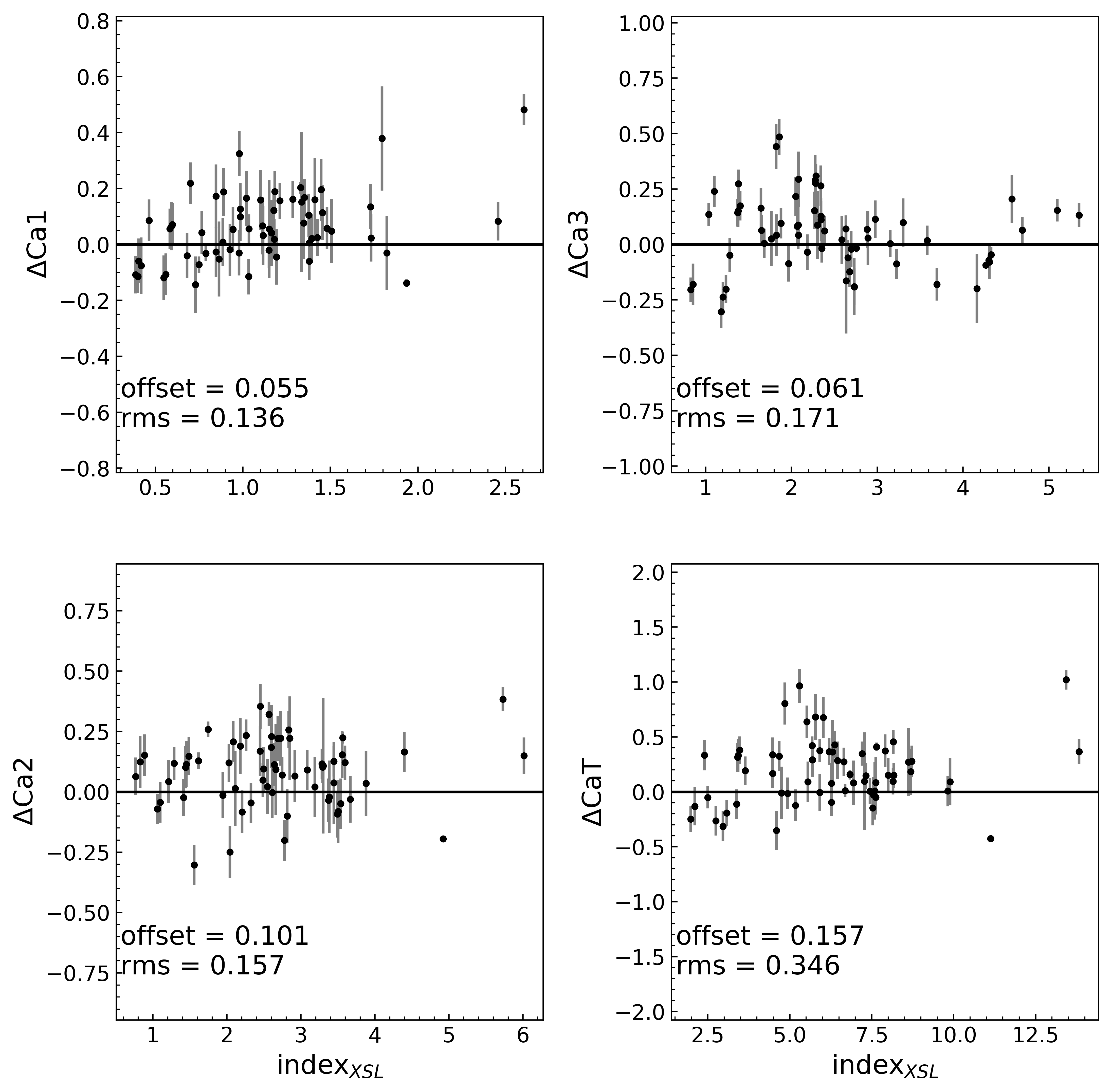}
    \caption{Calcium II Triplet line indices measured from the CaT library and XSL spectra, on the 5\,\AA\ LIS. Here $\Delta\mathrm{index} = \mathrm{index}_\mathrm{XSL} - \mathrm{index}_\mathrm{CaT}$. The grey bars are uncertainties drawn from a Monte Carlo sampling of the XSL spectra.}
    \label{fig:CaT_XLS_Lick}
\end{figure}{}

The Calcium II Triplet library \citep{CaT, CaT2, CaT3} consists of 706 stars and was developed for the empirical calibration of the Ca\,\textsc{ii} triplet and for stellar population synthesis modelling. The library covers the 834.8--902.0\,nm range at 1.5\,\AA\ (FWHM) spectral resolution. The Ca\,\textsc{ii} triplet is a very prominent feature in the near-IR spectrum of cool stars and has wide applications in stellar population studies, due to its sensitivity to surface gravity \citep[][Sect. 2.2 and references within]{CaT}. Due to the importance of this feature, we measure Ca1, Ca2, Ca3, and CaT absorption-line indices \citep{CaT} in the 5\,\AA\ LIS from the XSL spectra and compare these with those measured from the CaT spectra.  There are 76 spectra of 63 stars in common with the CaT library, of which 63 spectra belong to non-variable stars and are corrected for slit-losses. The comparison in Fig.~\ref{fig:CaT_XLS_Lick} shows a good agreement for the Ca\,\textsc{ii} triplet lines.

\subsection{Comparison with (E-)IRTF}

\begin{figure}
    \centering
    \includegraphics[width = 0.5\textwidth]{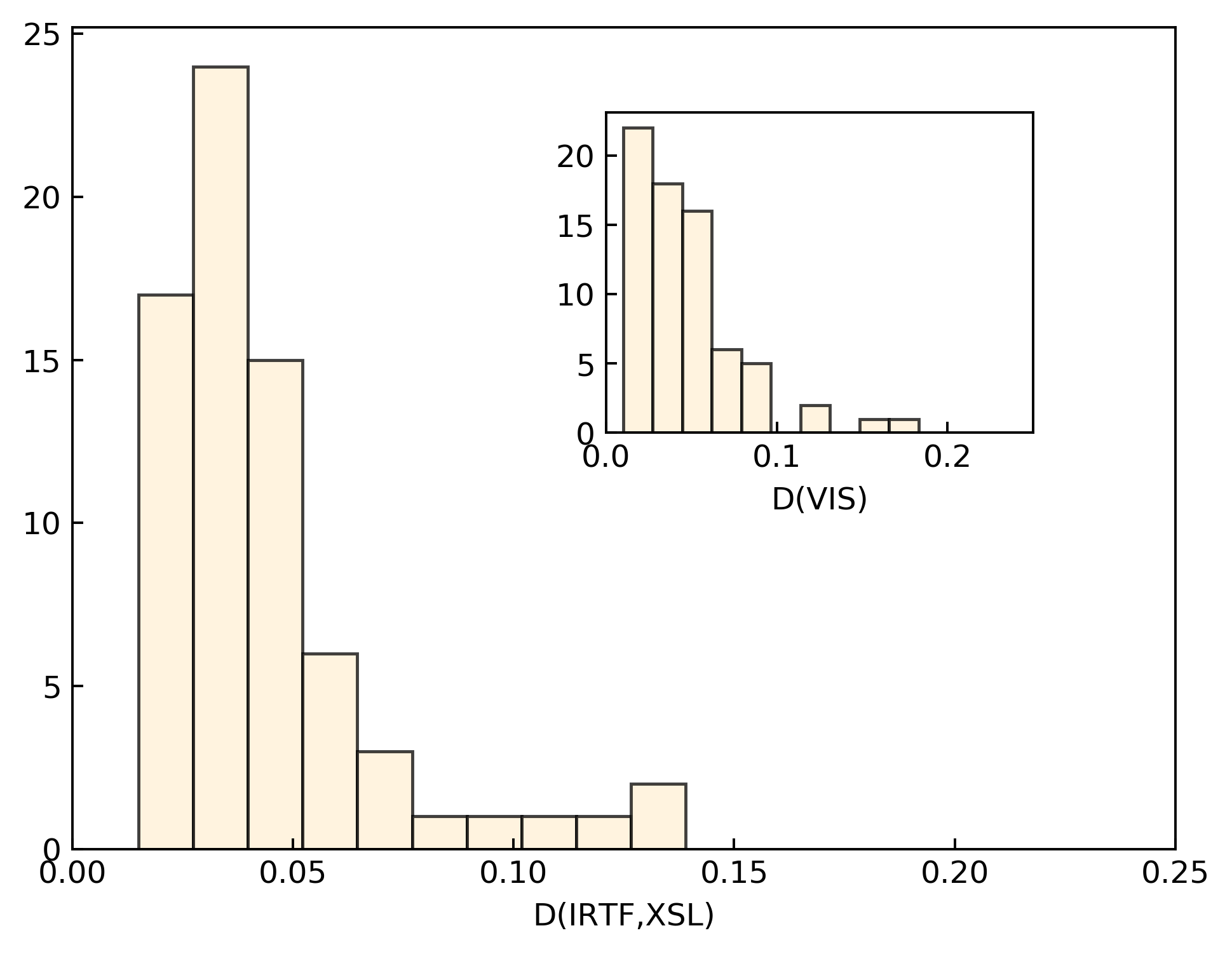}
    \caption{Histogram of the $D$-statistic between the IRTF spectra and their XSL counterparts for the full IRTF wavelength range (with problematic areas masked out) and in the XSL VIS portion of the spectrum (inset).}
    \label{fig:D_IRTF_XSL}
\end{figure}{}

The IRTF Spectral Library \citep{IRTF} and the extended-IRTF \citep{EIRTF} are collections of 0.8--5.0\,$\mu$m and 0.7--2.5\,$\mu$m spectra observed at a resolving power of $R=2000$ with SpeX at the NASA Infrared Telescope Facility (IRTF) on Mauna Kea. The original IRTF library covers mainly solar-metallicity late-type stars (but also AGB stars, carbon and S stars, L and T dwarfs, and some planets), and the E-IRTF expands the metallicity coverage. XSL DR3 has 25 spectra of 20 stars in common with IRTF and 60 spectra of 48 stars in common with E-IRTF. After cleaning the dataset of variable stars and XSL spectra with slit flux losses, 71 spectra can be compared. 

We refer to Section 4.5.3 of \citet{DR2} for the $J$, $H$ and $K_s$ photometry colour comparisons of these spectra. As the overlap between the XSL VIS spectral range and that of (E-)IRTF is small, we did not compare the colours across the X-shooter arms. 

We calculate the $D$-statistic at IRTF resolution for the wavelength range of E-IRTF, with the telluric and dichroic areas masked. This is also done separately for the VIS region of the XSL spectra, to see if there are noticeable issues with the NIR scaling factors. Comparisons are shown in Fig.~\ref{fig:D_IRTF_XSL}. We find that 55 of the 71 spectra match better than $D = 0.05$, and an additional 12 spectra match better than $D = 0.1$. That leaves 4 spectra with unsatisfactory matches. We investigate these spectra in Fig.~\ref{fig:IRTFstars}, where the ratios between XSL and (E-)IRTF spectra in the bottom panel reveals large slope differences between HD 160365 (X0141), HD 65583 (X0566), and HD 213042 (X0430). \citet{EIRTF} saw similar slope differences between some IRTF and (E)-IRTF stars, but we see differences with stars from both IRTF and E-IRTF. Moreover, \citet{DR2} found signs of discontinuity between the $J$ and the $H$ band, although their comparison between the XSL and the theoretical spectra did not show errors larger than a few percent. Most of these four spectra also show significant residuals from the telluric correction of XSL. 
\begin{figure}
    \centering
    \includegraphics[width = 0.5\textwidth]{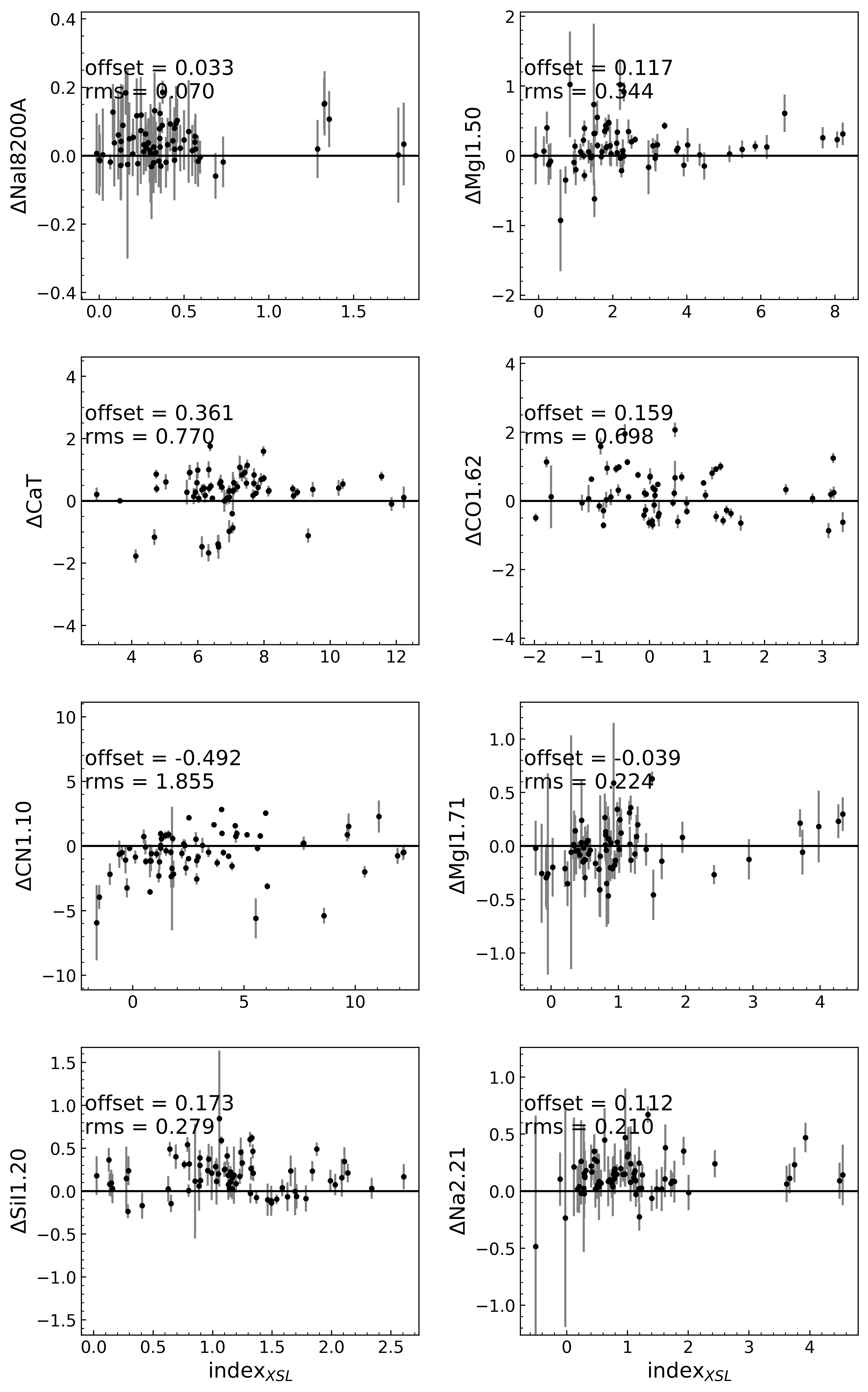}
    \caption{Indices measured on the 14 \AA{} LIS from (E-)IRTF and XSL spectra. Here $\Delta\mathrm{index} = \mathrm{index}_\mathrm{XSL} - \mathrm{index}_\mathrm{(E-IRTF)}$. The grey bars are uncertainties drawn from a Monte Carlo sampling of the XSL spectra.}
    \label{fig:IRTF_XLS_Lick}
\end{figure}{}

We also measured various NIR absorption-line indices on the 14\,\AA\ LIS from (E-)IRTF and XSL spectra, shown in Fig.~\ref{fig:IRTF_XLS_Lick}, with error bars taken from Monte Carlo samples of the XSL spectra. There is an overall good agreement between the indices shown, but some show noteworthy differences; for example, CaT shows erratic behaviour. Figure \ref{fig:IRTFstars} shows 10\,nm-scale features in the residual spectrum that could explain these discrepancies.

Figure \ref{fig:MILES_IRTF_XSL} shows four spectra in common between MILES, IRTF, and XSL. The IRTF reddened spectra have been extinction corrected using the XSL values. While the comparison with MILES shows small-scale residuals, the comparison with the IRTF show spectra with more global slope differences.



\section{Conclusions}

In this paper, we present the third data release of the X-shooter stellar library (XSL DR3), which consists of 830 spectra covering a wide range of stellar parameters, merged to the full wavelength range of X-shooter. The spectra were observed with the X-shooter spectrograph on ESO’s VLT, which has three spectral arms: UVB, covering 300--556\,nm; VIS, covering 533--1020\,nm; and NIR, covering 994--2480\,nm, with spectral resolutions of $\sigma=13,11,16\,\mathrm{km\,s}^{-1,}$ respectively. The XSL DR3 spectra cover a wavelength range of 350--2480\,nm and are corrected for the Galactic dust extinction. The DR3 dataset consists of the XSL DR2 data and spectra of 20 archival M-dwarf stars. 

We covered an extensive quality characterisation. We compared synthetic broadband colours, goodness of fit described by the $D$-statistic, and various absorption line indices of stars in common between XSL and four other libraries (MILES, CaT library, NGSL, and (E-)IRTF). We used Gaia EDR3 published BP and RP photometry of the XSL stars to compare with the colours measured from the DR3 spectra. Furthermore, we compare the XSL DR3 Galactic dust extinction values with those from L20 study and from \citet{Bayestar2019} 3D extinction maps. We find the following results: 
\begin{enumerate}
    \item The overlap areas (545--590\,nm and 994--1150\,nm) of X-shooter arms are affected by the dichroic artefact, which is not described by the error spectrum.
    The severity varies from spectrum to spectrum and is worse between UVB and VIS than between VIS and NIR. These areas in the spectrum should be used with extreme caution.
    
    \item We compared our extinction values with those from L20. We find a good match for warm ($T_\mathrm{eff} > 4000\,\mathrm{K}$) stars, with an rms scatter of 0.12\,mag. Furthermore, we found extinction values for 182 XSL stars that are situated in the mapped regions of \citet{Bayestar2019} and found the rms scatter of the differences to be 0.29\,mag. The comparison of the three methods points towards an uncertainty of 0.20\,mag for stars warmer than 4000\,K.
    
    \item We find an excellent agreement between the $(\mathrm{BP-RP})$ colours measured from the XSL DR3 spectra and from \citet{GaiaEDR3}, with zero median offset and an rms scatter of 0.037\,mag for warm ($T_\mathrm{eff} > 4000\,\mathrm{K}$) stars.
    
    \item We find good agreement between observed artificial broad-band colours of XSL DR3 and of MILES or NGSL spectra. When comparing 180 spectra in common with MILES, we find insignificant offsets of $0.001$/$-0.001$/$-0.004$\,mag and small rms scatters of $0.040$/$0.045$/$0.028$~mag in $(\mathrm{box1-box2})$, $(\mathrm{box2-box3}),$ and $(\mathrm{box3-box4})$ colours, respectively. We get similar values when comparing with 116 spectra in common with NGSL: offsets of $0.017$/$0.007$/$-0.006$\,mag and rms scatters of $0.022$/$0.023$/$0.018$\,mag in $(\mathrm{box1-box2})$, $(\mathrm{box2-box3}),$ and $(\mathrm{box3-box4})$. We note that in the UVB arm, the $(\mathrm{box1-box2})$ colours have a larger median offset of 0.017\,mag, but such offset is not seen in the MILES colours. The scatter in all colours in the comparison with NGSL are smaller than in the comparison with MILES, as we were comparing reddened spectra, which eliminates uncertainties arising from the extinction correction.
    
    \item We describe the goodness of fit between spectral libraries by a normalised rms deviation (the $D$-statistic). In the majority of cases, the spectra of the stars in common between XSL and MILES, NGSL and (E-)IRTF demonstrate a match better than 5\%: 144 out of 180 spectra in common with MILES, 110 out of 116 for NGSL, and
    55 out of 71 spectra for the (E-)IRTF library. The $1\sigma$ flux calibration errors of XSL correspond to $D\simeq0.03$, but this statistic is also sensitive to low flux levels and noise.
    
    \item We show a variety of absorption line indices and compare them with the values measured spectra in common with the MILES, CaT, and (E-)IRTF libraries. Overall, the indices match well. Comparisons with MILES show some systematic trends, but similar trends with MILES are also seen in \citet{LEMONY}.

\end{enumerate}

The X-shooter Spectral Library aims to be a benchmark stellar library in the optical and NIR spectral regions. The combination of high resolution, a large number of stars, and the 350--2480\,nm wavelength range makes XSL a useful tool for the optical and NIR studies of intermediate and old stellar populations. XSL emphasises cool stars, with close to 200 M-type stars, which tend to be scarce in stellar libraries but are critical in stellar population studies in the NIR.

\begin{acknowledgements}
AA acknowledges funding from the European Research Council (ERC) under the European Unions Horizon 2020 research and innovation programme (grant agreement No. 834148).
RFP acknowledges financial support from the European Union’s Horizon 2020 research and innovation programme under the Marie Sk\l{}odowska-Curie grant agreement No. 721463 to the SUNDIAL ITN network. A.V. and J.F-B acknowledge support through the RAVET project by the grant PID2019-107427GB-C32 from the Spanish Ministry of Science, Innovation and Universities (MCIU), and through the IAC project TRACES which is partially supported through the state budget and the regional budget of the Consejer\'ia de Econom\'ia, Industria, Comercio y Conocimiento of the Canary Islands Autonomous Community. PC acknowledges support from Conselho Nacional de Desenvolvimento Cient\'ifico e Tecnol\'ogico (CNPq) under grant 310041/2018-0 and from Funda\c{c}\~{a}o de Amparo \`{a} Pesquisa do Estado de S\~{a}o Paulo (FAPESP) process number 2018/05392-8. PSB acknowledges the financial support from the Spanish National Plan for Scientific and Technical Research and Innovation, through the grant PID2019-107427GB-C31. L.M. thanks FAPESP (grant 2018/26381-4) and CNPQ (grant 306359/2018-9) for partial funding of this research. 
\end{acknowledgements}


\bibliographystyle{aa} 
\bibliography{refer} 

\clearpage
\begin{appendix} 
\section{Stellar parameters of the additional M dwarfs}

We present the adopted stellar parameters, as determined in Sect.~\ref{Sec:Mdwarfs}, for the 20 M dwarfs added to XSL in Table~\ref{tab:MD_parameters}.

\begin{table}[!h]
    \centering
    \caption{20 M-dwarf stars added to XSL and their stellar parameters.}
    \begin{tabular*}{0.75\textwidth}{llrrrrr}
    \hline\hline
         \multicolumn{1}{c}{XSL}  & \multicolumn{1}{c}{Star} & \multicolumn{1}{c}{RA} & \multicolumn{1}{c}{DEC}  & \multicolumn{1}{c}{$T_\mathrm{eff}$} & \multicolumn{1}{c}{$\log g$} & \multicolumn{1}{c}{$[\mathrm{Fe} / \mathrm{H}]_\mathrm{e}$}  \\
         \multicolumn{1}{c}{ID} & \multicolumn{1}{c}{(HNAME)} & \multicolumn{1}{c}{(J2000)} & \multicolumn{1}{c}{(J2000)}  & \multicolumn{1}{c}{(K)} & \multicolumn{1}{c}{($\log \mathrm{cm\,s^{-2}}$)} & \multicolumn{1}{c}{(dex)}  \\
         \hline
MD001 & LP 731-58  & 10:48:13.14 & $-$11:20:25.4  & 2900   &5.38   &0.22  \\
MD002 & LTT 2544  & 06:24:04.23 & $-$45:56:14.0 &       3585   &4.78   &$-$0.23  \\
MD003 & HD 50281B & 06:52:17.77 & $-$05:11:23.3  &      3447   &4.84   &$-$0.39  \\
MD004 & * 18 Pup B & 08:10:34.31 &      $-$13:48:51.2  &        3468   &4.84   &$-$0.50  \\
MD005 & L 442-13 & 02:54:03.05 &        $-$35:54:54.8 & 3336   &4.87   &$-$0.31 \\
MD006 & G 272-119 &     01:54:21.38 & $-$15:43:50.50 & 3420 &           4.85 &$-$0.46 \\
MD007 & * omi02 Eri C & 04:15:19.91 &   $-$07:40:01.3  &        3018   &4.93   &$-$0.14 \\
MD008 & G 83-29 &       04:39:43.27 & +09:51:44.3         &     3346   &4.87   &$-$0.65  \\
MD009 & HD 31412B  & 04:55:54.52 &      +04:40:16.1       &     3472   &4.77   &$-$0.45  \\
MD010 & LP 659-4 & 05:58:17.37 &        $-$04:38:01.8     &     3012   &4.99   &$-$0.14  \\
MD011 & G 102-4  & 05:28:56.63& +12:31:52.9       &     3191   &4.88   &$-$0.40  \\
MD012 & HD 38529B & 05:46:19.22 &       +01:12:48.9   & 3428   &4.86   &$-$0.30  \\
MD013 & G 106-36  &06:17:10.54 &        +05:07:07.3       &     3262   &4.89   &$-$0.53  \\
MD014 & HD 46375B & 06:33:12.26 &       +05:27:54.6       &     3457   &4.82   &$-$0.28  \\
MD017 & GJ 768.1 B &     19:51:01.10&   +10:24:36.8  &  3275   &4.87   &$-$0.38  \\
MD018 & G 143-35 &20:11:12.94&  +16:11:11.1       &     3145   &4.91   &$-$0.04  \\
MD020 & LP 703-44& 23:41:45.13& $-$05:58:16.7     &     3255   &4.87   &$-$0.41  \\
MD021 & V* V645 Cen&    14:29:35.66&    $-$62:40:34.0   &       2900   &5.14   &$-$0.02  \\
MD022 & HD 125455B  &14:19:35.21&       $-$05:09:07.4     &     3144   &4.90   &$-$0.42  \\
MD023 & HD 115404B &13:16:52.23&        +17:00:59.5 &   3763   &4.77   &$-$0.62  \\
         \hline
    \end{tabular*}
    
    \label{tab:MD_parameters}
\end{table}
\clearpage

\section{XSL DR3 primary header keywords}

We present the FITS primary header keywords in Table~B.1.

\begin{table}[!h]
\caption{XSL keyword dictionary (primary header).}
\label{table:key_dico}
{\centering
\resizebox{2\linewidth}{!}{%
\begin{tabular}{l l l l  c}
\toprule\toprule
General topic & Data reduction step                     & Keyword &     Description  & Default value\\
\toprule
\multirow{3}{2.5cm}{General information} & & OBJECT &  Machine-readable name of the star & \\
& & HNAME &  Human-readable name of the star & \\
& & XSL\_ID & X-shooter Spectral Library unique identifier & \\
& & PROV\textit{i} & Originating raw science file(s) & \\
\midrule
\multirow{12}{2.5cm}{Data reduction} & 1D extraction & EXT\_AVG &       Two 1D-spectra averaged  & F \\
\cmidrule{2-5}
&  Response curve & CAL\_STAR & Spectro-photometric standard star target & \\
& & CAL\_RAW\textit{i} & Originating raw flux-standard file(s) & \\
\cmidrule{2-5}
& First flux calibration        & FLUX\_COR &   First flux calibration applied & F \\
\cmidrule{2-5}
& Flux-loss correction  & LOSS\_COR &   Correction for slit-losses applied in DR2 & F \\
&       & SPL\_COR &    Correction for slit-losses with a spline function applied in DR3  & F \\
& & STA\_RAW\textit{i} & Originating raw wide-slit file(s) & \\
\midrule
\multirow{15}{2.5cm}{Characterization and quality assurance}  & Rest-frame correction & REST\_COR & Correction to rest-frame applied & T\\
&                      & REST\_UVB & UVB $cz$ values$^{(1)}$ [in $\mathrm{km~s^{-1}}$]  & \\
&                      & REST\_VIS &VIS $cz$ values$^{(1)}$ [in $\mathrm{km~s^{-1}}$]  & \\
&                      & REST\_NIR &NIR $cz$ values$^{(1)}$ [in $\mathrm{km~s^{-1}}$]  & \\
\cmidrule{2-5}
&   Barycor values & BARY\_COR & Barycentric radial velocity correction value & \\ 
\cmidrule{2-5}
&   Merging UVB, VIS, NIR spectra & S\_U\_VAL & UVB scaling factor value & \\ 
&& S\_U\_ORI & UVB scaling factor reference &\\
&& S\_N\_VAL & NIR scaling factor value &\\
&& S\_N\_ORI & NIR scaling factor origin reference &\\
&& ARM\_ZERO & DR2 arm spectrum missing (zeros) & F\\
\cmidrule{2-5}
&   Galactic dust extinction & AV\_VAL & Total extinction in V & \\ 
&& AV\_ORI & AV origin reference (interp.;interp/spline;reference) & \\
\cmidrule{2-5}
& Quality flags &  SNR  &               Median signal-to-noise ratio & \\
\cmidrule{3-5}
& & WAVY\_UVB &         Wavy spectrum between 460 and 520 nm & F\\
& & &   (suspected residuals of the blaze function) & \\
\cmidrule{3-5}
& & HAIR\_VIS & Some narrow spikes in the VIS spectrum & F\\
& & NOIS\_VIS & Noisy VIS spectrum & F \\
& & WAVY\_VIS & Wavy in some part of the VIS spectrum  & F       \\
\cmidrule{3-5}
& & HAIR\_NIR & Some narrow spikes in the NIR spectrum & F \\
& & WAVY\_NIR & Wavy in some part of the NIR spectrum  & F       \\
\bottomrule
\end{tabular}
}
\\
} 
Notes to Table\,\ref{table:key_dico}. (1) The $cz$ values are barycentric.
\end{table}
\FloatBarrier
\clearpage

\section{Comments about individual objects or spectra}
\label{app:comments_stars}
In XSL DR3, we chose to include spectra of peculiar objects and some spectra with known observational artefacts, because they may still be useful to some users. We list the most obvious cases in Table~\ref{tab:peculiar} and Table~\ref{tab:peculiar2}. Many of these spectra should not be used in standard population synthesis applications. 


\begin{table}[!h]
\caption[]{Peculiar stars, abnormal spectra, or zero spectrum.}
\label{tab:peculiar}
\centering
\resizebox{2\linewidth}{!}{
\begin{tabular}{l| lll}
\toprule\toprule
Category & Spectrum (XSL\_ID) & Star (HNAME) & Comment \\ 
\toprule
\multirow{16}{1.5cm}{Peculiar}  & X0085, X0133 & HD\,96446 & He-rich \\
&X0116 & HD\,57060 & P-Cygni emission in H$\alpha$ \& HeI (1.083$\mu$m); known eclipsing binary \\
&X0190 & OGLEII\,DIA\,BUL-SC19\,2332 & O-rich mira, possible TiO emission near 1.25\,$\mu$m \citep{LanconIAU18} \\
&X0214, X0248 & HD\,190073 & Herbig Ae/Be star, numerous emission lines, \\
  & & & continuum contribution from cool component in the NIR arm \\
&X0357 & CD-69\,1618 & He-rich \\
&X0420 & SV*\,HV\,11223 & S-star signatures confirmed \citep[ZrO, 940\,nm; cf.][]{WBF83} \\
&X0424 & CL*\,NGC\,6522\,ARP\,3213 & Superimposed stars with very different velocities: \\
&& & One dominant in the UVB arm  ($cz \simeq -198\,\mathrm{km\,s^{-1}}$), the other \\
 && & in the NIR ($cz \simeq +30\,\mathrm{km\,s^{-1}}$). The UVB velocity is used for the \\
 && & barycentric correction in all arms.\\
&X0513 & CL*\,NGC\,419\,LE\,35 & C-star with contamination by nearby hotter star below 450\,nm \\   
&X0653 & HD\,172488 & Possibly He-rich \\
&X0765 & SV*\,HV\,11366 & S-star signatures confirmed \citep[ZrO, 940\,nm; cf.][]{WBF83} \\
&X0798 & SHV\,0527122-695006 & M giant with contamination by nearby hot star below 450\,nm \\
\toprule
\multirow{8}{1.5cm}{Spectral type discussion}  &X0478, X0675 & V874\,Aql & O-rich LPV (not C-star as was stated in \citealt{Nassau57}) \\
&X0587 & IRAS\,10151-6008? & O-rich star (not C-star as assumed in \citealt{Whitelock06}). \\
&& & We suspect Whitelock et al. observed 2MASS\,10165028-6023549, \\
&& & while we observed 2MASS\,10165173-6023466.\\
&X0660 & OGLE\,204664c4? & G8V/K0V star. This name was taken from \cite{Zoccali08}, \\
&& & but does not directly relate to the OGLE survey catalogs. \\
&& & Prefer Gaia DR2 4049054206130689024. \\
&X0876 & TU\,Car & O-rich LPV (as already suggested in \citealt{Aaronson89}), not C-star \\
\toprule
\multirow{15}{1.5cm}{Abnormal} & X0061 & SHV\,0525478-690944 & C star. NIR poor
below 1.3\,$\mu$m. \\
 &X0085 & HD\,96446 & Wavy NIR continuum (prefer X0133) \\ 
&X0144 & OGLEII\,DIA\,BUL-SC26\,0532 &  A few artificial waves in the VIS arm \\
& & & (residuals from stripes in raw images)\\
&X0196 & HD\,194453  & K-band continuum wavy \\
&X0208 & HD\,179821 & K-band continuum wavy \\
&X0214 & HD\,190073 & K-band continuum wavy (prefer X0248)\\
&X0221 & HD\,175640  & K-band continuum wavy \\ 
&X0237 & SHV\,0510004-692755 & A few artificial waves in the VIS arm, as for X0144. \\
&X0269 & HD\,163346 &  A few artificial waves in the VIS arm, as for X0144. \\
&X0304 & HD\,37828 & Continuum poor between 1 and 1.4\,$\mu$m (prefer X0097) \\ 
&X0338 & R~Cha & Slightly affected by saturation in the $H$-band [1600--1700 nm] \\
& & & and $K$-band  [2130--2230 nm] (NIR arm) \\
& X0450 & HD~166991 & Artifacts in $K$-band (NIR arm) \\
&X0758, X0766, X0790 & Feige\,110 & S/N ratio poor in NIR, use average \\
& X0778 & HD~11397 & Artifacts in $K$-band (NIR arm) \\
& X0834 & HD~42143 & Bad telluric correction (O$_2$) in $A$-band (VIS arm) \\
& X0878 & HD~83632 & Artifacts in $H$-band (NIR arm) \\
\toprule
\multirow{8}{1.5cm}{Stellar param.$^{(1)}$} 
 &X0156 & GJ\,644C & From L20 \\
 &X0216, X0278 & HD\,216219 & From \citet{Cenarro2007} \\ 
 &X0243 & HD\,128801 & From \citet{Cenarro2007} \\ 
 &X0418 & HD\,2796 & From \citet{Cenarro2007} \\ 
 &X0430 & HD\,213042 & From \citet{Cenarro2007} \\ 
 &X0452 & HD\,179315 & From\citet{Luck2011} \\ 
 &X0754, X0755 & CL*\,NGC\,330\,ROB\,A3 & From \citet{Hill1999} \\ 

\toprule
\bottomrule
\end{tabular}
}
\end{table}
\newpage \noindent Table~\ref{tab:peculiar} is Table B.1. of \citet[]{DR2} with supplemental information. The note to Table \ref{tab:peculiar} states the following. (1) By default, we assumed stellar parameters from \cite{Arentsen2019}, \citet{Gonneau2017}, or those determined in this work. Exceptions are listed here.

\begin{table*}[!h]
\caption[]{Zero spectrum.}
\label{tab:peculiar2}
\centering
\resizebox{0.75\linewidth}{!}{
\begin{tabular}{l| lll}
\toprule\toprule
Category & Spectrum (XSL\_ID) & Star (HNAME) & Comment \\ 
\toprule
\multirow{15}{1.5cm}{Partially zero spectrum} 
& X0020 & ISO-MCMSJ005714.4-730121 & UVB \\
& X0029 &  ISO-MCMS J005304.7-730409 & UVB \\
& X0036 &  HD 39801 & NIR \\
& X0052 &  T Cae & NIR \\
& X0100 & SHV 0528537-695119 & UVB \\
& X0127 & CD-31 4916 & NIR \\
& X0130, X0131 & LHS 2065 & UVB \\
& X0145 & OGLEII DIA BUL-SC41 3443 & UVB \\
& X0150 & HD 101712 & NIR \\ 
& X0154 & BMB 286 & UVB \\      
& X0209 & GJ 752B & UVB \\
& X0233 & HD 165438 & NIR \\
& X0234 & IRAS 15060+0947 & NIR \\
& X0253 & OGLEII DIA BUL-SC22 1319 & UVB \\
& X0296 & OGLEII DIA BUL-SC13 0324 & UVB \\
& X0312 & HD 52005 & NIR \\
& X0316 & HD 76221 & NIR \\
& X0321 & HD 99648 & NIR \\
& X0330 & HD 35601 & NIR \\
& X0332 & BS 4463 & NIR \\
& X0336 & HD 63302 & NIR \\
& X0337 & RU Pup & NIR \\
& X0339 & HD 82734 & NIR \\
& X0340 & Y Hya & NIR \\
& X0341 & EV Car & NIR \\
& X0342 & BS 3923 & NIR \\
& X0343 & BS 4104 & NIR \\
& X0344 & CPD-5703502 & NIR \\
& X0353, X0605, X0606, X0635 &  [ABC89] Cir18 & UVB \\
& X0516 & SHV 0502469-692418 & UVB \\
& X0544 & SHV 0527072-701238 & UVB \\
& X0574 & [ABC89] Pup42 & UVB \\
& X0612 & HD 104307 & NIR\\
& X0653 & HD 172488 & UVB \\
& X0704 & NGC 6838 1037 & NIR \\
& X0759 &  HV\,12149 & UVB \\
& X0784 & SHV 0448341-691510 & NIR \\
& X0812 & SHV 0531398-701050 & NIR \\
& X0818, X0820 & SHV 0535237-700720& NIR \\
& X0830 & HD 36395 & NIR \\
& X0892 & LHS 318 & NIR \\      
\bottomrule
\end{tabular}
}
\end{table*}
\FloatBarrier
\clearpage
\onecolumn
\begin{multicols}{2}
\section{Examples of spectra in common between different spectra libraries}

In this appendix, we present comparison of the XSL DR3 spectra with interpolated XSL, MILES, NGSL, and (E-)IRTF spectra.

\subsection{Comparison with the interpolated XSL spectra}

\begin{figure*}[!hb]
\centering
 \subfigure{\includegraphics[width=0.85\textwidth]{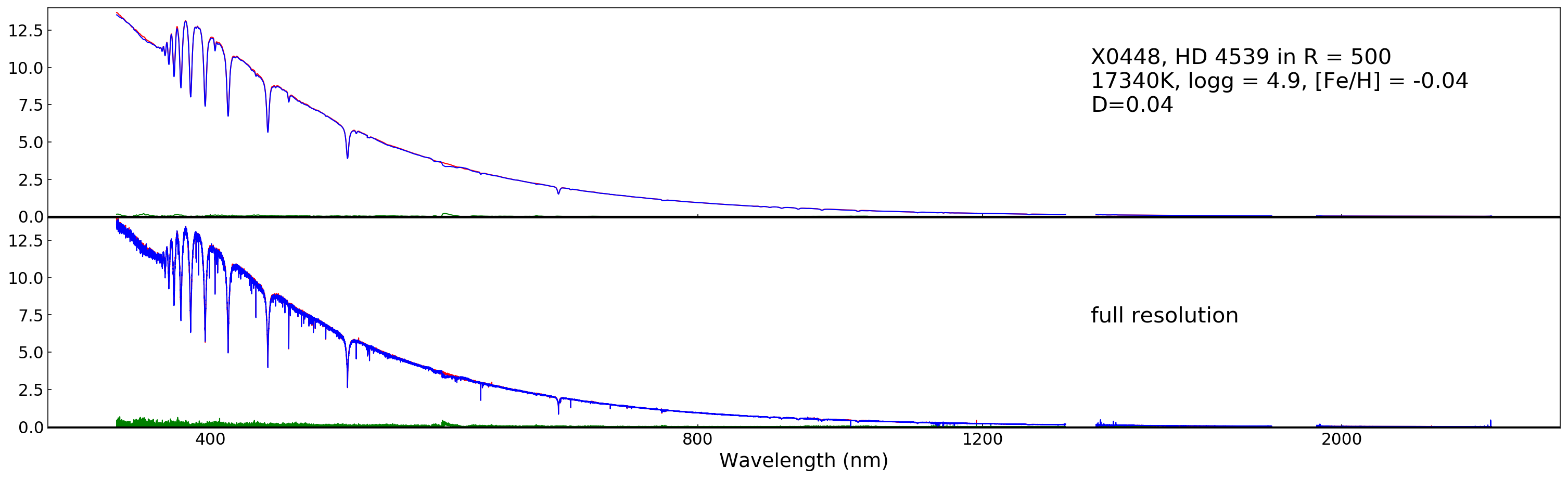}}
 \subfigure{\includegraphics[width=0.85\textwidth]{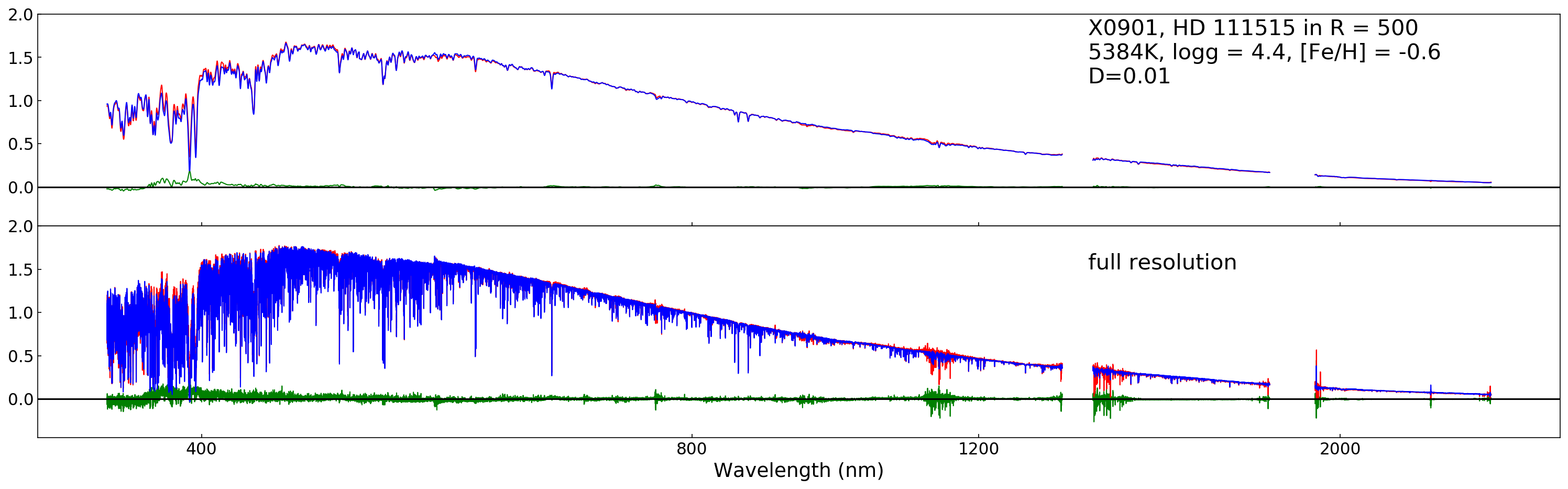}}
 \subfigure{\includegraphics[width=0.85\textwidth]{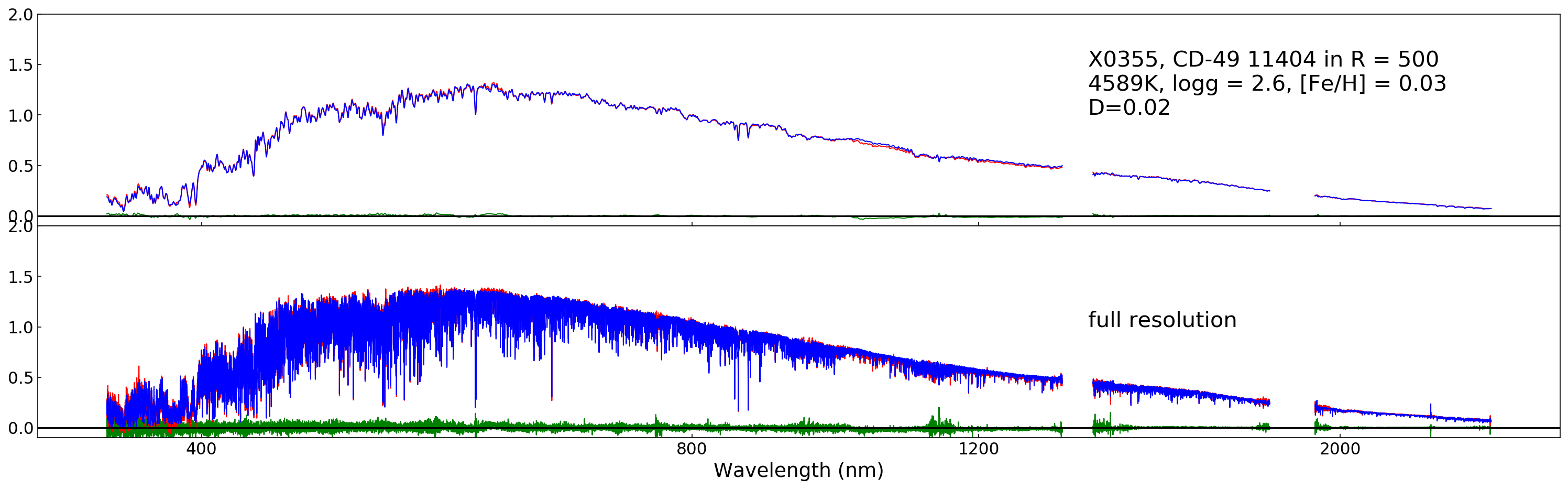}}
 \subfigure{\includegraphics[width=0.85\textwidth]{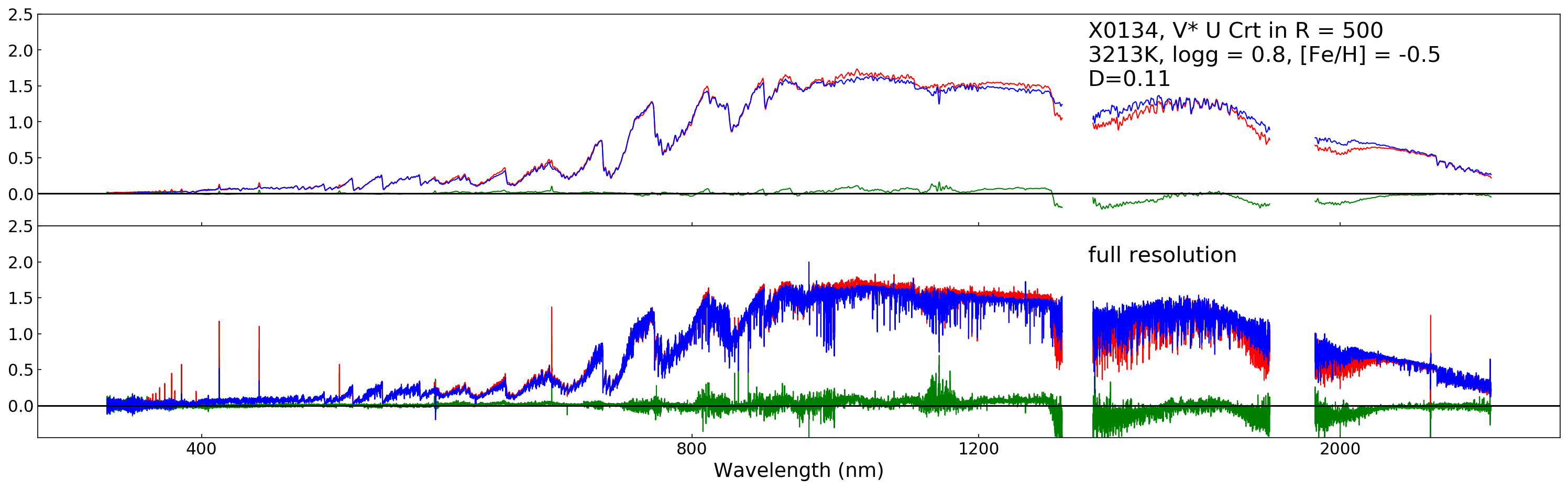}}
 \caption{Comparisons between XSL spectra (red) and interpolated spectra (blue). }
\label{fig:interp_xsl_spectra}
\end{figure*}

\begin{figure*}[!h]
\centering     
\subfigure{\includegraphics[width=0.95\textwidth]{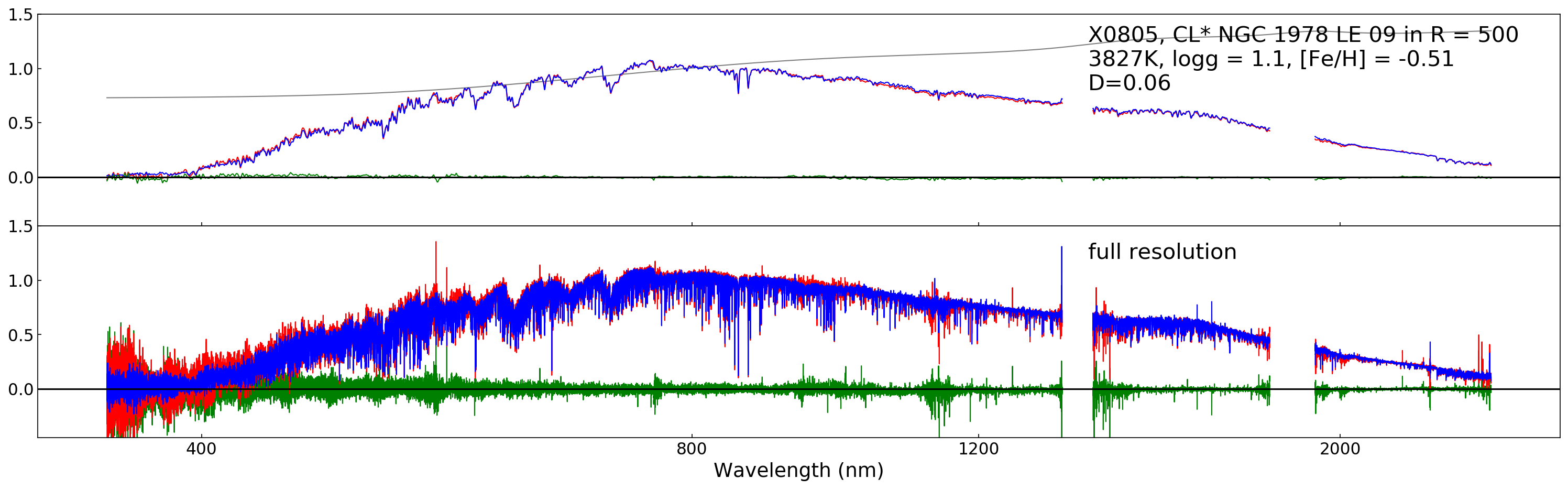}}
\subfigure{\includegraphics[width=0.95\textwidth]{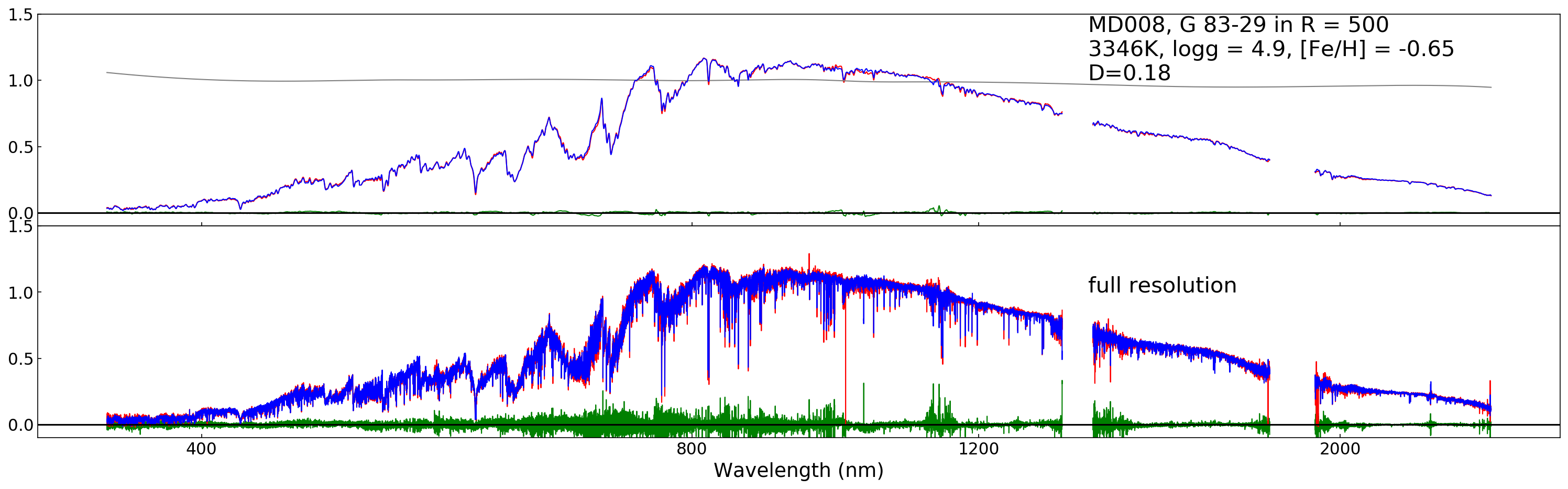}}
\subfigure{\includegraphics[width=0.95\textwidth]{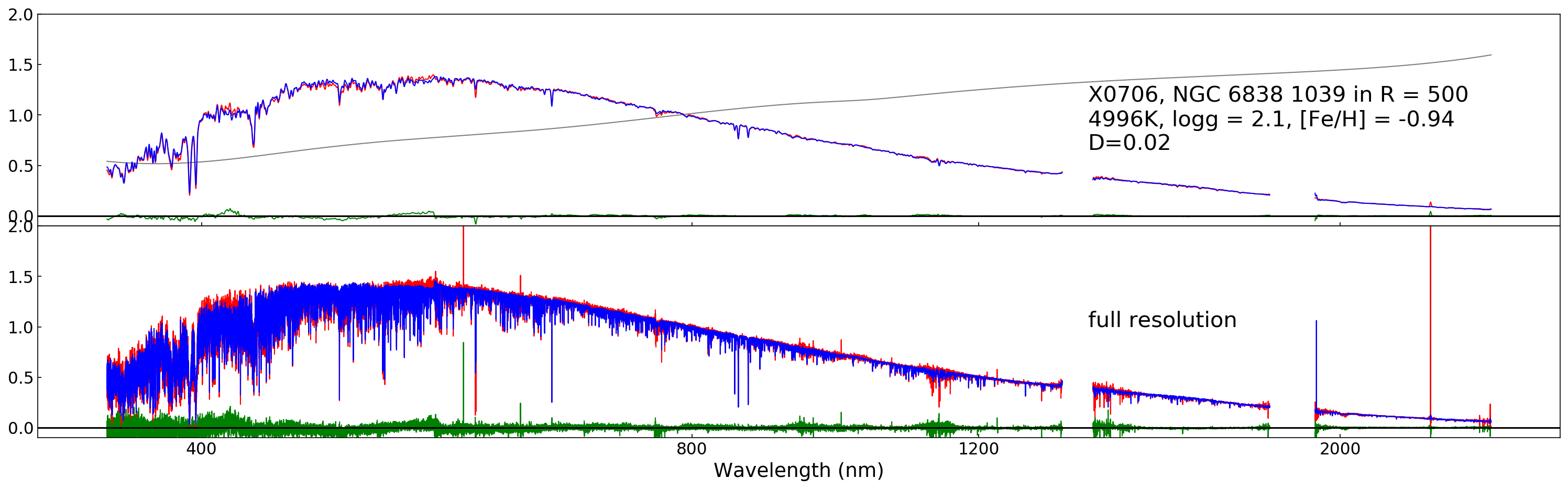}}
\subfigure{\includegraphics[width=0.95\textwidth]{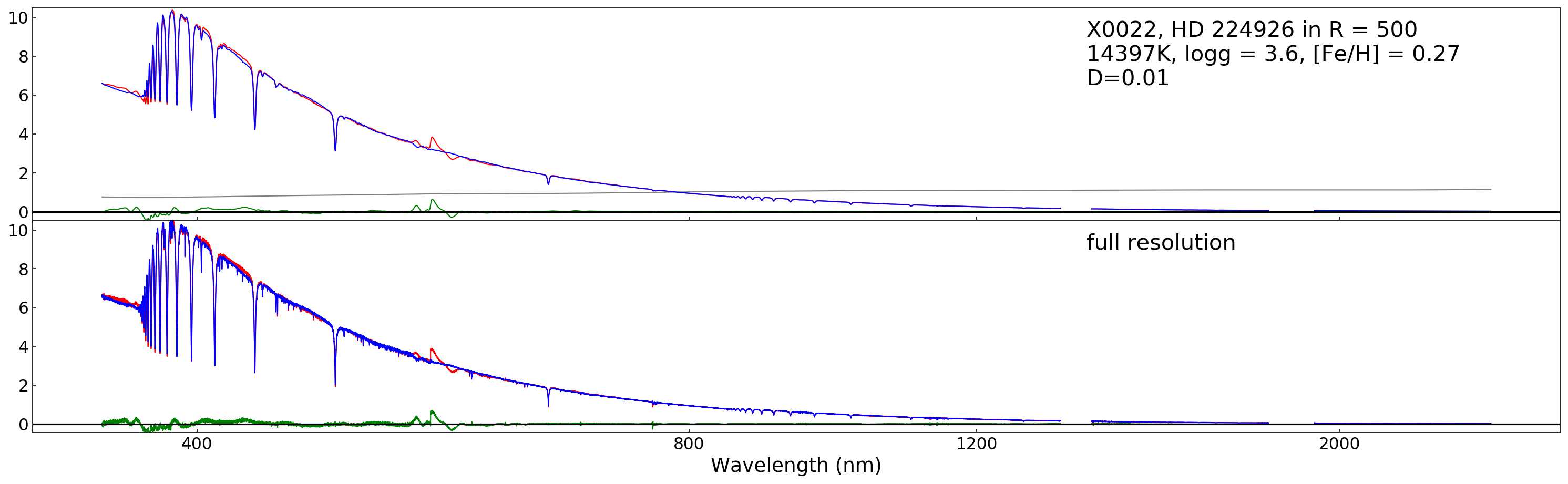}}
\caption{Comparisons between XSL spectra (red) and interpolated spectra (blue). These stars are corrected for slit flux loss and Galactic dust extinction with a cubic spline function (grey). }
\label{fig:interp_xsl_spectra_slitcorr}
\end{figure*}

Here, we present a comparison of observed XSL DR3 spectra with interpolated XSL spectra with the same stellar parameters, before (Fig.~\ref{fig:interp_xsl_spectra}) and after (Fig.~\ref{fig:interp_xsl_spectra_slitcorr}) slit loss and extinction correction.
\end{multicols}

\clearpage
\begin{multicols}{2}
\subsection{Comparison between XSL and NGSL stars}

Here, we present a comparison of observed XSL DR3 spectra with NGSL spectra of the same stars.
\end{multicols}
\begin{figure}[!h]
\centering     
\subfigure[]{\includegraphics[width=0.95\textwidth]{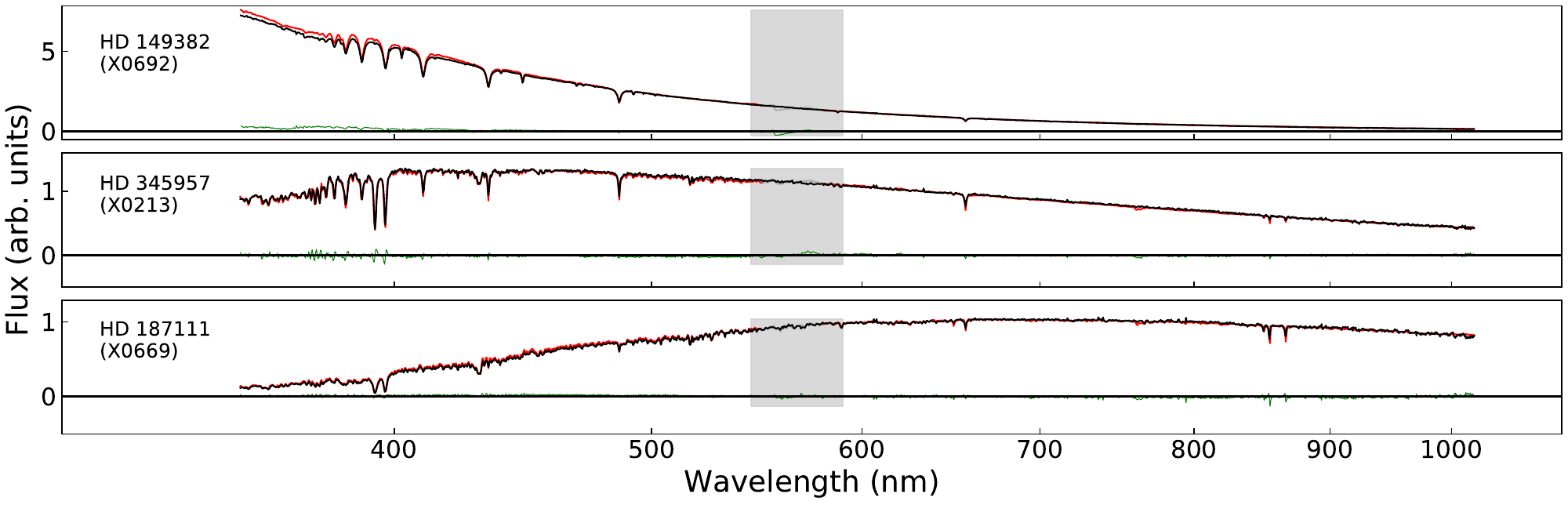}}
\subfigure[]{\includegraphics[width=0.95\textwidth]{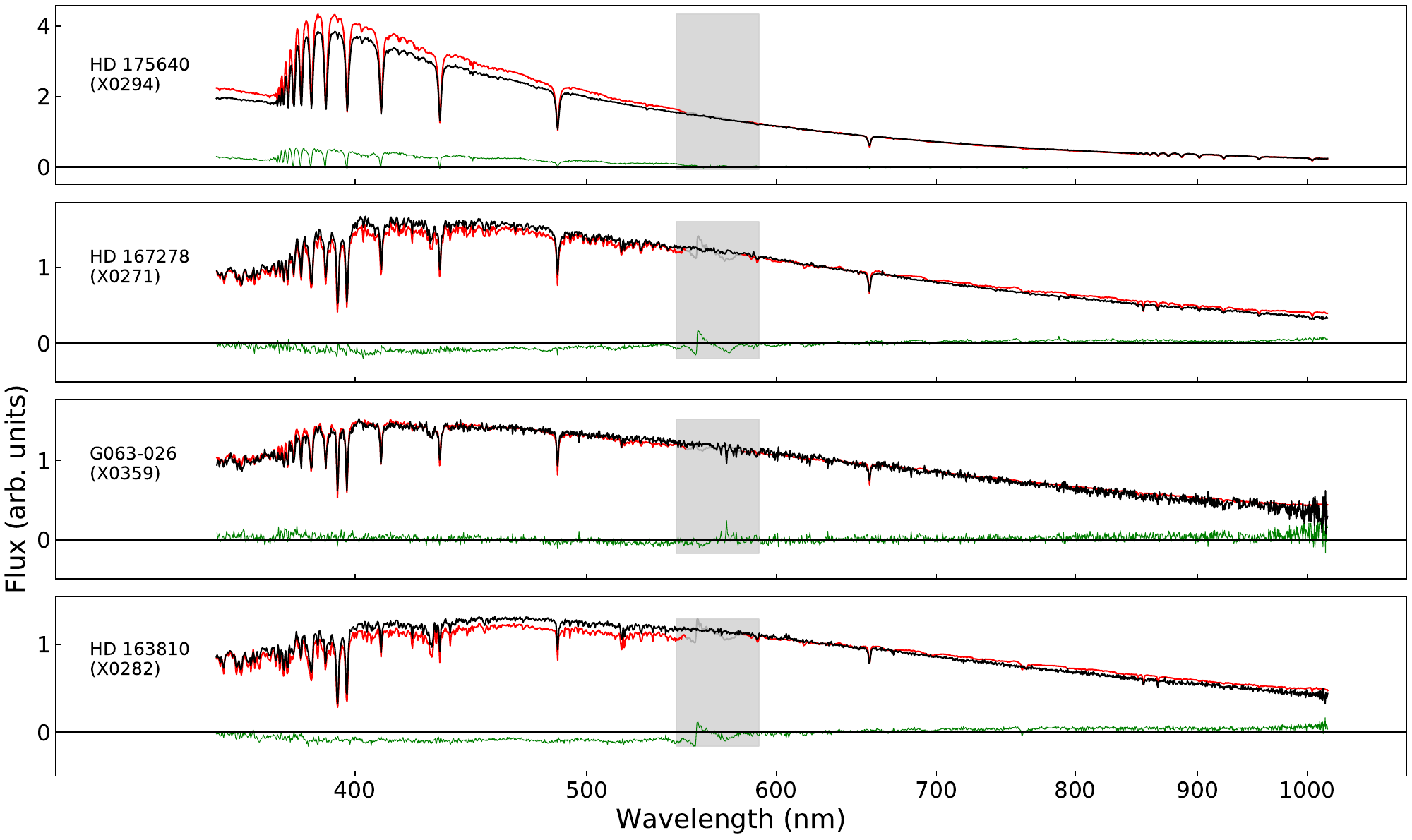}}

\caption{(a) Three stars in common between XSL (red) and NGSL (black), smoothed to the NGSL resolution. \textit{Top}: HD 149382. \textit{Middle:} HD 345957. \textit{Bottom:} HD 187111. Residual spectra are in green.
(b) Four stars that have the largest mismatch between NGSL (black) and XSL (red). The grey shaded areas on both sub-figures mark the UVB and VIS merging regions with possible dichroic contamination.}
    \label{fig:NGSLstars}
\end{figure}
\FloatBarrier
\clearpage
\begin{multicols}{2}

\subsection{Comparison between XSL and (E-)IRTF stars}

Here, we present a comparison of observed XSL DR3 spectra with (E-)IRTF spectra of the same stars.
\end{multicols}
\begin{figure}[!h]
    \centering
    \subfigure[]{\includegraphics[width=0.9\textwidth]{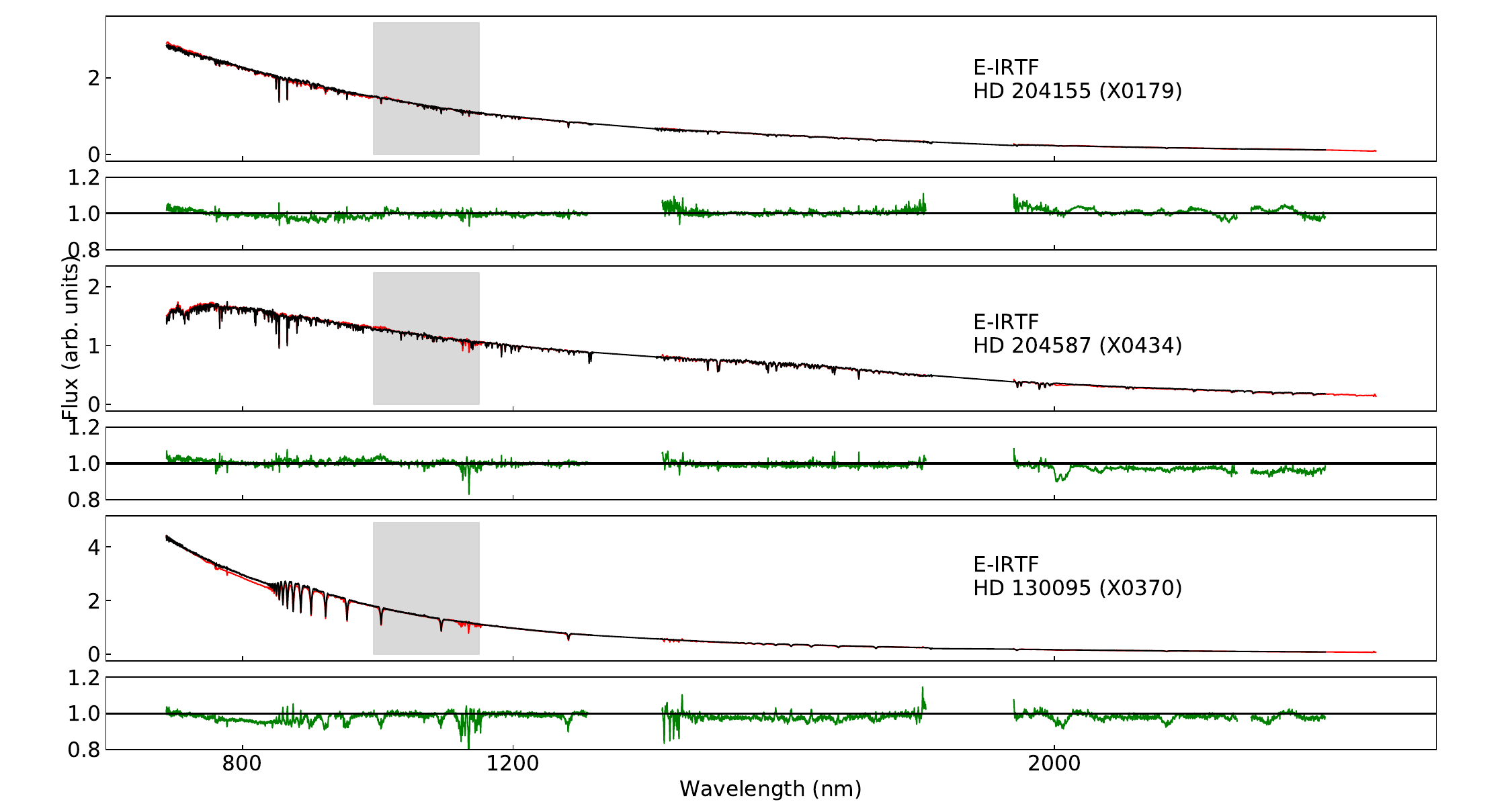}}
    \subfigure[]{\includegraphics[width=0.9\textwidth]{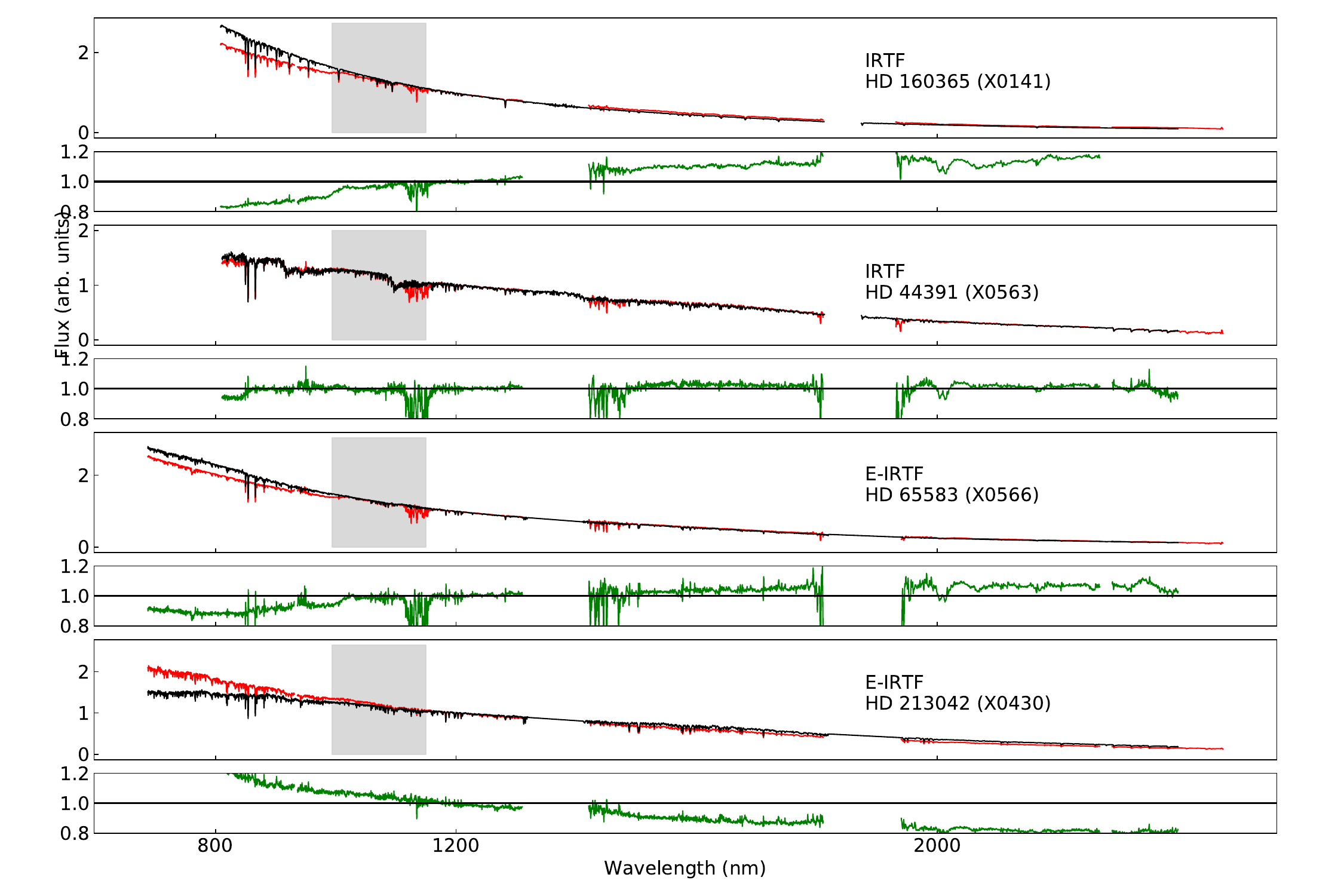}}
    \caption{(a) Three stars in common with XSL (red) and IRTF (black), smoothed to IRTF resolution. (b) Four stars that have the largest mismatch between the IRTF and XSL. In this case, green represents the XSL/IRTF ratio.}
    \label{fig:IRTFstars}
\end{figure}
\FloatBarrier
\clearpage
\begin{multicols}{2}
\subsection{Common stars in XSL, MILES and IRTF stars}

Finally, we present a comparison of four observed XSL DR3 spectra with MILES and IRTF spectra of the same stars.
\end{multicols}
\begin{figure}[!h]
    \centering
    \includegraphics[width = 0.85\textwidth]{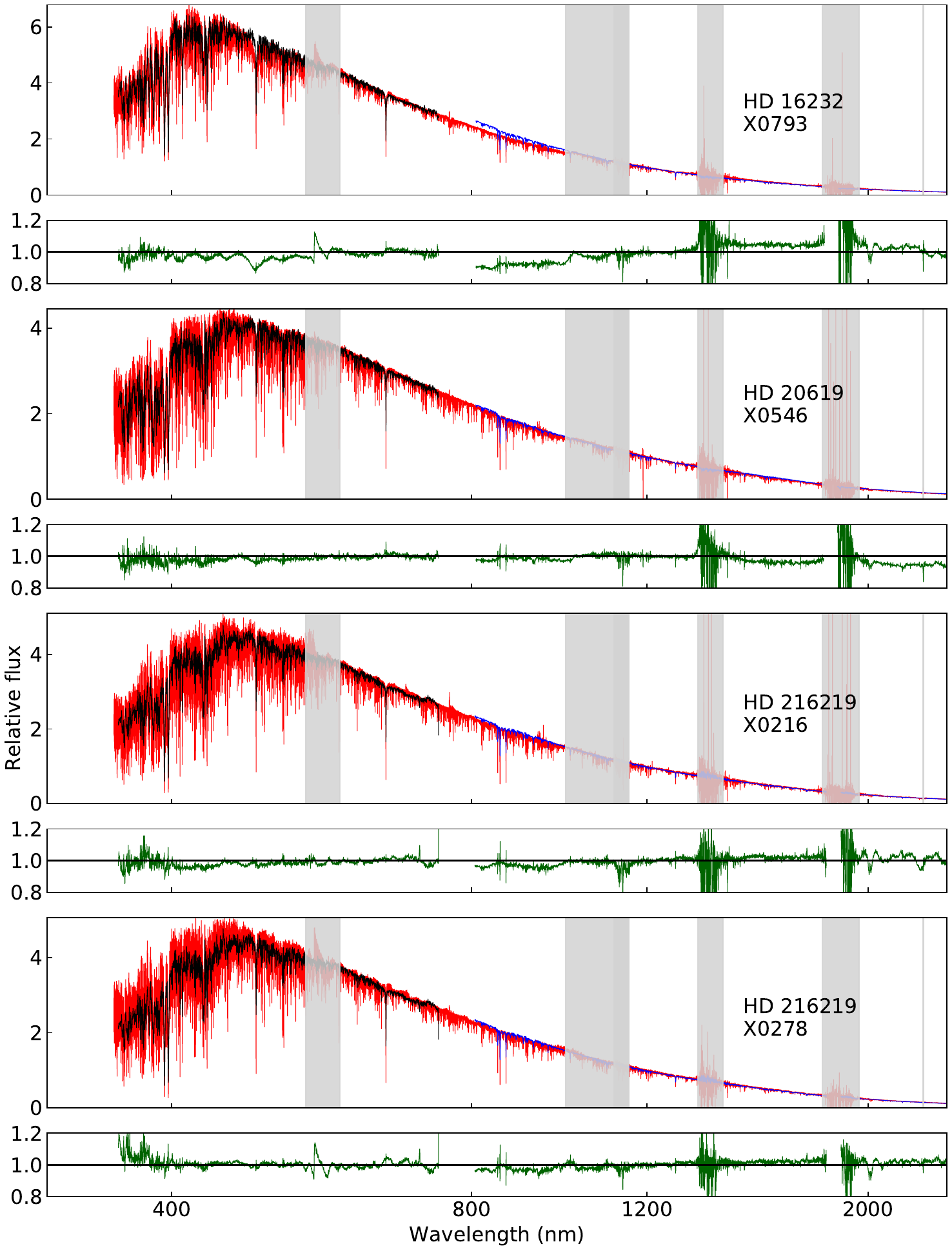}
    \caption{Four spectra in common between MILES (black), IRTF (blue) and XSL (red). Below each comparison, we plot the ratio of the XSL spectra (smoothed to the appropriate resolution) to IRTF and to MILES.}
    \label{fig:MILES_IRTF_XSL}
\end{figure}{}

\end{appendix}
\end{document}